\documentclass[10pt]{article}
\usepackage[T1]{fontenc}
\usepackage[utf8]{inputenc}
\usepackage{lmodern}

\usepackage[a4paper,margin=1in]{geometry}
\usepackage{setspace}
\usepackage{amsmath,amssymb,amsthm}

\usepackage{graphicx}
\usepackage{booktabs}

\usepackage{listings}
\usepackage[authoryear]{natbib}
\usepackage[hidelinks]{hyperref}

\usepackage{float}
\usepackage{siunitx}
\usepackage{array}
\usepackage[flushleft]{threeparttable}
\usepackage{caption}
\usepackage{subcaption}
\usepackage{titlesec}

\usepackage{xcolor}

\usepackage{tabularx}
\usepackage{makecell}

\usepackage{graphicx}
\usepackage{subcaption} 

\usepackage{dblfloatfix} 
\newcolumntype{L}{>{\raggedright\arraybackslash}X}

\newcommand{\keywords}[1]{%
  \par\noindent\textbf{Keywords: }#1\par
}

\numberwithin{equation}{section}

\renewcommand{\arraystretch}{1.25}

\title{%
    Gimbal Regression\\
    \large Orientation-Adaptive Local Linear Regression under Spatial Heterogeneity
}
\author{
    Yuichiro Otani
}
\date{\today}

\begin{document}
\maketitle

\begin{abstract}
    Local regression is widely used to explore spatial heterogeneity by fitting neighborhood-specific models and mapping coefficient surfaces. In realistic spatial sampling, however, neighborhoods are often anisotropic or effectively low-dimensional, leading to ill-conditioned local normal equations and coefficient variation driven by numerical artifacts rather than substantive heterogeneity. Such failures are not reliably detected by predictive error and are frequently obscured when estimation procedures embed implicit tuning or iterative optimization without exposing local diagnostics.

\noindent
This paper proposes \emph{Gimbal Regression} (GR), a deterministic, geometry-aware local regression framework designed for stable and auditable local estimation. GR is formulated as a realized estimator map: a reproducible composition from neighborhood data to explicit geometric objects (bearing-based direction, value-based second-moment orientation, geometry-based anisotropy ratio), to a diagonal directional weight field with deterministic safeguards (one-shot effective sample size correction and uniform fallback), and finally to a closed-form local solve. Orientation quantities define a reference frame for weight evaluation only; GR does not rotate the regression design matrix and does not posit a stochastic spatial dependence model.

\noindent
Theoretical results are stated conditional on the realized neighborhood and realized branch of the piecewise weight map. Under this conditioning, the local estimator is a deterministic linear operator and admits finite-perturbation (operator-norm) stability bounds. Simulation experiments verify activation properties under controlled anisotropy and isotropy regimes and confirm no-harm behavior when directional structure is non-identifiable. Empirical applications demonstrate computational predictability, diagnostic transparency, and robust stability relative to common local regression baselines.

\noindent
GR is positioned not as a predictive replacement for model-based geostatistics or machine learning methods, but as a deterministic, interpretable, and diagnostically explicit local modeling framework whose numerical behavior is visible, reproducible, and conditionally stable.
\end{abstract}
\keywords{
    geographically weighted regression
    numerical stability
    local regression
    anisotropy
}

\setcounter{section}{0}
\newpage
\section{Introduction}
Local regression is widely used to study spatial heterogeneity by fitting a model within a neighborhood of each location and mapping the resulting coefficient surfaces. In realistic spatial sampling, however, anisotropic or effectively low-dimensional neighborhoods can make local design matrices nearly collinear, yielding ill-conditioned normal equations and coefficient surfaces driven by numerical artifacts. Such failures are not reliably detected by prediction error and may be obscured when estimation is wrapped inside tuning loops or iterative procedures that do not expose local numerical diagnostics.

This paper proposes \emph{Gimbal Regression} (GR), a deterministic, geometry-aware local regression framework designed for \emph{stable and auditable} local estimation with explicit diagnostics. GR is not a minor adjustment of anisotropic kernels or GWR-style weighting; it specifies local regression as an \emph{auditable estimator map} in which orientation selection, directional weight construction, and safeguard branches are fixed as a reproducible mapping. Intermediate geometric and numerical quantities are treated as first-class outputs.

The contributions of this paper are as follows: First, we introduce an auditable estimator map with realized branches. GR fixes the full weighting-and-safeguard procedure as a \emph{realized piecewise estimator map}, so what is analyzed and reported matches the computational object that is actually executed, rather than an idealized weighting rule. Second, we develop a geometry-as-diagnostics viewpoint rather than treating geometry as merely a kernel choice. GR elevates local geometric and numerical quantities---orientation, anisotropy, effective sample size, and conditioning---to primary outputs, enabling practitioners to locate where local estimation is well-posed versus ill-posed instead of relying on implicit optimizer behavior. Third, we target stability at the level of the local normal equations, formulating stability as finite-perturbation control of the local solve conditional on the realized weights. This emphasis makes the numerical behavior of the local normal equations explicit, instead of optimizing prediction-driven global objectives.

\subsection{Method overview}
GR is computed in a single deterministic pass at each target location $s_i$. A fixed neighborhood rule selects local data, from which GR constructs a diagonal directional weight matrix using a bearing-based orientation and a value-based orientation (local second moments). These orientations only set the reference frame for weight evaluation---they do not rotate the design matrix or define a GLS covariance model. Coefficients follow from a closed-form solve, with a one-shot effective sample size (ESS) correction and uniform fallback as deterministic safeguards; stability statements are aligned with the realized branch and realized weights.

\subsection{Stability viewpoint and interpretation}
The theory matches the estimator that is actually computed. Because weight construction can be response-dependent, we do not claim unconditional optimality. Conditional on the realized neighborhood and the realized branch of the piecewise weight map (including safeguards), the local estimator is a deterministic linear operator and admits finite-perturbation stability bounds. In practice, numerical stability is assessed from conditioning diagnostics of the realized local normal equations (e.g., the condition number), with ESS and safeguard activation as complementary indicators. Because ill-conditioned solves can make apparent coefficient variation numerically driven rather than substantive, substantive interpretation of spatial coefficient variation is restricted to locations where these diagnostics indicate a well-posed, well-conditioned local solve.

\subsection{Computational structure}
GR is computationally predictable at scale. Once the neighborhood rule and tuning constants are fixed, the estimator map requires no iterative optimization (e.g., likelihood maximization, EM loops, or backfitting). Per location, the work is dominated by neighborhood identification, one-pass weight evaluation (with one-shot ESS correction), and a small $p\times p$ solve. Because safeguard branches are explicit, computational regimes and failure modes are auditable rather than implicit artifacts of tuning.

\subsection{Paper organization}
Chapter~2 reviews related work. Chapter~3 introduces the problem setup and notation. Chapters~4--6 present the GR estimator map, its conditional stability properties, and its computational structure. Chapters~7--8 report simulation and empirical studies. Chapter~9 discusses scope and diagnostic-first interpretation. Chapter~10 concludes.
\section{Background and Positioning}
This chapter situates Gimbal Regression (GR) within the landscape of spatial modeling approaches and clarifies the specific gap it targets: stable and auditable \emph{local linear} estimation under heterogeneous and anisotropic neighborhood geometry.

\subsection{Local regression as a tool for spatial heterogeneity}
Local regression approaches fit location-specific models using observations in a neighborhood around each target location \citep{ClevelandDevlin1988,Loader1999,FotheringhamBrunsdonCharlton2002}. Their appeal is interpretability: local coefficient surfaces provide a direct description of spatially varying covariate effects under a specified locality rule \citep{BrunsdonFotheringhamCharlton1996,BrunsdonFotheringhamCharlton1998,FotheringhamYangKang2017}. In many applied settings, such surfaces are used as exploratory or diagnostic objects rather than as globally optimal predictors.

A recurring practical issue is that local estimation problems can be ill-conditioned even when neighborhoods are large. Local collinearity is often driven by geometry (e.g., neighborhoods concentrated along a corridor, river, road network, coastline, or other effectively one-dimensional structure) and by strong directional alignment between covariate variation and spatial trend terms. In these regimes, local coefficient estimates can become numerically sensitive \citep{BelsleyKuhWelsch1980,WheelerTiefelsdorf2005,Higham2002,TrefethenBau1997} and difficult to interpret.

\subsection{Directional weighting and anisotropic kernels}
Directional or anisotropic weighting is a natural response to anisotropic spatial configurations \citep{MardiaJupp2000,Fisher1993}. A common approach replaces an isotropic distance kernel by an anisotropic kernel whose level sets are oriented ellipses. Such constructions can be interpreted either as deterministic rules that redistribute influence across directions or as components of a generative dependence model when tied to covariance assumptions.

\subsection{Dependence modeling and prediction-focused alternatives}
Model-based geostatistical approaches introduce stochastic dependence through covariance structures \citep{Cressie1993,DiggleRibeiro2007,BanerjeeCarlinGelfand2014,CressieWikle2011} and provide predictors and uncertainty quantification under those assumptions. Flexible nonlinear learners can also deliver strong predictive performance, often absorbing spatial structure implicitly through features or architectures.

\subsection{Positioning of Gimbal Regression}

Gimbal Regression (GR) is positioned as a geometry-aware local regression framework that is distinct from both directional weighting heuristics and fully specified dependence-modeling approaches. Like anisotropic kernel methods, GR responds to heterogeneous and directionally structured neighborhoods by modulating local influence as a function of spatial orientation. However, unlike model-based approaches, this modulation is not derived from or interpreted as a stochastic dependence structure. Instead, GR treats directionality as a local geometric property that informs how weights are evaluated within a neighborhood while preserving the interpretability of local linear estimation.

In GR, directional information enters only through a locally oriented metric used to evaluate kernel weights, producing a diagonal weight matrix. Orientation therefore acts as a reference frame for weight evaluation, while influence redistribution occurs entirely through the diagonal weights themselves. The regression design matrix is not rotated, and the column space of the local design remains unchanged. In this sense, GR does not encode cross-location covariance or impose a generative spatial dependence model; it remains a form of weighted local linear regression whose weights reflect the realized neighborhood geometry.

GR is complementary to dependence-modeling and prediction-focused approaches. Its primary output is a set of local linear coefficient estimates accompanied by diagnostics describing the realized neighborhood geometry, the realized weight field, and the numerical behavior of the local solve. When the scientific objective is explicit dependence modeling, uncertainty quantification under a generative spatial model, or maximal predictive accuracy in highly nonlinear regimes, GR should be viewed as a diagnostic baseline or an interpretable local component rather than as a replacement for those approaches.

A central motivation for GR is that interpretability in local regression can fail when neighborhood geometry or weighting schemes produce numerically degenerate estimation problems. In such settings, local normal equations can become ill-conditioned and coefficient surfaces may reflect numerical artifacts rather than substantive spatial variation. GR therefore incorporates explicit numerical safeguards as part of the estimator definition. These include deterministic isotropy and identifiability rules for angular quantities and a one-shot effective sample size (ESS) correction that broadens weights and triggers a uniform fallback when effective information is insufficient. Because these rules are specified as part of the estimator map (Appendix~A), the realized estimator remains reproducible and diagnostically transparent.

Formulating the procedure as a deterministic estimator map has practical implications for interpretation and computation. Under a fixed neighborhood rule and fixed tuning constants, the estimator is single-valued and exactly reproducible. When weight construction depends on the response, inferential statements are made conditional on the realized neighborhood and on the realized branch of the piecewise mapping. Finally, the absence of estimator-defined iteration implies that the computational structure of the method is predictable, which supports straightforward parallel evaluation across locations.

\subsection{Summary}
The positioning above clarifies that GR is not proposed as a dependence model, nor as a prediction-optimized learner, but as a deterministic, geometry-aware extension of local linear regression designed to stabilize and audit local estimation under heterogeneous and anisotropic neighborhood structure. Its contribution lies in making neighborhood geometry, directional influence redistribution, and numerical stability explicit components of the estimator map. The following chapter formalizes this map and establishes its structural properties before turning to stability analysis and empirical evaluation.
\section{Notation and Problem Setup}

This chapter introduces the notation used throughout the paper and formalizes the local estimation setting in which Gimbal Regression (GR) is defined. The focus is the per-location local linear solve, its associated normal equations, and the diagnostic objects used to assess numerical well-posedness under heterogeneous and anisotropic neighborhood geometry.

\subsection{Observations, locations, and neighborhoods}
Let $\{(s_i,x_i,y_i)\}_{i=1}^n$ denote observations, where $s_i\in\mathbb R^d$ is a spatial location, $x_i\in\mathbb R^p$ is a covariate vector (including an intercept when appropriate), and $y_i\in\mathbb R$ is a response. For each target location $s_i$, let $\mathcal N_i\subset\{1,\dots,n\}$ be the index set of neighbors used for local estimation, with cardinality $n_i=|\mathcal N_i|$. In later implementations, $\mathcal N_i$ is a $K$-nearest-neighbor set in geographic distance, but the estimator map is defined conditional on any deterministic neighborhood rule.

\subsection{Distances, bearings, and tangent-plane displacements}
Let $d_{ij}=d(s_i,s_j)$ be a fixed distance metric. For directional constructions, neighborhood geometry is represented in a local tangent-plane coordinate system. Let $\mathbf\Delta_{ij}\in\mathbb R^2$ denote the East--North displacement from $s_i$ to $s_j$ under a fixed convention, and let $\theta_{ij}\in(-\pi,\pi]$ denote the corresponding bearing angle. When useful, distances may be expressed in normalized form using a fixed scale $u>0$:
\begin{equation}
    z_{ij}=\frac{d_{ij}}{u}.
\end{equation}
In Chapters~4--6, kernels are written in raw distance units with bandwidth parameters; the normalized form is obtained by substituting $d_{ij}=uz_{ij}$ and scaling bandwidths accordingly.

\subsection{Local design matrices and normal equations}

For each target $i$, define the local response vector and local design matrix by stacking neighborhood observations:
\begin{equation}
    \begin{aligned}
        \mathbf y_i = (y_j)_{j\in\mathcal N_i}\in\mathbb R^{n_i}, \qquad
        \mathbf X_i = (x_j^\top)_{j\in\mathcal N_i}\in\mathbb R^{n_i\times p}.
    \end{aligned}
\end{equation}
A generic diagonal weighting rule yields a diagonal, positive semidefinite matrix
\begin{equation}
    \mathbf W_i=\mathrm{diag}(w_{ij})\in\mathbb R^{n_i\times n_i},\qquad w_{ij}\ge 0,\ \ j\in\mathcal N_i.
\end{equation}
Local numerical behavior is governed by normal-equation matrices. In particular, the unweighted Gram matrix is
\begin{equation}
    \mathbf G_i=\mathbf X_i^\top \mathbf X_i,
\end{equation}
and the weighted Gram matrix is
\begin{equation}
    \label{eq:gram-wieghted}
    \mathbf G_i(\mathbf W)=\mathbf X_i^\top \mathbf W_i \mathbf X_i.
\end{equation}
\subsection{Motivating failure mode: anisotropy and local ill-conditioning}
Local regression can become numerically unstable when neighborhoods are anisotropic or effectively low-dimensional, so that columns of $\mathbf X_i$ become nearly collinear within $\mathcal N_i$ and the associated normal equations approach singularity. Two recurring drivers are geometric degeneracy and directional alignment. In geometric degeneracy, neighborhood points concentrate along a narrow corridor in the tangent plane, which reduces effective geometric support. In directional alignment, dominant local variation occurs along an oblique direction relative to the fixed coordinate representation used to form $\mathbf X_i$, which amplifies near-collinearity among regressors and trend components.

Diagonal weighting rescales rows but does not change the column space of $\mathbf X_i$. Therefore, stability must be treated as a first-order diagnostic objective: methods intended for interpretation should expose when local estimation is well-posed versus numerically fragile, rather than relying on prediction error or implicit tuning behavior.

\subsection{Estimator-map viewpoint and realized conditioning}
GR is formulated as an auditable estimator map: a deterministic composition from neighborhood data to intermediate geometric objects and then to a closed-form local estimate. The full weight construction includes identifiability checks, isotropy declarations, a one-shot ESS safeguard, and a uniform fallback, and is specified as a \emph{realized piecewise map} (Appendix~\ref{app:bearing-resultant-isotropy-detection}--\ref{app:realized-local-estimator-and-well-posedness-convention}).\\

With this notation and estimator-map viewpoint established, the next chapter constructs the orientation, metric, and weight definitions that instantiate the local estimator.
\section{Gimbal Regression Framework}
This chapter specifies Gimbal Regression as a deterministic local estimator map. The framework is designed to stabilize and diagnose local regression under spatial heterogeneity by (i) defining a locally adaptive angular reference, (ii) constructing a locally adaptive directional weight field, and (iii) solving a closed-form local system in which influence is deterministically modulated without positing a stochastic dependence model. The construction separates bearing-based orientation (a geometric reference direction derived from the neighborhood configuration) from value-based orientation (a second-moment orientation derived from observed scalar pairs). These orientations are combined only through weight evaluation; they do not rotate the regression design matrix and they are not introduced as a GLS-type covariance specification. 

\subsection{Spatial Normalization}
\label{sec:Spatial Normalization}
Let $s_i\in\mathbb R^d$ denote spatial locations and let $d(s_i,s_j)$ be a distance metric. Distances are normalized by a positive scale parameter $u>0$:
\begin{equation}
    z_{ij}=\frac{d(s_i,s_j)}{u}.
\end{equation}

This normalization removes absolute spatial scale and yields a dimensionless distance variable. All subsequent constructions may be expressed either in raw distance units or in the normalized form $z_{ij}$; the model specification is invariant up to the choice of scale. For clarity of exposition, subsequent weight and metric expressions are written in raw distance units $d_{ij}$ with bandwidth $h$; the fully normalized form is obtained by substituting $d_{ij}=u z_{ij}$ and scaling $h$ accordingly.

\subsection{Bearing-Based Orientation}
\label{sec:Bearing-Based Orientation}
Directional geometry is summarized through a local reference direction derived from the bearing field. Let $\theta_{ij}$ denote the bearing angle from location $s_i$ to $s_j$. Bearings and displacement vectors are computed in a local tangent-plane approximation (e.g., East–North coordinates), which is accurate for sufficiently small neighborhoods relative to the Earth’s radius; for larger neighborhoods, the same construction applies by replacing the approximation with geodesic bearings and geodesic displacement coordinates.

A dominant local direction $\phi_i$ is defined as the resultant direction of distance-decayed angular contributions:
\begin{equation}
    \phi_i=\arg\left(\sum_{j\in\mathcal N_i}\omega_{ij}e^{\mathrm{i}\theta_{ij}}\right),\qquad \omega_{ij}=\exp\left(-\frac{d_{ij}^2}{h^2}\right),
\end{equation}
where $h>0$ is a distance bandwidth and $d_{ij}=d(s_i,s_j)$.

The dominant direction $\phi_i$ is well-defined only when the local bearing field is directionally imbalanced. Let the (normalized) resultant length be
\begin{equation}
    r_i=\frac{\left|\sum_{j\in\mathcal N_i}\omega_{ij}e^{\mathrm{i}\theta_{ij}}\right|}{\sum_{j\in\mathcal N_i}\omega_{ij}}.
\end{equation}

If $r_i$ falls below a fixed threshold, the local bearing geometry is treated as angularly isotropic and $\phi_i$ is treated as non-identifiable. Operationally, when this isotropy condition is triggered we set $\phi_i=0$ and in subsequent metric construction we take $R(\phi_i)=\mathbf I$, i.e., the bearing-based component is deactivated and the induced weights revert to their bearing-neutral form.

The bearing-based orientation is represented by the planar rotation operator $R(\phi_i)$ and is used only as a reference direction in metric construction. It does not directly rotate the regression coordinate system in the local design matrix.

\subsection{Value-Based Orientation}
\label{sec:Value-Based Orientation}
Value-based orientation summarizes local anisotropy in observed scalar pairs through second-moment diagonalization. For any pair of locally observed scalar random variables $(a,b)$, define the rotation angle
\begin{equation}
    \theta^\ast(a,b)=\frac{1}{2}\,\operatorname{atan2}\left(\operatorname{Var}(b)-\operatorname{Var}(a),\,2\operatorname{Cov}(a,b)\right).
\end{equation}

Here, $\operatorname{atan2}(\cdot,\cdot)$ denotes the two-argument inverse tangent defined in Appendix~\ref{app:value-based-orientation}, which uniquely determines the quadrant. In implementation, when the second-moment structure is numerically non-identifiable (i.e., both arguments passed to $\operatorname{atan2}$ are below a fixed tolerance), we set $\theta^\ast=0$ according to the piecewise rule in Appendix~\ref{app:value-based-orientation}.

This angle is the rotation that diagonalizes the local second-moment matrix of $(a,b)$. The construction is not claimed to be unique; it is adopted as a minimal deterministic summary of second-order anisotropy. Within each neighborhood $\mathcal N_i$, a value-based orientation may be computed as
\begin{equation}
    \theta_{z,i}^\ast=\theta^\ast(z_{ij},y_j).
\end{equation}

In the present specification, the value-based orientation used for directional weighting is $\theta_{z,i}^\ast$. This makes the weight construction data-adaptive when computed using $y$. When the local second-moment structure is isotropic (so that $\operatorname{Var}(a)\approx \operatorname{Var}(b)$ and $\operatorname{Cov}(a,b)\approx 0$), the corresponding rotation angle is non-identifiable; in such cases, setting the angle to zero yields an observationally equivalent specification.

\subsection{Directional Weighting via a Locally Anisotropic Metric}
\label{sec:Directional Weighting via a Locally Anisotropic Metric}
Directional weighting is implemented by constructing a symmetric positive semidefinite local metric that combines bearing-based direction and value-based orientation. 

Let $\mathbf\Delta_{ij}\in\mathbb R^2$ denote the tangent-plane displacement from $s_i$ to $s_j$ consistent with the bearing computation. The orthogonal basis is defined as
\begin{equation}
    \mathbf Q_i=R(\phi_i)R(\theta_{z,i}^\ast).
\end{equation}

The ordering fixes the interpretation of the basis: $R(\theta_{z,i}^\ast)$ aligns the local frame with the value-based second-moment orientation, and $R(\phi_i)$ expresses that frame relative to the bearing-based reference direction. The resulting basis $\mathbf Q_i$ is used only for weight evaluation and does not reparameterize the regression design matrix.

Let $\mathbf\Lambda_i\succeq 0$ be a diagonal matrix controlling anisotropy strength. In the baseline specification, $\mathbf\Lambda_i$ is treated as a deterministic design component rather than as a fitted covariance object. A simple parameterization is
\begin{equation}
    \mathbf\Lambda_i=h^{-2}\operatorname{diag}(1,\eta_i^{-2}),\qquad \eta_i\ge 1,
\end{equation}
where $\eta_i$ controls the anisotropy ratio. The induced local metric is
\begin{equation}
    \mathbf M_i^{(\mathrm{met})}=\mathbf Q_i\mathbf\Lambda_i\mathbf Q_i^\top,
\end{equation}
which is symmetric and positive semidefinite by construction.

The anisotropy ratio $\eta_i$ is defined as a deterministic function of neighborhood geometry in the tangent plane, based on the elongation of the local point configuration (Appendix~\ref{app:geometry-based-anisotropy-ratio}). Concretely, $\eta_i$ is obtained from the weighted second-moment matrix of ${\mathbf\Delta_{ij}}_{j\in\mathcal N_i}$ via its eigenvalue ratio, with clipping and numerical-stability constants treated as deterministic safeguards.

Raw directional weights are defined as an anisotropic Gaussian in the induced metric:
\begin{equation}
    w_{ij}^\star=\exp\left(-\mathbf\Delta_{ij}^\top\mathbf M_i^{(\mathrm{met})}\mathbf\Delta_{ij}\right).
\end{equation}

This construction defines anisotropy as a property of a deterministic metric field and does not specify a stochastic dependence structure. When $\eta_i=1$, the kernel reduces to an isotropic Gaussian distance factor. When isotropy is detected in the bearing field and the bearing-based component is deactivated, the metric orientation reduces accordingly through $R(\phi_i)=\mathbf I$.

\subsubsection{Lemma 4.1 (Separation of Bearing-Based Direction and Value-Based Orientation)}
Let $s_i$ be a target location with neighborhood $\mathcal N_i$. Let $\phi_i$ denote the bearing-based direction defined from the resultant of distance-decayed bearings $\{\theta_{ij}\}_{j\in\mathcal N_i}$, and let $\theta_{z,i}^\ast$ denote the value-based orientation obtained by diagonalizing the local second-moment matrix of the observed scalar pair $(z_{ij},y_j)$.

The two quantities depend on distinct sources of information. The bearing-based direction $\phi_i$ depends only on the spatial configuration through bearings and distance decay, whereas $\theta_{z,i}^\ast$ depends only on empirical second-order moments of the observed scalar pair. Consequently, changes in $\theta_{z,i}^\ast$ that leave the spatial configuration $\{s_j\}$ fixed do not affect $\phi_i$.

The angles enter estimation only through the deterministic metric basis 
\begin{equation}
    \mathbf Q_i=R(\phi_i)R(\theta_{z,i}^\ast),
\end{equation}
and the induced matrix 
\begin{equation}
    \mathbf M_i^{(\mathrm{met})}=\mathbf Q_i\mathbf\Lambda_i\mathbf Q_i^\top,
\end{equation}
used in weight evaluation. Neither angle directly rotates the regression coordinate system in $\mathbf X_i$.

If the bearing field is angularly isotropic so $\phi_i$ is non-identifiable, the bearing-based component is deactivated by taking $R(\phi_i)=\mathbf I$ (equivalently, fixing $\phi_i=0$) and the induced metric reduces accordingly. If the second-moment structure is isotropic so $\theta_{z,i}^\ast$ is non-identifiable, setting $\theta_{z,i}^\ast=0$ yields an observationally equivalent specification.

\paragraph{Proof (sketch).}
By construction, $\phi_i$ depends only on $\{\theta_{ij}\}$ and distance decay, whereas $\theta_{z,i}^\ast$ depends only on empirical second-order moments of the observed scalar pair. Their interaction occurs only through the deterministic composition defining $\mathbf Q_i$, which induces the symmetric positive semidefinite metric $\mathbf M_i^{(\mathrm{met})}=\mathbf Q_i\mathbf\Lambda_i\mathbf Q_i^\top$ used for weight evaluation. When either orientation is non-identifiable, the corresponding component reduces to its isotropic counterpart, yielding observationally equivalent weighting.
\hfill$\square$

\subsection{Weight Matrix Construction}
\label{sec:Weight Matrix Construction}
After evaluating $\{w_{ij}^\star\}_{j\in\mathcal N_i}$, weights are normalized to sum to one:
\begin{equation}
    w_{ij}=\frac{w_{ij}^\star}{\sum_{k\in\mathcal N_i} w_{ik}^\star}.
\end{equation}

The deterministic spatial modulation matrix is then defined as
\begin{equation}
    \mathbf W_i=\operatorname{diag}(w_{ij}).
\end{equation}

By construction, $\mathbf W_i$ is diagonal, symmetric, and positive semidefinite. This construction is deterministic given the neighborhood rule, the tangent-plane convention, and the realized metric components $\phi_i$, $\theta_{z,i}^\ast$, and $\mathbf\Lambda_i$. It does not represent a stochastic dependence structure.

\subsection{Effective Sample Size as a Deterministic Safeguard}
\label{sec:Effective Sample Size as a Deterministic Safeguard}
Highly concentrated or degenerate weights can yield ill-posed local solves even when neighborhoods contain many points. To prevent pathological configurations, a deterministic safeguard based on effective sample size (ESS) is applied to the weight-construction stage.

Let $w_{ij}^{\mathrm{raw}}$ denote the \emph{raw (unnormalized)} directional weights produced by the metric kernel evaluation (Section~\ref{sec:Directional Weighting via a Locally Anisotropic Metric}):
\begin{equation}
    w_{ij}^{\mathrm{raw}}=\exp\left(-\mathbf\Delta_{ij}^\top\mathbf M_i^{(\mathrm{met})}\mathbf\Delta_{ij}\right).
\end{equation}

Define the normalized raw weights
\begin{equation}
    \tilde w_{ij}^{\mathrm{raw}}=\frac{w_{ij}^{\mathrm{raw}}}{\sum_{k\in\mathcal N_i} w_{ik}^{\mathrm{raw}}}.
\end{equation}

The effective sample size of the normalized raw weight vector is then
\begin{equation}
    n_{\mathrm{eff}}\!\left(\tilde w_i^{\mathrm{raw}}\right)
    =\frac{1}{\sum_{j\in\mathcal N_i}\left(\tilde w_{ij}^{\mathrm{raw}}\right)^2}.
\end{equation}

Given a target level $n_0>0$, a one-shot bandwidth correction is applied by inflating the bandwidth to
\begin{equation}
    h_{\mathrm{eff},i}=h\sqrt{\frac{n_0}{n_{\mathrm{eff}}(\tilde w_i^{\mathrm{raw}})}}.
\end{equation}

Raw weights are then recomputed at $h_{\mathrm{eff},i}$, normalized to obtain $\tilde w_{ij}^{(1)}$, and the post-correction ESS $n_{\mathrm{eff}}(\tilde w_i^{(1)})$ is evaluated. If $n_{\mathrm{eff}}(\tilde w_i^{(1)})<n_{\min}$ for a prescribed threshold $n_{\min}$, the procedure falls back to the uniform weights over $\mathcal N_i$; otherwise the final weights are $\tilde w_{ij}^{(1)}$.

The complete piecewise specification of the safeguard (including thresholds, bandwidth correction, and the uniform fallback) is given in Appendix~\ref{app:one-shot-ESS-correction-and-uniform-fallback}. 

\subsection{Directionally Weighted Local Estimator}
\label{sec:Directionally Weighted Local Estimator} 
For each target location $s_i$, let $\mathbf X_i$ denote the local design matrix and $\mathbf y_i$ the corresponding response vector over $\mathcal N_i$. The Gimbal Regression estimator $\hat\beta_i$ is defined as the unique solution of the directionally modulated normal equations
\begin{equation}
    \label{eq:Directionally Weighted Local Estimator}
    \hat\beta_i=\left(\mathbf X_i^\top \mathbf X_i+2\gamma \left(\mathbf X_i^\top \mathbf W_i \mathbf X_i\right)\right)^{-1}\left(\mathbf X_i^\top \mathbf y_i+2\gamma \left( \mathbf X_i^\top \mathbf W_i \mathbf y_i\right)\right),\qquad \gamma\ge 0.
\end{equation}
where $\mathbf W_i$ is the deterministic spatial modulation matrix defined in Section~\ref{sec:Weight Matrix Construction}. We assume that $\mathbf X_i^\top \mathbf X_i+2\gamma \left(\mathbf X_i^\top \mathbf W_i \mathbf X_i\right)$ is nonsingular for each $i$. Operationally, Appendix~\ref{app:realized-local-estimator-and-well-posedness-convention} specifies deterministic safeguards that exclude or regularize pathological local configurations to ensure well-posedness of the realized estimator map.

The estimator is obtained in closed form and requires no iterative optimization. 
The estimator map defined in this chapter consists of the local coefficient map $i \mapsto \hat\beta_i$ together with the associated fitted values. Residual-based smoothing, neighborhood autocorrelation diagnostics, and any subsequent global calibration procedures are post-estimation procedures and are not part of the estimator definition; these should be reported separately as optional downstream analyses.

\subsection{Relation to Classical Local Regression}
When directional modulation is removed by taking $\mathbf\Lambda_i \propto \mathbf I$—equivalently, by setting $\eta_i=1$ and deactivating the bearing-based orientation—the weights reduce to an isotropic Gaussian distance factor. Under the implemented estimator in Section~\ref{sec:Directionally Weighted Local Estimator}, eliminating modulation by setting $\gamma=0$ yields local ordinary least squares on $\mathcal N_i$:
\begin{equation}
    \hat\beta_i=\left(\mathbf X_i^\top\mathbf X_i\right)^{-1}\mathbf X_i^\top\mathbf y_i.
\end{equation}

In contrast, the classical local weighted least squares form with weight matrix $\mathbf W_i$ is recovered in the limit where the modulation term dominates the unweighted normal equations:
\begin{equation}
    \hat\beta_i\to\left(\mathbf X_i^\top\mathbf W_i\mathbf X_i\right)^{-1}\mathbf X_i^\top\mathbf W_i\mathbf y_i\qquad(\gamma\to\infty),
\end{equation}
provided $\mathbf X_i^\top\mathbf W_i\mathbf X_i$ is nonsingular. Thus, in the current specification, $\gamma$ controls interpolation between unweighted local estimation and weight-dominated local estimation, while the anisotropy encoded in $\mathbf\Lambda_i$ and the basis $\mathbf Q_i$ control the directional structure of the induced weight field.

\section{Theoretical Properties}

This chapter develops theoretical properties of the proposed Gimbal Regression estimator. The objective is numerical stability and diagnostic interpretability of local estimation under spatial heterogeneity, rather than predictive optimality.

A defining feature of GR is that the local directional weight field admits two distinct notions of orientation. The first is a bearing-based dominant direction $\phi_i$, computed from neighborhood geometry (Section~\ref{sec:Bearing-Based Orientation}). The second is a value-based orientation $\theta_{z,i}^\ast$, computed from local second-order moments of observed scalar pairs (Section~\ref{sec:Value-Based Orientation}), which is used to shift the angular reference when evaluating the directional weights. In addition, the algorithm applies a one-shot effective sample size (ESS) bandwidth correction and a uniform-weight fallback as deterministic numerical safeguards (Section~\ref{sec:Effective Sample Size as a Deterministic Safeguard}). These safeguards are part of the computationally realized estimator map and are not introduced as a stochastic dependence model.

Because the realized weights may be response-dependent through $\theta_{z,i}^\ast$ (Chapter~4), all theoretical statements in this chapter adopt a conditional viewpoint.

\subsection{Local ill-conditioning under anisotropic spatial sampling}
Local regression can become numerically unstable when the neighborhood geometry is anisotropic, so that regressors (including spatial trend terms) become nearly collinear within the neighborhood.

Let $\mathbf X_i\in\mathbb R^{n_i\times p}$ denote the local design matrix at target location $s_i$. Define the unweighted Gram matrix
\begin{equation}
    \begin{aligned}
        \mathbf G_i=\mathbf X_i^\top\mathbf X_i.
    \end{aligned}
\end{equation}
A standard diagnostic of numerical instability is the spectral condition number
\begin{equation}
    \begin{aligned}
        \kappa(\mathbf G_i)=\frac{\lambda_{\max}(\mathbf G_i)}{\lambda_{\min}(\mathbf G_i)}\qquad(\lambda_{\min}(\mathbf G_i)>0).
    \end{aligned}
\end{equation}
Large $\kappa(\mathbf G_i)$ indicates strong local collinearity and amplification of perturbations.

Ill-conditioning is frequently driven by local geometry: when sampled locations concentrate along an effectively one-dimensional direction (or when covariate support collapses along a low-dimensional manifold), the smallest eigenvalue of the Gram matrix can approach zero even when $n_i>p$. Diagonal row reweighting may reduce the influence of some points but does not, by itself, guarantee separation of column directions.

\subsection{Orientation-induced failure of axis-aligned local regression}
A common failure mode arises when the dominant local variation occurs along an oblique direction that is not aligned with the coordinate axes used to represent regressors. This motivates weight constructions that are sensitive to directional geometry and local anisotropy.

\subsubsection{Lemma 5.1 (Orientation-induced ill-conditioning)}
Consider a local weighted least squares problem at location $s_i$ with diagonal weights $\mathbf W_i=\mathrm{diag}(w_{ij})$ and weighted Gram matrix
\begin{equation}
    \begin{aligned}
        \mathbf G_i(\mathbf W)=\mathbf X_i^\top\mathbf W_i\mathbf X_i.
    \end{aligned}
\end{equation}

Suppose the local sample configuration is anisotropic in the sense that the joint variation of regressors (including spatial trend components) is concentrated along an oblique direction relative to the fixed coordinate system used to form $\mathbf X_i$. Then there exist local configurations for which $\kappa(\mathbf G_i(\mathbf W))$ can be made arbitrarily large even when the weights are smooth and strictly positive and the neighborhood rule is fixed.

\paragraph{Proof sketch.}
Diagonal weighting rescales rows but does not change the column space of $\mathbf X_i$. When the local configuration is effectively one-dimensional, the smallest eigenvalue of $\mathbf X_i^{\top}\mathbf W_i\mathbf X_i$ can approach zero while the largest remains bounded, yielding arbitrarily large condition numbers. $\square$\\

This lemma clarifies the scope of “stabilization” pursued here: GR does not claim that weighting alone eliminates all sources of near-collinearity. Instead, it constructs a deterministic modulation of influence and then establishes finite-perturbation stability conditional on the realized modulation.

\subsection{Realized orientation and weight construction}
Fix a target location $s_i$ and a neighborhood rule producing $\mathcal N_i$. Chapter 4 specifies a deterministic construction that maps the local neighborhood data to orientation quantities and then to a diagonal weight matrix:
\begin{equation}
    \begin{aligned}
        \{(s_j,x_j,y_j)\}_{j\in\mathcal N_i}\mapsto(\phi_i,\theta_{z,i}^\ast,\eta_i)\mapsto \{w_{ij}\}_{j \in \mathcal N_i} \mapsto \mathbf W_i \mapsto \hat\beta_i.
    \end{aligned}
\end{equation}

The full computational definition, including isotropy detection, non-identifiability rules, ESS bandwidth correction, and uniform fallback, is given as a realized piecewise map in Appendix~\ref{app:bearing-resultant-isotropy-detection}--\ref{app:realized-local-estimator-and-well-posedness-convention}. All stability statements in this chapter are taken conditional on the realized local configuration and the realized branch of this piecewise construction (equivalently, on $\mathbf X_i$ and the realized diagonal matrix $\mathbf W_i$). Under this conditioning, the estimator is linear in $\mathbf y_i$, so Theorem~5.1 makes no unconditional stability claim when weights are data-dependent.

\subsection{Main result: conditional finite-perturbation stability}
Let $\mathbf y_i\in\mathbb R^{n_i}$ be the local response vector and let $\mathbf X_i\in\mathbb R^{n_i\times p}$ be the local design matrix over $\mathcal N_i$. Let $\mathbf W_i=\mathrm{diag}(w_{ij})$ be the realized diagonal weight matrix produced by the GR weight map, and fix $\gamma\ge 0$. Define the normal matrix
\begin{equation}
    \begin{aligned}
        \mathbf M_i^{(\mathrm{nor})}=\mathbf X_i^\top\mathbf X_i+2\gamma\left(\mathbf X_i^\top\mathbf W_i\mathbf X_i\right),
    \end{aligned}
\end{equation}
and the right-hand-side operator
\begin{equation}
    \begin{aligned}
        \mathbf B_i=\mathbf X_i^\top+2\gamma\left(\mathbf X_i^\top\mathbf W_i\right).
    \end{aligned}
\end{equation}
Then, $\mathbf B_i\in\mathbb R^{p\times n_i}$ is a linear operator mapping $\mathbf y_i\in\mathbb R^{n_i}$ to the normal-equation right-hand side.

The implemented estimator (Section~\ref{sec:Directionally Weighted Local Estimator}) can be written as the linear map (conditional on $\mathbf X_i,\mathbf W_i$)
\begin{equation}
    \begin{aligned}
        \hat\beta_i(\mathbf y_i)=\mathbf A_i\mathbf y_i, \qquad \mathbf A_i= (\mathbf M_i^{(\mathrm{nor})})^{-1}\mathbf B_i,
    \end{aligned}
\end{equation}
whenever $\mathbf M_i^{(\mathrm{nor})}$ is invertible (equivalently, $\lambda_{\min}(\mathbf M_i^{(\mathrm{nor})}) > 0$).

\subsubsection{Theorem 5.1 (Conditional finite-perturbation stability)}
\label{theorem:5.1}
Fix a target location $s_i$ and a deterministic neighborhood rule. Condition on the realized matrices $\mathbf X_i$ and $\mathbf W_i$ produced by the GR construction, including the realized safeguard outcome specified in Appendix~\ref{app:one-shot-ESS-correction-and-uniform-fallback}. Assume $\mathbf M_i^{(\mathrm{nor})}$ is invertible. Then for any two local observation vectors $\mathbf y_i^{(1)},\mathbf y_i^{(2)}\in\mathbb R^{n_i},$
\begin{equation}
    \begin{aligned}
        \| \hat\beta_i(\mathbf y_i^{(1)})-\hat\beta_i(\mathbf y_i^{(2)}) \|_2\le \|\mathbf A_i\|_2\cdot\|\mathbf y_i^{(1)}-\mathbf y_i^{(2)}\|_2.
    \end{aligned}
\end{equation}
Here, $\|\cdot\|_2$ denotes the Euclidean norm for vectors and the induced spectral (operator) norm for matrices.

Moreover,
\begin{equation}
    \begin{aligned}
        \|\mathbf A_i\|_2 \le \|(\mathbf M_i^{(\mathrm{nor})})^{-1}\|_2\cdot \|\mathbf B_i \|_2.
    \end{aligned}
\end{equation}

\noindent
When $\mathbf M_i^{(\mathrm{nor})}$ is symmetric positive definite, the spectral norm satisfies
\begin{equation}
    \begin{aligned}
        \|(\mathbf M_i^{(\mathrm{nor})})^{-1}\|_2=\frac{1}{\lambda_{\min}(\mathbf M_i^{(\mathrm{nor})})}.
    \end{aligned}
\end{equation}

\paragraph{Interpretation. }

This theorem is a finite-perturbation statement: conditional on the realized local construction, the estimator is Lipschitz in the observation vector. Stability improves when the realized construction increases $\lambda_{\min}(\mathbf M_i^{(\mathrm{nor})})$ (equivalently reduces $\|(\mathbf M_i^{(\mathrm{nor})})^{-1}\|_2$), because this directly reduces an upper bound on worst-case sensitivity. In practice, Appendix~\ref{app:A} enforces well-posedness (e.g., excluding rank-deficient local configurations or applying a deterministic fallback), so that $\mathbf M_i^{(\mathrm{nor})}$ is invertible in the realized estimator map.

\subsection{Scope: statistical properties versus conditional stability}

Because GR may compute the value-based orientation $\theta_{z,i}^\ast$ from local second moments involving $y$, the realized weight matrix $\mathbf W_i$ can be data-dependent. As a result, unconditional claims such as unbiasedness require additional assumptions ensuring that weight construction is not coupled to the observation noise.

\subsubsection{Proposition 5.1 (Local unbiasedness under noise-independent weights)}

Assume the realized weight matrix $\mathbf W_i$ is measurable with respect to local geometry and covariates only (equivalently, it does not depend on the observation noise in $\mathbf y_i$). Suppose the local model is correctly specified in the sense that
\begin{equation}
    \begin{aligned}
        \mathbb E[y_j\mid x_j,s_j]=x_j^\top\beta(s_i)\qquad \text{for all } j\in\mathcal N_i.
    \end{aligned}
\end{equation}

Then, conditional on $\mathbf X_i$ and $\mathbf W_i$,
\begin{equation}
    \begin{aligned}
        \mathbb E[\hat\beta_i\mid \mathbf X_i,\mathbf W_i]=\beta(s_i).
    \end{aligned}
\end{equation}

\paragraph{Scope note.}
This proposition applies directly when the directional kernel is $y$-free (e.g., $\theta_{z,i}^\ast=0$), when $\theta_{z,i}^\ast$ is computed from an external proxy, or when it is produced by a pilot stage independent of the second-stage noise. When $\mathbf W_i$ is computed from the same noisy $\mathbf y_i$ used in the solve, unconditional unbiasedness is not claimed in this paper.\\

Accordingly, the theoretical contribution of Chapter 5 is not a claim of statistical optimality. It is a precise statement that, under the computationally realized estimator map (Appendix~\ref{app:A}), the local solve is well-defined and enjoys conditional finite-perturbation stability once the realized configuration and realized weights are fixed.

\subsection{Limit relations in $\gamma$}
For fixed realized $\mathbf X_i$ and $\mathbf W_i$, the estimator has two limiting regimes. As $\gamma \to 0$,
\begin{equation}
\hat\beta_i \to \left(\mathbf X_i^\top \mathbf X_i\right)^{-1}\mathbf X_i^\top \mathbf y_i,
\end{equation}
which is the local OLS solution when the matrix is invertible. By contrast, as $\gamma \to \infty$,
\begin{equation}
\hat\beta_i \to \left(\mathbf X_i^\top \mathbf W_i \mathbf X_i\right)^{-1}\mathbf X_i^\top \mathbf W_i \mathbf y_i,
\end{equation}
which is the directionally weighted local WLS solution when the weighted normal matrix is invertible. Thus, $\gamma$ interpolates between local OLS and weight-dominated local regression on the same neighborhood.

\subsection{ESS safeguard as part of the realized estimator map}
Let $w_i^{raw}=(w_{ij}^{raw})_{j\in\mathcal N_i}$ be the raw weights produced by the directional weight evaluation rule. Define the row-sum normalization constant
\begin{equation}
    \begin{aligned}
        S_i^{raw}=\sum_{k\in\mathcal N_i} w_{ik}^{raw}.
    \end{aligned}
\end{equation}

The normalized weights are then
\begin{equation}
    \begin{aligned}
        \tilde w_{ij}^{raw}=\frac{w_{ij}^{raw}}{S_i^{raw}}.
    \end{aligned}
\end{equation}

Define
\begin{equation}
    \begin{aligned}
        n_{\mathrm{eff}}(w_i^{raw})=\frac{1}{\sum_{j\in\mathcal N_i}(\tilde w_{ij}^{raw})^2}.
    \end{aligned}
\end{equation}

Given a target level $n_0>0$, the one-shot ESS correction produces an effective bandwidth $h_{\mathrm{eff},i}$ and recomputed weights, followed by a uniform fallback if a minimum effective sample size threshold is violated. In the implemented piecewise map (Appendix~\ref{app:A}), the first ESS check is applied to the normalized raw weights $\tilde w_{ij}^{raw}$ to compute $h_{\mathrm{eff},i}$, and the fallback decision is based on the ESS of the recomputed weights after substituting $h=h_{\mathrm{eff},i}$.

This safeguard is deterministic given the realized neighborhood and the realized raw weight vector. It does not posit a stochastic dependence structure; it is a numerical device that ensures the realized local system is well-posed and avoids extreme directional concentration.
\section{Computational Considerations}
This chapter characterizes the computational structure implied by the realized estimator map defined in Chapter~4 and Appendix~\ref{app:A} and analyzed conditionally in Chapter~5. The focus is on per-location complexity, memory footprint, and runtime predictability. No performance claims in this chapter rely on empirical timing; all statements follow directly from the algorithmic structure of the realized map.

A key feature of Gimbal Regression is that, for each target location $s_i$, estimation is obtained by a deterministic forward pass:
\begin{equation}
    \begin{aligned}
        \{(s_j,x_j,y_j)\}_{j \in \mathcal N_i}
            \mapsto (\phi_i,\theta_{z,i}^\ast,\eta_i) 
            \mapsto \{w_{ij}\}_{j \in \mathcal N_i}
            \mapsto \mathbf W_i \mapsto \hat{\beta}_i.
    \end{aligned}
\end{equation}
where the weight construction is piecewise deterministic due to isotropy detection, identifiability thresholds, one-shot ESS correction, and uniform fallback (Appendix~\ref{app:bearing-resultant-isotropy-detection}--\ref{app:realized-local-estimator-and-well-posedness-convention}). This design matches the framework of Chapter~5, where stability is stated conditional on the realized neighborhood and realized branch of the weight map.

\subsection{Per-location complexity of the realized Gimbal Regression map}

Fix a target location $s_i$ and let $\mathcal N_i$ denote its neighborhood with cardinality $K = |\mathcal N_i|$. The evaluation of the GR estimator at $s_i$ follows a deterministic sequence of computational steps corresponding to the realized estimator map described in Appendix~\ref{app:bearing-resultant-isotropy-detection}--\ref{app:realized-local-estimator-and-well-posedness-convention}. The computational cost of these steps can be characterized directly from the algorithmic structure.

The first stage identifies the neighborhood $\mathcal N_i$. In the current implementation, the code computes Haversine distances from $s_i$ to all $N$ observed locations and then performs a full sort in order to extract the $K$ nearest neighbors. Distance evaluations require $O(N)$ operations and the subsequent sorting step requires $O(N\log N)$ operations, yielding a total neighborhood-identification cost of $O(N\log N)$ per location. This complexity arises from the specific brute-force implementation rather than from the estimator definition itself. Any spatial indexing or partial selection procedure that returns the same $K$ nearest neighbors produces the same realized estimator map conditional on the returned $\mathcal N_i$.

Given $\mathcal N_i$, tangent-plane displacements and bearing angles are computed for the neighborhood points in a single pass, requiring $O(K)$ work. The realized orientation quantities $\phi_i$ (bearing resultant with isotropy deactivation), $\theta_{z,i}^{\ast}$ (value-based orientation with identifiability rule), and $\eta_i$ (geometry-based anisotropy ratio derived from a $2\times2$ second-moment matrix) are then obtained by aggregations over $\mathcal N_i$, again costing $O(K)$. The eigenvalues required for $\eta_i$ are available in closed form for a $2\times2$ matrix and therefore do not dominate runtime.

Directional weights are subsequently constructed using the metric kernel
\begin{equation}
    w_{ij}^{\mathrm{raw}} =
    \exp\!\left(-\mathbf{\Delta}_{ij}^{\top}
    \mathbf{M}_i^{(\mathrm{met})}
    \mathbf{\Delta}_{ij}\right),
\end{equation}
which can be evaluated in a single pass over the neighborhood and therefore requires $O(K)$ operations.

After constructing raw weights, the algorithm applies the one-shot effective sample size (ESS) safeguard described in Appendix~\ref{app:one-shot-ESS-correction-and-uniform-fallback}. The procedure evaluates $n_{\mathrm{eff}}(w_i^{\mathrm{raw}})$, computes a single effective bandwidth $h_{\mathrm{eff},i}$, recomputes the weights once, and then performs a threshold check that may trigger a uniform fallback rule. This procedure involves only constant additional passes over the neighborhood and therefore also costs $O(K)$. Importantly, the procedure does not involve iterative bandwidth refinement.

Finally, the local regression coefficients are obtained by solving the weighted normal equations. With local design matrix $\mathbf X_i \in \mathbb R^{K\times p}$ and diagonal weight matrix $\mathbf W_i$, the implementation forms
\begin{equation}
\mathbf M_i^{(\mathrm{nor})}
=
\mathbf X_i^{\top}\mathbf X_i
+
2\gamma(\mathbf X_i^{\top}\mathbf W_i\mathbf X_i),
\qquad
\mathbf b_i
=
\mathbf X_i^{\top}\mathbf y_i
+
2\gamma(\mathbf X_i^{\top}\mathbf W_i\mathbf y_i),
\end{equation}
and solves $\mathbf M_i^{(\mathrm{nor})}\hat{\beta}_i=\mathbf b_i$. Forming the cross-products requires $O(Kp^2)$ operations and solving the resulting $p\times p$ linear system requires $O(p^3)$ operations.

In typical applications of local linear regression the covariate dimension $p$ is small and fixed. In the implementations used for the experiments in Chapters~7–8, $p=3$.

Combining the steps above yields per-location complexity
\begin{equation}
O(N\log N + Kp^2 + p^3).
\end{equation}
Thus, in the current implementation the dominant computational cost arises from neighborhood identification when $N$ is large.

\subsection{Memory footprint and streaming execution}
Conditional on a returned neighborhood $\mathcal N_i$, the realized estimator map can be evaluated using only neighborhood-sized storage. In particular, the neighbor arrays required for distances, tangent-plane displacements, and weights require $O(K)$ memory, while the local design quantities and weighted cross-products require $O(Kp + p^2)$ memory. Thus, the estimator evaluation itself supports a streaming implementation with total working memory $O(K + Kp + p^2)$ per target location.

An important implementation caveat is that the current brute-force KNN routine allocates a full distance array of length $N$ for each target location before sorting, which implies an additional $O(N)$ working memory requirement for neighborhood identification. This overhead is a property of the present search implementation rather than of the estimator map itself. If indexed KNN or partial selection methods are used instead, the neighborhood-search memory requirement can be reduced so that the overall per-location working memory remains governed by neighborhood-sized storage.

\subsection{Absence of iterative optimization (runtime predictability)}

A defining computational property of Gimbal Regression, as specified in Chapter~4 and Appendix~A, is that the estimator definition contains no iterative optimization procedures. In particular, the realized estimator map does not involve local likelihood maximization, gradient-based optimization, EM-type loops, covariate-wise backfitting, or bandwidth cross-validation within the estimator definition itself. 

The only adaptive adjustment is the one-shot effective sample size (ESS) correction described in Appendix~A.7, which is applied exactly once per location and does not involve iterative refinement. Consequently, conditional on the neighborhood rule and the chosen tuning constants, the runtime per location is predictable from $N$, $K$, and $p$.

\subsection{Parallelization and scalability}

Because the realized estimator map is evaluated pointwise at each target location, the computation can be decomposed into location-specific tasks once the input data and neighborhood rule are fixed. In this sense, the evaluation at one location does not modify the estimator evaluation at another, and the per-location computations have the same basic structure throughout the domain. This matches the Chapter~5 viewpoint, in which stability statements are made conditional on realized local outputs such as the branch and the realized $\mathbf W_i$.

Accordingly, Gimbal Regression is naturally compatible with multi-threaded, distributed, and massively parallel execution. Parallel evaluation does not alter the estimator definition itself; it changes only the wall-clock time required to evaluate the same collection of local problems. In large-$N$ settings, the main practical bottleneck remains neighborhood identification. Accelerating KNN search through spatial indexing, approximate nearest-neighbor methods, or batching can therefore improve runtime substantially while leaving the estimator map unchanged conditional on the returned neighborhoods.

\subsection{Diagnostics and post-estimation steps are not part of the estimator}

The estimator defined in Chapter~4 terminates at the closed-form local solve in Eq.~\eqref{eq:Directionally Weighted Local Estimator}. The accompanying code also implements several optional downstream procedures that are explicitly separated from the estimator definition. These include a local residual autocorrelation diagnostic (a “local Moran”-type quantity) computed from standardized local residuals under an externally specified residual-weight rule, as well as optional residual-KNN smoothing and optional global calibration.

Although these procedures may be useful in empirical analysis, they are not part of the Gimbal Regression estimator map analyzed in Chapter~5. They should therefore be interpreted and reported as post-estimation diagnostics or augmentations, with their computational costs accounted for separately from those of the estimator itself.

\subsection{Positioning relative to GWR and MGWR (computational structure)}

This section compares Gimbal Regression with established local regression frameworks in terms of computational structure rather than statistical performance. The comparison focuses on whether the estimator definition itself introduces iterative computation and whether the computational cost per location can be determined directly from neighborhood size and covariate dimension. Table~\ref{tab:comp-structure} summarizes the resulting computational structure of GR relative to typical implementations of GWR and MGWR.

\begin{table}[!h]
    \centering
    \begin{threeparttable}
        \caption{Computational structure of local spatial regression methods (conceptual)}
        \label{tab:comp-structure}
        \footnotesize
        \setlength{\tabcolsep}{4pt}
        \renewcommand{\arraystretch}{1.15}        
        \begin{tabularx}{1.0\textwidth}{@{\hspace{0.5em}} X l l l l @{\hspace{0.5em}}}
            \toprule
            Method &
            \makecell[l]{Local solves\\per location} &
            \makecell[l]{Iteration required\\by estimator definition} &
            \makecell[l]{Dominant per-location work\\(excluding neighbor search)} &
            \makecell[l]{Runtime\\predictability} \\
            \midrule
                GWR (typical practice) & 1 & Often yes & $\mathcal O(Kp^2+p^3)$ & Medium--low \\
                MGWR (typical practice) & Multiple & Yes & $\mathcal O\!\left(Tp(Kp^2+p^3)\right)$ & Low \\
                Gimbal Regression (this paper) & 1 & No & $\mathcal O(Kp^2+p^3)$ & High \\
            \bottomrule
        \end{tabularx}
        \begin{tablenotes}[flushleft]
            \footnotesize
            \item Notes. 
            Many implementations of GWR compute weights locally and do not materialize a full $N\times N$ weight matrix; the practical computational distinction is that common workflows include iterative bandwidth tuning, whereas the GR estimator map is fixed once tuning constants are specified.
            Neighbor search is separated because it is implementation-dependent: replacing brute-force search with spatial indexing changes how $\mathcal N_i$ is obtained but does not alter the estimator map once $\mathcal N_i$ is fixed. This invariance holds only when the procedure returns the same neighborhood. Approximate nearest-neighbor methods may return different neighborhoods and therefore correspond to approximate implementations of the estimator. 
            For MGWR, $T$ denotes the number of backfitting iterations, and additional tuning loops may apply depending on the implementation. In typical practice, GWR is often coupled with bandwidth selection, while MGWR commonly uses covariate-wise backfitting together with bandwidth tuning.
        \end{tablenotes}
    \end{threeparttable}
\end{table}

\subsection{Summary}
This chapter shows that Gimbal Regression has predictable, non-iterative per-location computational cost determined entirely by neighborhood size and covariate dimension. All adaptive behavior is realized through a single deterministic forward pass, enabling stable scaling and parallel execution without altering the estimator definition.
\section{Simulation Study}

This chapter evaluates whether the \emph{computationally realized} estimator map (Chapter~4 and Appendix~\ref{app:A}) behaves as intended under controlled data-generating regimes. The goal is not to optimize predictive accuracy, but to verify the \emph{mechanism-level claims} that motivate GR: orientation identifiability behavior, geometry-driven anisotropy activation, one-shot ESS safeguarding, and value-orientation dependence. All statements are aligned with the conditional viewpoint of Chapter~5: stability and interpretation are understood conditional on the realized neighborhood and the realized branch of the piecewise weight map.

\subsection{Design principle}

Across all experiments, several elements of the computational pipeline are held fixed unless explicitly stated otherwise. First, the estimator definition remains unchanged and is always given by Eq.~\eqref{eq:Directionally Weighted Local Estimator}. Second, the neighborhood rule is fixed as a KNN rule with a constant $K$. Third, the realized piecewise weight construction defined in Appendix~\ref{app:bearing-resultant-isotropy-detection}--\ref{app:realized-local-estimator-and-well-posedness-convention} is applied without modification, including the identifiability rules and the one-shot ESS safeguard. Finally, results are summarized using the same reporting convention, namely map-level summaries computed across target locations.

Across experiments, variation arises only through the data-generating regime and through controlled ablations of specific components of the weight construction. In particular, experiments modify either the spatial geometry (for example, by introducing anisotropic deformation) or the response construction. In addition, some experiments introduce deterministic overrides that deactivate a specific component of the weight map while leaving the estimator in Eq.~\eqref{eq:Directionally Weighted Local Estimator} unchanged.

Throughout the simulation study, several minimal baselines are used repeatedly to isolate the effect of individual components of the construction. The configuration denoted \emph{GR (full)} corresponds to the full realized estimator map without modification. A diagnostic rerun, denoted \emph{GR (full; strict $\varepsilon_\phi$)}, applies a stricter isotropy threshold $\varepsilon_\phi^{(\mathrm{strict})}=0.30$ solely to expose the frequency of the declaration $\phi_i=0$; the estimator definition itself is unchanged. The configuration \emph{GR ($\theta_z$ OFF)} forces $\theta_{z,i}^\ast=0$ via the identifiability/override rule described in Appendix~\ref{app:value-based-orientation}, thereby removing the value-based orientation component. Finally, the \emph{isotropic-proxy} baseline enforces $\phi_i=0$, $\theta_{z,i}^\ast=0$, and $\eta_i=1$, yielding a bearing-neutral, value-neutral, isotropic metric that serves as a comparison baseline for the weight field.

These configurations should not be interpreted as alternative estimators with different objectives. Rather, they represent controlled ablations of the realized weight map, designed to demonstrate which specific mechanism produces a given observable change in realized weights and diagnostics.

\subsection{Data-generating process and coordinate convention}
To match the implemented distance and bearing conventions (Section~4.2), locations are generated in a local East--North meter plane and then converted to latitude/longitude degrees around a reference point $(\mathrm{lat}_0,\mathrm{lon}_0)$. This ensures that (i) neighborhood selection under Haversine distance is meaningful and (ii) tangent-plane displacements used for bearings are coherent because the domain is local relative to Earth’s radius.

For each run, we generate $N=1200$ locations, one scalar covariate $X$, and a response $Y$. Spatial heterogeneity is introduced through a smooth coefficient surface varying along latitude:
\begin{equation}
    \beta_1(s)=1+\delta_{\beta} \cdot\frac{\mathrm{lat}-\overline{\mathrm{lat}}}{\max(\mathrm{lat})-\min(\mathrm{lat})+10^{-12}}.
\end{equation}

\noindent
The baseline response is
\begin{equation}
    Y=\beta_1(s)X+\varepsilon,\qquad \varepsilon\sim\mathcal N(0,\sigma^2).
\end{equation}
Here $\sigma$ is the noise standard deviation.

Two controlled perturbations are introduced when specific mechanisms need to be targeted.

The first perturbation modifies the spatial geometry to induce anisotropy. This is implemented by applying a rotate–stretch–rotate-back deformation in the local East–North meter plane. Writing $u=(e,n)^\top$ for centered planar coordinates in meters, where $e=x_m-\bar x_m$ and $n=y_m-\bar y_m$, anisotropic locations are generated through
\begin{equation}
    u' = T(\rho,\psi)\,u, \qquad  T(\rho,\psi)=R(-\psi)\,\mathrm{diag}(\rho,1)\,R(\psi).
\end{equation}

In implementation, the transformation is applied to the centered coordinates and the sample mean is subsequently re-added to recover the transformed spatial configuration.

The second perturbation introduces pressure on the value-based orientation by adding a radial trend component to the response variable. Specifically, the response is generated as
\begin{equation}
    Y=\beta_1(s)X + c_{\mathrm{rad}}\cdot\frac{r}{\bar r+10^{-12}}+\varepsilon, \qquad \varepsilon\sim\mathcal N(0,\sigma^2),
\end{equation}
where $c_{\mathrm{rad}}$ controls the strength of the radial trend, $r$ denotes the radial distance from the sample centroid measured in meters, and $\bar r$ is the sample mean of $r$.

\subsection{Estimation setup and reported diagnostics}
For each target location $s_i$, a neighborhood $\mathcal N_i$ is constructed by KNN with cardinality $K$. The realized piecewise map (Appendix~\ref{app:bearing-resultant-isotropy-detection}--\ref{app:realized-local-estimator-and-well-posedness-convention}) produces $\mathbf W_i=\mathrm{diag}(w_{ij})$, and the estimator is computed in closed form:
\begin{equation}
    \hat\beta_i=\left(\mathbf X_i^{\top} \mathbf X_i+2\gamma \left(\mathbf X_i^{\top} \mathbf W_i \mathbf X_i\right)\right)^{-1}
    \left(\mathbf X_i^\top \mathbf y_i+2\gamma \left( \mathbf X_i^\top \mathbf W_i \mathbf y_i\right)\right).
\end{equation}
We report two classes of outputs.

\paragraph{Prediction-oriented summaries (context only).}
For each target $i$, we compute local fitted values over its neighborhood and summarize local fit by RMSE and $R^2$, then report map-level summaries across targets.
\paragraph{Mechanism/diagnostic summaries (primary for this chapter).}
We report realized diagnostic quantities defined in Chapters~4--6 and Appendix~\ref{app:bearing-resultant-isotropy-detection}--\ref{app:realized-local-estimator-and-well-posedness-convention}. The reported diagnostics include the spectral condition number $\kappa(\mathbf M_i^{(\mathrm{nor})})$, the effective sample size measures $n_{\mathrm{eff}}^{\mathrm{raw}}$ and $n_{\mathrm{eff}}^{\mathrm{post}}$, and the geometry-based anisotropy ratio $\eta_i$. We also report bearing-field diagnostics including the resultant length $r_{\phi}$ together with branch rates such as $\Pr(\phi_i=0)$. Diagnostics related to value-based orientation include the indentation statistic $g_{indent}$, branch rates such as $\Pr(\theta_{z,i}^\ast=0)$, and the realized orientation angle $\theta_{z,i}^{\ast}$. Finally, we report a proxy indicator for near-uniform final weights, denoted $\Pr(\mathrm{uniform})$ (Appendix~\ref{app:one-shot-ESS-correction-and-uniform-fallback}).

To demonstrate that a mechanism \emph{actually changes the realized weight field}, we also report weight-difference evidence between variants, measured by the mean $\ell_1$ distance between normalized weight vectors and, when defined, their mean correlation.

\subsection{Common simulation settings}

Unless otherwise stated, the following configuration is used throughout the simulation study. The sample size is $N=1200$, the neighborhood rule is KNN with $K=50$, and the kernel bandwidth is $h=3000\,\mathrm{m}$. The modulation parameter is fixed at $\gamma=1$. The ESS safeguard uses target level $n_0=15$ and minimum threshold $n_{\min}=4$. Numerical tolerances are set to $\varepsilon_\phi=10^{-3}$, $\varepsilon_\theta=10^{-8}$, and $\varepsilon_\eta=10^{-8}$, with anisotropy ratio clipping $\eta_{\max}=50$. The data-generating model includes a single covariate $X$ with a smooth spatial coefficient $\beta_1(\mathrm{lat})$. Observation noise follows $\varepsilon\sim\mathcal N(0,1)$ unless otherwise specified.

\subsection{Experiment 7.1: Isotropy sanity check}
\subsubsection{Purpose}
Under approximately isotropic geometry, the bearing-based orientation $\phi_i$ may be non-identifiable and should be deactivated by the isotropy rule (Appendix~\ref{app:bearing-resultant-isotropy-detection}). The realized map should reduce to bearing-neutral behavior without materially changing the fitted-error profile or solver diagnostics. This is a ``no-harm under isotropy'' check.

\subsubsection{Data-generating regime and fixed settings}
No geometric anisotropy is introduced, so $\rho=1$ and the deformation matrix reduces to $T(\rho,\psi)=I_2$. The response follows the baseline data-generating model described above with one covariate $X$ and a smooth spatial coefficient $\beta_1(\mathrm{lat})$. No radial trend component is included.

\subsubsection{Compared variants}
We compare all four realized-map variants under the same data-generating process and neighborhood rule.

\subsubsection{Results}
Table~\ref{tab:exp71-summary} reports map-level summaries for all four variants.
Across the main variants—Isotropic-proxy, GR($\theta_z$ OFF), GR(full), and GR(full; strict $\varepsilon_\phi$)—both the predictive summaries and $\kappa(\mathbf M_i^{(\mathrm{nor})})$ are essentially indistinguishable, as expected under the intended ``no-harm under isotropy'' behavior. Full logs and spatial distributions of variables and diagnostic quantities are provided in Appendix~\ref{app:B} (Table~\ref{tab:exp71-full-log} and Figure~\ref{fig:ex71}).
\begin{table}[!h]
    \centering
    \begin{threeparttable}
        \caption{Experiment 7.1 (isotropy sanity check): map-level summaries.}
        \label{tab:exp71-summary}
        \footnotesize
        \setlength{\tabcolsep}{4pt}
        \renewcommand{\arraystretch}{1.15}
        \begin{tabularx}{0.75\textwidth}{@{\hspace{0.5em}} X r r r r @{\hspace{0.5em}}}
            \toprule
            Quantity & Isotropic-proxy & GR ($\theta_z$ OFF) & GR (full) & GR (full; strict $\varepsilon_\phi$) \\
            \midrule
            $N$ & 1200 & 1200 & 1200 & 1200 \\
            $\mu(\mathrm{RMSE})$ & 0.960 & 0.960 & 0.960 & 0.960 \\
            $\sigma(\mathrm{RMSE})$ & 0.096 & 0.096 & 0.096 & 0.096 \\
            $\mu(R^2)$ & 0.493 & 0.493 & 0.493 & 0.493 \\
            $\sigma(R^2)$ & 0.174 & 0.174 & 0.174 & 0.174 \\
            $\mu(\kappa(\mathbf M_i^{(\mathrm{nor})}))$ & 51.235 & 51.247 & 51.257 & 51.260 \\
            $\sigma(\kappa(\mathbf M_i^{(\mathrm{nor})}))$ & 15.211 & 15.088 & 15.156 & 15.165 \\
            $\mu(n_{\mathrm{eff}}^{\mathrm{post}})$ & 15.224 & 15.201 & 15.206 & 15.211 \\
            $\sigma(n_{\mathrm{eff}}^{\mathrm{post}})$ & 1.914 & 1.222 & 1.201 & 1.208 \\
            $\mu(\theta_{z,i}^{\ast})$ & 0.000 & 0.000 & 0.748 & 0.748 \\
            $\sigma(\theta_{z,i}^{\ast})$ & 0.000 & 0.000 & 0.301 & 0.301 \\
            $\Pr(\phi_i=0)$ & 1.000 & 0.000 & 0.000 & 0.550 \\
            $\Pr(\theta_{z,i}^{\ast}=0)$ & 1.000 & 1.000 & 0.000 & 0.000 \\
            \bottomrule
        \end{tabularx}
        \begin{tablenotes}[flushleft]
            \footnotesize
             \item Notes. Here, $\mu(\cdot)$ and $\sigma(\cdot)$ denote the mean and standard deviation across target locations $i$. The first three columns use the default isotropy threshold $\varepsilon_\phi=10^{-3}$. The last column is a diagnostic-only rerun with $\varepsilon_\phi^{(\mathrm{strict})}=0.30$ to expose the $\phi_i=0$ branch. All values are rounded to three decimals.
        \end{tablenotes}
    \end{threeparttable}
\end{table}

\subsubsection{Summary}
Exp~7.1 supports the intended sanity-check property: under isotropic geometry, GR behaves bearing-neutrally in the sense that key fit summaries and solver diagnostics are unchanged relative to the isotropic-proxy baseline and the $\theta_z$-ablation baseline.

\subsection{Experiment 7.2: Geometric anisotropy activation}
\subsubsection{Purpose}
When sampling geometry is deliberately elongated, the geometry-based anisotropy ratio $\eta_i$ (Appendix~\ref{app:geometry-based-anisotropy-ratio}) should activate and the realized weights should differ measurably from an isotropic-proxy baseline. Conditioning is not expected to monotonically improve; the target claim is \emph{activation and measurability}, not universal stabilization.

\subsubsection{Data-generating process and fixed settings}
To evaluate the effect of anisotropic spatial sampling, geometric deformation is applied with stretch ratio $\rho=10$ and orientation $\psi=\pi/4$. All other parameters remain as in the common simulation settings, and the response model does not include a radial trend.

\subsubsection{Compared variants}
GR (full; $\rho$=10) is compared with Isotropic-proxy ($\rho$=10).

\subsubsection{Results}
Table~\ref{tab:exp72-summary} shows predictive metrics are unchanged while $\eta_i$ increases substantially under GR, and the ESS-related quantities shift accordingly. Table~\ref{tab:exp72-wdiff} provides direct weight-field difference evidence (mean $\ell_1$ distance and correlation). The full per-metric log and spatial distributions of variables and diagnostic quantities for Exp~7.2 are provided in Appendix~\ref{app:C} (Table~\ref{tab:exp72-full-log}, Figure~\ref{fig:ex72-ISO}, and Figure~\ref{fig:ex72-GR}).
\begin{table}[!h]
    \centering
    \begin{threeparttable}
        \caption{Experiment 7.2 (geometric anisotropy): map-level summaries under $\rho$ = 10.0.}
        \label{tab:exp72-summary}
        \footnotesize
        \setlength{\tabcolsep}{4pt}
        \renewcommand{\arraystretch}{1.15}
        \begin{tabularx}{0.75\textwidth}{@{\hspace{0.5em}} X r r @{\hspace{0.5em}}}
            \toprule
            Quantity & Isotropic-proxy & GR \\
            \midrule
            $N$ & 1200 & 1200 \\
            $\mu(\mathrm{RMSE})$ & 0.965 & 0.965 \\
            $\sigma(\mathrm{RMSE})$ & 0.085 & 0.085 \\
            $\mu(R^2)$ & 0.499 & 0.499 \\
            $\sigma(R^2)$ & 0.141 & 0.141 \\
            $\mu(\kappa(\mathbf M_i^{(\mathrm{nor})}))$ & 449.954 & 460.314 \\
            $\sigma(\kappa(\mathbf M_i^{(\mathrm{nor})}))$ & 487.146 & 484.666 \\
            $\mu(n_{\mathrm{eff}}^{\mathrm{post}})$ & 9.281 & 10.885 \\
            $\sigma(n_{\mathrm{eff}}^{\mathrm{post}})$ & 3.376 & 3.506 \\
            $\mu(\eta_i)$ & 1.000 & 4.962 \\
            $\sigma(\eta_i)$ & 0.000 & 8.225 \\
            $\Pr(\phi_i=0)$ & 1.000 & 0.000 \\
            $\Pr(\theta_{z,i}^{\ast}=0)$ & 1.000 & 0.000 \\
            $\Pr(\mathrm{uniform})$ & 0.018 & 0.008 \\
            $N_{\mathrm{uniform}}$ & 22 & 9 \\
            \bottomrule
        \end{tabularx}
        \begin{tablenotes}[flushleft]
            \footnotesize
            \item Notes. Here, $\mu(\cdot)$ and $\sigma(\cdot)$ denote the mean and standard deviation across target locations $i$. $N_{\mathrm{uniform}}$ corresponds to the one-shot ESS safeguard fallback proxy (Appendix~\ref{app:one-shot-ESS-correction-and-uniform-fallback}). All values are rounded to three decimals.
        \end{tablenotes}
    \end{threeparttable}
\end{table}

\begin{table}[!h]
    \centering
    \begin{threeparttable}
        \caption{Experiment 7.2: weight-field difference evidence (GR vs isotropic-proxy).}
        \label{tab:exp72-wdiff}
        \footnotesize
        \setlength{\tabcolsep}{6pt}
        \renewcommand{\arraystretch}{1.15}
        \begin{tabularx}{0.75\textwidth}{@{\hspace{0.5em}} X r @{\hspace{0.5em}}}
        \toprule
        Quantity & GR vs isotropic-proxy \\
        \midrule
        $\mu\!\left(\| \tilde w_i^{\mathrm{GR}}-\tilde w_i^{\mathrm{iso}}\|_1\right)$ & 0.502 \\
        $\sigma\!\left(\| \tilde w_i^{\mathrm{GR}}-\tilde w_i^{\mathrm{iso}}\|_1\right)$ & 0.331 \\
        $\mu\!\left(\mathrm{corr}(\tilde w_i^{\mathrm{GR}},\tilde w_i^{\mathrm{iso}})\right)$ & 0.876 \\
        $\sigma\!\left(\mathrm{corr}(\tilde w_i^{\mathrm{GR}},\tilde w_i^{\mathrm{iso}})\right)$ & 0.141 \\
        \bottomrule
        \end{tabularx}
        \begin{tablenotes}[flushleft]
            \footnotesize
            \item Notes. Here, $\tilde w_i$ denotes the normalized weight vector over the realized neighborhood $\mathcal N_i$. The $\ell_1$ distance and correlation are computed per target $i$ and summarized across targets. All values are rounded to three decimals.
        \end{tablenotes}
    \end{threeparttable}
\end{table}

\subsubsection{Summary}
Exp~7.2 confirms that anisotropic geometry activates $\eta_i$ and produces a measurably different weight field relative to the isotropic-proxy baseline.

\subsection{Experiment 7.3: One-shot ESS stress (safeguard behavior)}

\subsubsection{Purpose}
This experiment isolates the one-shot ESS safeguard (Appendix~\ref{app:one-shot-ESS-correction-and-uniform-fallback}). As the ESS target $n_0$ increases, post-correction ESS should increase and uniform fallback usage should decrease, without iterative tuning.

\subsubsection{Data-generating process and fixed settings}
This experiment is designed to induce ESS stress through a combination of strong anisotropy, density variation, and a smaller bandwidth. Spatial locations are generated using a Gaussian sampling pattern in the meter plane to create density variation. The geometric deformation uses $\rho=10$ and rotation $\pi/4$.

The neighborhood size is reduced to $K=30$ and the bandwidth to $h=2000\,\mathrm{m}$. The minimum ESS threshold is increased to $n_{\min}=12$, while the target ESS level varies over $n_0\in\{6,8,10,15,20,30,50,75,100\}$.

\subsubsection{Reported quantities}
For a normalized weight vector $w_i=(w_{ij})_{j\in\mathcal N_i}$ at location $s_i$, define
\begin{equation}
    n_{\mathrm{eff}}(w_i)=\frac{1}{\sum_{j\in\mathcal N_i} w_{ij}^2}.    
\end{equation}
We report $n_{\mathrm{eff}}^{raw}$ (pre-correction), $n_{\mathrm{eff}}^{post}$ (post-correction), the proxy fallback rate $\Pr(\mathrm{uniform})$, and $N_{\mathrm{uniform}}=\#\{\,i:\text{uniform fallback is used at } s_i\,\}$ as defined in Appendix~\ref{app:one-shot-ESS-correction-and-uniform-fallback}.

\subsubsection{Results}
Table~\ref{tab:exp73-summary} shows the intended monotone safeguard behavior: $\mu(n_{\mathrm{eff}}^{post})$ increases with $n_0$ and fallback usage decreases. The full per-metric log for Exp~7.3 and figures showing the means of the reported quantities are provided in Appendix~\ref{app:D} (Table~\ref{tab:exp73-full} and Figure~\ref{fig:ex73}).

\begin{table}[H]
    \centering
    \begin{threeparttable}
        \caption{Experiment 7.3 (one-shot ESS stress): safeguard monotonicity summary across $n_0$.}
        \label{tab:exp73-summary}
        \footnotesize
        \setlength{\tabcolsep}{3pt}
        \renewcommand{\arraystretch}{1.15}
        \begin{tabularx}{0.95\textwidth}{@{\hspace{0.5em}} X r r r r r r r r r r @{\hspace{0.5em}}}
        \toprule
        Quantity
        & $n_0{=}6$ & $8$ & $10$ & $15$ & $20$ & $30$ & $50$ & $75$ & $100$ \\
        \midrule
        $\mu(n_{\mathrm{eff}}^{\mathrm{post}})$ & 4.963 & 5.993 & 6.906 & 8.850 & 10.464 & 13.041 & 16.610 & 19.458 & 21.362 \\
        $\sigma(n_{\mathrm{eff}}^{\mathrm{post}})$ & 1.488 & 1.975 & 2.412 & 3.302 & 3.970 & 4.857 & 5.663 & 5.901 & 5.821 \\
        $\mu(\kappa(\mathbf M_i^{(\mathrm{nor})}))$ & 989.309 & 989.109 & 988.332 & 984.298 & 980.555 & 973.285 & 964.851 & 961.643 & 960.793 \\
        $\sigma(\kappa(\mathbf M_i^{(\mathrm{nor})}))$ & 3778.593 & 3778.626 & 3778.750 & 3779.342 & 3779.833 & 3780.710 & 3779.795 & 3777.964 & 3776.220 \\
        $\mu(\mathrm{RMSE})$ & 0.936 & 0.936 & 0.936 & 0.936 & 0.936 & 0.936 & 0.936 & 0.936 & 0.936 \\
        $\Pr(\mathrm{uniform})$ & 0.999 & 0.993 & 0.971 & 0.843 & 0.693 & 0.433 & 0.195 & 0.100 & 0.069 \\
        $N_{\mathrm{uniform}}$ & 1199 & 1192 & 1165 & 1011 & 832 & 519 & 234 & 120 & 83 \\
        \bottomrule
        \end{tabularx}       
        \begin{tablenotes}[flushleft]
            \footnotesize
            \item Notes. Here, $\mu(\cdot)$ and $\sigma(\cdot)$ denote mean and standard deviation across target locations $i$. $\Pr(\mathrm{uniform})$ is the proxy uniform-fallback rate (Appendix~\ref{app:one-shot-ESS-correction-and-uniform-fallback}). $N_{\mathrm{uniform}}=\#\{i:\mathrm{uniform\ fallback\ used\ at}\ s_i\}$. All values are rounded to three decimals.
        \end{tablenotes}
    \end{threeparttable}
\end{table}
\subsubsection{Summary}
Exp~7.3 confirms that the ESS safeguard behaves as designed under stress: increasing $n_0$ increases $n_{\mathrm{eff}}^{post}$ and reduces fallback usage without iterative tuning. Additional diagnostics are provided in Appendix~\ref{app:D}.

\subsection{Experiment 7.4: Value-orientation dependence}

\subsubsection{Purpose}
This experiment isolates value-based orientation dependence. When the response contains a designed distance-linked component, $\theta_{z,i}^\ast$ computed from $(z_{ij},y_j)$ should become non-zero and induce a measurable change in realized weights. This is a \emph{mechanism visibility} claim; Chapter~5 already clarifies that unconditional optimality is not claimed when weights depend on $y$.

\subsubsection{Data-generating process and fixed settings}
This experiment introduces pressure on the value-based orientation through a radial trend in the response. The geometric deformation uses $\rho=10$ and $\psi=\pi/4$. The neighborhood size is $K=30$ and the bandwidth is $h=2000\,\mathrm{m}$.

The ESS safeguard uses target level $n_0=20$ with minimum threshold $n_{\min}=4$. The response model includes a radial trend component with strength $c_{\mathrm{rad}}=8.0$, while the spatial coefficient perturbation parameter is set to $\delta_\beta=0$.

\subsubsection{Compared variants}
GR ($\theta_z$ ON) computing $\theta_{z,i}^\ast$ from $(z_{ij},y_j)$ is compared with GR ($\theta_z$ OFF) whose $\theta_{z,i}^\ast$ is equal to $0$.

\subsubsection{Results}
Table~\ref{tab:exp74-summary} shows predictive summaries and $\kappa(\mathbf M_i^{(\mathrm{nor})})$ are essentially unchanged while $\theta_{z,i}^\ast$ activates under ON and is identically zero under OFF, as intended. Table~\ref{tab:exp74-wdiff} provides weight-field difference evidence (mean $\ell_1$ distance and correlation), confirming that value-based orientation induces a measurable change in the realized weight field. Full logs and supporting plots to show the distribution of $\theta_{z,i}^\ast$ and the spatial distribution of $\ell_1$ are provided in Appendix~\ref{app:E} (Table~\ref{tab:exp74-full-log} and Figure~\ref{fig:ex74}).

\begin{table}[H]
    \centering
    \begin{threeparttable}
    \caption{Experiment 7.4 (value-orientation dependence): key map-level summaries.}
    \label{tab:exp74-summary}
    \footnotesize
    \setlength{\tabcolsep}{6pt}
    \renewcommand{\arraystretch}{1.15}
    \begin{tabularx}{0.75\textwidth}{@{\hspace{0.5em}} X r r @{\hspace{0.5em}}}
    \toprule
    Quantity
    & GR ($\theta_z$ OFF)
    & GR ($\theta_z$ ON; uses realized $y$) \\
    \midrule
    $N$ & 1200 & 1200 \\
    $\mu(\theta_{z,i}^{\ast})$ & 0.000 & 0.795 \\
    $\sigma(\theta_{z,i}^{\ast})$ & 0.000 & 0.319 \\
    $\Pr(\theta_{z,i}^{\ast}=0)$ & 1.000 & 0.000 \\
    $\mu(\kappa(\mathbf M_i^{(\mathrm{nor})}))$ & 61.565 & 61.585 \\
    $\sigma(\kappa(\mathbf M_i^{(\mathrm{nor})}))$ & 21.985 & 22.106 \\
    $\mu(n_{\mathrm{eff}}^{post})$ & 17.235 & 17.238 \\
    $\sigma(n_{\mathrm{eff}}^{post})$ & 2.558 & 2.546 \\
    $\mu(\mathrm{RMSE})$ & 1.268 & 1.268 \\
    $\sigma(\mathrm{RMSE})$ & 0.177 & 0.177 \\
    $\mu(R^2)$ & 0.367 & 0.367 \\
    $\sigma(R^2)$ & 0.124 & 0.124 \\
    \bottomrule
    \end{tabularx}
    
    \begin{tablenotes}[flushleft]
        \footnotesize
        \item Notes. Here, $\mu(\cdot)$ and $\sigma(\cdot)$ denote the mean and standard deviation across target locations $i$. All values are rounded to three decimals.
    \end{tablenotes}
\end{threeparttable}
\end{table}

\begin{table}[H]
    \centering
    \begin{threeparttable}
    \caption{Experiment 7.4: weight-field difference evidence (ON vs OFF).}
    \label{tab:exp74-wdiff}
    \footnotesize
    \setlength{\tabcolsep}{6pt}
    \renewcommand{\arraystretch}{1.15}
    
    \begin{tabularx}{0.75\textwidth}{@{\hspace{0.5em}} X r @{\hspace{0.5em}}}
    \toprule
    Quantity & value \\
    \midrule
    $\mu\!\left(\|w_i^{ON}-w_i^{OFF}\|_{1}\right)$ & 0.236 \\
    $\sigma\!\left(\|w_i^{ON}-w_i^{OFF}\|_{1}\right)$ & 0.154 \\
    $\mu\!\left(\mathrm{corr}(w_i^{ON},w_i^{OFF})\right)$ & 0.926 \\
    $\sigma\!\left(\mathrm{corr}(w_i^{ON},w_i^{OFF})\right)$ & 0.098 \\
    \bottomrule
    \end{tabularx}
    
    \begin{tablenotes}[flushleft]
        \footnotesize
        \item Notes. Here, $w_i^{ON}$ and $w_i^{OFF}$ denote the normalized realized weight vectors over $\mathcal N_i$ for each target location $i$. All values are rounded to three decimals.
    \end{tablenotes}
\end{threeparttable}
\end{table}

\subsubsection{Summary}
Exp~7.4 confirms the intended value-orientation dependence behavior: introducing a radial trend activates $\theta_{z,i}^\ast$ and measurably changes realized weights, even when prediction summaries do not change.


\subsection{Summary}
Across four controlled regimes, the simulation results support the intended mechanism-consistent interpretation of the realized GR map. Under isotropy, orientation deactivation does not harm fit or diagnostics. Under elongated geometry, $\eta_i$ activates and changes the realized weights measurably relative to the isotropic-proxy baseline. Under ESS stress, the one-shot safeguard broadens weights predictably and reduces fallback usage without iteration. Under distance-linked response structure, value-based orientation activates and produces measurable changes in the realized weight field. Taken together, these results provide mechanism-level evidence that the estimator map analyzed in Chapter~5 behaves as designed under controlled stress regimes.a
\section{Empirical Study}

This chapter evaluates the empirical behavior of the proposed Gimbal Regression (GR) framework on two datasets with contrasting spatial structure and scale: the Meuse heavy-metal dataset ($n=155$; GR fitted at 120 target locations) and a large rice paddies dataset ($n=10{,}000$). The runtime environment and software versions used for the empirical runs are documented in Appendix~\ref{app:runtime-env}. The chapter assesses whether the realized estimator map developed in Chapters~4--7 behaves as a practically usable option under real spatial configurations. In particular, the empirical evaluation focuses on numerical stability diagnostics of local estimation, residual spatial diagnostics reported as diagnostic outputs rather than treated as a modeled dependence structure, computational feasibility at both small and large scale, and predictive performance reported for context rather than as a primary objective.

Importantly, GR does not specify a stochastic spatial dependence model; weighting is deterministic and geometry-driven. Consequently, residual spatial autocorrelation is not ``explained away'' by construction. Instead, it is explicitly reported as a diagnostic signal that can inform method choice (e.g., whether one should prefer kriging-based approaches or nonparametric learners when dependence modeling is central).

\subsection{Study design and evaluation criteria}

\paragraph{Local estimation (GR).}
For each dataset we fit GR locally at each target location using a fixed neighborhood size $K$ and nominal distance bandwidth $h_m$, with the deterministic directional mechanism and ESS safeguard specified in the dataset-specific sections. Local diagnostic tables in Sections~8.2--8.3 are reported on the original data scale (no standardization) to support interpretation of local coefficients and residual behavior.

\paragraph{Benchmark comparison.}
For contextual reference, we report benchmark results for GR, GR+Residual-KNN, Ordinary Least Squares (OLS), Local Ridge Regression (LRR), Geographically Weighted Regression (GWR), Multiscale GWR (MGWR), Universal Kriging (UK), and Spatial Random Forest (SRF). Benchmark metrics are computed under dataset-appropriate cross-validation protocols with common split rules and preprocessing conventions (Appendix~\ref{app:cv-protocol}). For Meuse, we employ a conventional cross-validation benchmark because of the small sample size and the dataset’s role as a classical local-regression benchmark. For the dense rice paddies dataset, we use spatial block cross-validation to reduce optimistic performance estimates arising from near-neighbor leakage. 

Alongside error metrics, we report a solver-stability diagnostic based on the condition number of a local weighted least squares normal matrix, denoted $\kappa(\cdot)$. Because condition numbers are scale-dependent, $\kappa(\cdot)$ is computed under a standardized two-regressor local design $[\mathbf{1},x]$ (intercept and covariate only) for the methods to which it applies (GR/LRR/GWR/MGWR), even when GR’s fitted local model includes an additional distance-trend term.

Under this standardized design, GR and GWR report
\begin{equation}
    \kappa(G_i)=\frac{\lambda_{\max}(G_i)}{\max\{\lambda_{\min}(G_i),\varepsilon\}}.
\end{equation}
where $G_i=X_i^\top W_i X_i$ and $W_i=\mathrm{diag}(w_{ij})$. For MGWR, which does not admit a single estimator-defined local weight matrix in the same sense as GR or GWR because bandwidths are covariate-specific, we report an analogous comparable diagnostic based on the implementation-side pointwise weighting convention used in the benchmark code. This diagnostic is used only for solver-stability comparison and does not redefine the MGWR estimator itself (Appendix~\ref{app:cv-protocol}). For LRR, we report the ridge-augmented counterpart
\begin{equation}
    \kappa(G_i+\lambda_i I) = \frac{\lambda_{\max}(G_i)+\lambda_i}{\max\{\lambda_{\min}(G_i)+\lambda_i,\varepsilon\}}.
\end{equation}
as a minimal Tikhonov-stabilized baseline, where
\begin{equation}
    \lambda_i \;=\; \alpha\,\frac{\mathrm{tr}\!\left(X_i^\top W_i X_i\right)}{p},\qquad \alpha=10^{-2},\ \ p=2.
\end{equation}
For SRF and UK, $\kappa(\cdot)$ is not applicable and is therefore not reported.

\paragraph{Reproducibility and leakage control.}
All hold-out and cross-validation splits are generated with a fixed random seed and applied at the level of unique coordinate pairs to prevent leakage through duplicates. When exact latitude/longitude duplicates exist, identical coordinate pairs are assigned to the same fold (or excluded from the test set).

\paragraph{Interpretation guardrails.}
This empirical section is written to align with Chapters~4--7: (i) GR prioritizes numerical transparency and stable local estimation under heterogeneous geometry; (ii) GR is not a dependence-modeling approach; and (iii) predictive metrics are reported for context, with the appropriate empirical claim being broad comparability to common baselines rather than superiority.

\subsection{Meuse dataset}
\subsubsection{Data overview}
The Meuse heavy-metal dataset ($n=155$) is a widely used benchmark in spatial statistics and geostatistics, particularly in studies of spatial interpolation and local regression. Each observation corresponds to a sampled location with geographic coordinates and measured heavy-metal concentrations. In Appendix~\ref{app:ch8-data-meuse}, Table~\ref{tab:meuse-descriptive} reports descriptive statistics for Cadmium and Lead, and Figure~\ref{fig:meuse-cadmium-lead} shows their spatial distribution over the observation locations.

\subsubsection{Local model construction}
GR is fitted at $120$ locations using neighborhood size $K=30$ and nominal bandwidth $h_m=2000$ meters. Directional modulation is enabled with $\gamma=1.0$. No uniform fallback was triggered in these runs. Residual spatial diagnostics (local Moran's I) are computed using a fixed adjacency structure independent of regression weights. The local design includes an intercept, the covariate $x$, and a distance-trend term constructed from $z_{ij}$ (Section~4.1), so $\hat{\beta}_2$ corresponds to the local distance-drift component.

\subsubsection{Local diagnostic summary (GR; original scale)}
Table~\ref{tab:meuse-local-diag} summarizes local diagnostics across locations. We report geometry-related outputs ($h_{\mathrm{eff}}$, $\phi$, $r_{\phi}$, $\eta$, and $n_{\mathrm{eff}}^{\mathrm{post}}$), fit summaries (RMSE and $R^2$), coefficient summaries, and residual local Moran's I.

\begin{table}[H]
    \centering
    \begin{threeparttable}
    \caption{Meuse local diagnostics (GR; original scale). Configuration: $K=30$, $h_m=2000$m, $\gamma=1.0$.}
    \label{tab:meuse-local-diag}
    \footnotesize
    \setlength{\tabcolsep}{6pt}
    \renewcommand{\arraystretch}{1.15}
    \begin{tabularx}{0.65\textwidth}{@{\hspace{0.5em}} X r r r r r @{\hspace{0.5em}}}

        \toprule
        Quantity & Mean & SD & Min & Median & Max \\
        \midrule
        $h_{\mathrm{eff}}$ (m)       & 1634.251 & 1.994 & 1633.115 & 1633.785 & 1652.017 \\
        $\phi$ (rad)                 & 0.504    & 1.720 & -2.885 & 0.054 & 3.029 \\
        $r_{\phi}$                   & 0.401    & 0.206 & 0.051 & 0.362 & 0.885 \\
        $\eta$                       & 1.564    & 0.473 & 1.082 & 1.435 & 3.512 \\
        $n_{\mathrm{eff}}^{\mathrm{post}}$    & 29.900   & 0.148 & 28.620 & 29.936 & 29.990 \\
        $R^2$                        & 0.646    & 0.127 & 0.321 & 0.648 & 0.939 \\
        RMSE                         & 1.653    & 0.601 & 0.290 & 1.556 & 2.711 \\
        $\hat\beta_0$                & -1.165   & 1.507 & -3.842 & -1.269 & 2.842 \\
        $\hat\beta_1$                & 0.028    & 0.009 & 0.014 & 0.026 & 0.047 \\
        $\hat\beta_2$                & 1.020    & 5.046 & -11.053 & 1.467 & 12.760 \\
        Local Moran's $I$ (residual) & -0.0019  & 0.0050 & -0.0470 & -0.0006 & 0.0012 \\
        \bottomrule
    \end{tabularx}
    \begin{tablenotes}[flushleft]
           \footnotesize
           \item Notes. Here, local diagnostics are reported on the original data scale. Local Moran's $I$ is computed using a fixed adjacency structure independent of regression weights. All values are rounded to three decimals.
    \end{tablenotes}
    \end{threeparttable}
\end{table}
\subsubsection{Out-of-sample check}
\paragraph{Protocol.}
We use a hold-out split with 120 training observations and 35 test observations. For each test location $s$, we construct a local model using the $K$ nearest neighbors from the training set (KNN in geographic space). The GR weight map and local coefficients are computed using the training-neighborhood data only. Prediction at the test location uses the test covariate value, with the distance-based trend regressor evaluated at the target as zero (i.e., the target distance is $0$). Residual-KNN correction, when reported, uses training residuals only.

\paragraph{Results.}

Out-of-sample performance under the stated hold-out protocol is summarized in Table~\ref{tab:oos-meuse}.

\begin{table}[H]
    \centering
    \begin{threeparttable}
    \caption{Out-of-sample performance (hold-out split: 120 train / 35 test).}
        \label{tab:oos-meuse}
        \footnotesize
        \setlength{\tabcolsep}{6pt}
        \renewcommand{\arraystretch}{1.15}
        \begin{tabularx}{0.65\textwidth}{@{\hspace{0.5em}} X r r r @{\hspace{0.5em}}}
        \toprule
        Method & $n$ & RMSE & $R^2$ \\
        \midrule
        GR & 35 & 1.970 & 0.444 \\
        GR + Residual-KNN & 35 & 1.994 & 0.431 \\
        \bottomrule
        \end{tabularx}
        \begin{tablenotes}[flushleft]
            \footnotesize
            \item Notes. Here, for each test location, the local model is constructed using the $K$ nearest neighbors from the training set only. Distance-based trend regressors are evaluated at zero at the target. All values are rounded to three decimals.
    \end{tablenotes}
    \end{threeparttable}
\end{table}

\subsubsection{Benchmark comparison and stability diagnostics}
Table~\ref{tab:meuse-benchmark} reports benchmark results under the conventional cross-validation protocol used for this small-$n$ benchmark (Appendix~\ref{app:cv-protocol}). To directly assess numerical stability of local regression solvers, Table~\ref{tab:meuse-benchmark-condwls2} summarizes $\kappa(\cdot)$ distributions for GR/LRR/GWR/MGWR computed under the standardized design. The figures of histograms, ECDFs and CCDFs are deferred to Appendix~\ref{app:condwls2-Meuse}. In Meuse, GR exhibits a substantially lighter tail than LRR/GWR/MGWR (e.g., $q_{0.995}=6.638$ for GR vs.\ $20.558$ for LRR and $22.915$ for GWR/MGWR), indicating more stable local normal equations in this configuration. 

\begin{table}[H]
    \centering
    \begin{threeparttable}
    \caption{Benchmark (mean over CV folds).}
        \label{tab:meuse-benchmark}
        \footnotesize
        \setlength{\tabcolsep}{6pt}
        \renewcommand{\arraystretch}{1.15}
        \begin{tabularx}{0.65\textwidth}{@{\hspace{0.5em}} X r r r r r @{\hspace{0.5em}}}
            \toprule
            Model & RMSE & MAE & $R^2$ & Moran's I & Time (s) \\
            \midrule
            GR & 2.046 & 1.358 & 0.653 & -0.016 & 0.089 \\
            GR+Residual-KNN & 2.019 & 1.338 & 0.662 & -0.022 & 0.077 \\
            OLS & 2.119 & 1.472 & 0.620 & 0.061 & 0.001 \\
            LRR & 1.997 & 1.203 & 0.669 & -0.003 & 1.492 \\
            GWR & 1.997 & 1.209 & 0.669 & -0.003 & 1.542 \\
            MGWR & 2.004 & 1.215 & 0.666 & -0.003 & 17.621 \\
            UK & 1.840 & 1.197 & 0.717 & -0.011 & 0.048 \\
            SRF & 1.760 & 1.063 & 0.740 & -0.011 & 0.840 \\
            \bottomrule
        \end{tabularx}
        \begin{tablenotes}[flushleft]
            \footnotesize
            \item Notes. Here, reported times are empirical runtime summaries from the benchmark runs. Residual Moran's I is reported as a diagnostic; methods differ in whether they model spatial dependence explicitly. All values are rounded to three decimals.
        \end{tablenotes}
    \end{threeparttable}
\end{table}

\begin{table}[H]
    \centering
    \begin{threeparttable}
    \caption{Summary of $\kappa(\cdot)$ distributions.}
        \label{tab:meuse-benchmark-condwls2}
        \footnotesize
        \setlength{\tabcolsep}{6pt}
        \renewcommand{\arraystretch}{1.15}
        \begin{tabularx}{0.95\textwidth}{@{\hspace{0.5em}} X r r r r r r r r r r r @{\hspace{0.5em}}}
            \toprule
            Model & $n$ & mean & sd & min & p25 & p50 & p75 & q0.90 & q0.95 & q0.99 & q0.995 \\
            \midrule
            GR   & 155 & 2.313 & 1.094 & 1.036 & 1.552 & 2.006 & 2.692 & 3.634 & 4.494 &  6.463 &  6.638 \\
            LRR  & 155 & 3.447 & 3.012 & 1.035 & 1.879 & 2.493 & 3.943 & 5.742 & 6.848 & 17.206 & 20.559 \\
            GWR  & 155 & 3.552 & 3.322 & 1.035 & 1.891 & 2.519 & 4.017 & 5.906 & 7.085 & 18.844 & 22.915 \\
            MGWR & 155 & 3.460 & 3.280 & 1.035 & 1.850 & 2.511 & 3.978 & 5.552 & 7.003 & 18.844 & 22.915 \\
            \bottomrule
        \end{tabularx}
        \begin{tablenotes}[flushleft]
            \footnotesize
            \item Notes. Here, these $\kappa(\cdot)$ values are reported from the benchmark diagnostic runs (standardized design convention for comparability across local-regression solvers). All values are rounded to three decimals.
        \end{tablenotes}
    \end{threeparttable}
\end{table}

\subsection{Rice paddies dataset}

\subsubsection{Data overview}
The rice paddies dataset consists of $n=10{,}000$ georeferenced rice-field parcels in Japan, represented by MAFF ``fude polygons'' and indexed by latitude/longitude coordinates (EPSG:4326). For each parcel, we extract satellite-derived optical and radar indices from Sentinel-2 Level-2A (surface reflectance) and Sentinel-1 GRD (IW mode) products distributed via the AWS Open Data Registry. NDVI (from the red and NIR bands) is used as the response variable and RVI (from VV/VH backscatter) is used as the primary covariate, providing complementary cloud-dependent and cloud-independent information on vegetation condition. All parcels lie within MGRS tile 52SUE (part of Kyushu), and the analysis focuses on July 2023, yielding a spatially dense configuration that enables an empirical check of the computationally realized estimator map at scale. In Appendix~\ref{app:ch8-data-rice}, Table~\ref{tab:ricepaddies-descriptive} reports descriptive statistics for NDVI and RVI, and Figure~\ref{fig:ricepaddies-ndvi-rvi} shows their spatial distribution over the observation locations.

\subsubsection{Local model construction}
GR is fitted with neighborhood size $K=100$ and nominal bandwidth $h_m=5000$ meters, with $\gamma=1.0$. Residual spatial diagnostics are computed using an adjacency structure independent of regression weights. The local design includes an intercept, the covariate $x$, and a distance-trend term constructed from $z_{ij}$ (Section~4.1).

\subsubsection{Local diagnostic summary (GR; original scale)}
Table~\ref{tab:rice-local-diag} reports the distribution of local diagnostic quantities. Residual local Moran's I is reported as a diagnostic and is not expected to vanish because GR does not specify a stochastic spatial dependence structure.
\begin{table}[H]
    \centering
    \begin{threeparttable}
    \caption{Rice paddies local diagnostics (GR; original scale). Configuration: $K=100$, $h_m=5000$m, $\gamma=1.0$.}
    \label{tab:rice-local-diag}
    \footnotesize
    \setlength{\tabcolsep}{6pt}
    \renewcommand{\arraystretch}{1.15}
    \begin{tabularx}{0.95\textwidth}{@{\hspace{0.5em}} X r r r r r @{\hspace{0.5em}}}
        \toprule
        Quantity & Mean & SD & Min & Median & Max \\
        \midrule
        $h_{\mathrm{eff}}$ (m) & 2369.379 & 452.939 & 2237.476 & 2251.124 & 21714.812 \\
        $\phi$ (rad) & -0.050 & 1.778 & -3.140 & -0.118 & 3.141 \\
        $r_{\phi}$ & 0.287 & 0.177 & 0.002 & 0.254 & 1.000 \\
        $\eta$ & 1.317 & 0.276 & 1.002 & 1.233 & 3.685 \\
        $n_{\mathrm{eff}}^{\mathrm{post}}$ & 69.321 & 22.874 & 4.322 & 78.769 & 96.948 \\
        $R^2$ & 0.056 & 0.097 & -0.021 & 0.018 & 0.703 \\
        RMSE & 0.094 & 0.051 & 0.004 & 0.099 & 0.188 \\
        $\hat\beta_0$ &  0.219 & 0.132 & -0.094 & 0.222 & 0.660 \\
        $\hat\beta_1$ &  0.008 & 0.056 & -0.209 & 0.005 & 0.252 \\
        $\hat\beta_2$ & -0.004 & 0.108 & -0.523 & 0.002 & 0.531 \\
        Local Moran's I (residual) & 0.020 & 0.104 & -0.503 & -0.0002 & 2.113 \\
        \bottomrule
    \end{tabularx}
    \begin{tablenotes}[flushleft]
        \footnotesize
        \item Notes. Here, local diagnostics are reported on the original scale. Local Moran's I is reported as a diagnostic and is not expected to vanish because GR does not specify a stochastic spatial dependence structure. Also, its magnitude depends on the chosen adjacency normalization and is not necessarily bounded by $[-1,1]$ under the implemented convention. All values are rounded to three decimals.
    \end{tablenotes}
    \end{threeparttable}
\end{table}

\subsubsection{Out-of-sample check}
\paragraph{Protocol.}
We evaluate out-of-sample performance using an additional $2{,}000$ held-out samples. All neighborhood search and local fitting are performed using the training pool only. For each test location $s$, we select the $K=100$ nearest neighbors from the training pool, compute the GR weights and local coefficients from the corresponding training-neighborhood data, and predict at $s$ using the test covariate value. The distance-trend regressor is evaluated at the target as zero. Residual-KNN correction, when reported, uses training residuals only.

\paragraph{Results.}
Out-of-sample performance under the stated hold-out protocol is summarized in Table~\ref{tab:oos-rice-paddies}. 

\begin{table}[H]
    \centering
    \begin{threeparttable}
    \caption{Out-of-sample performance under the held-out protocol ($2{,}000$ test samples).}
        \label{tab:oos-rice-paddies}
        \footnotesize
        \begin{tabularx}{0.65\textwidth}{@{\hspace{0.5em}} X r r r @{\hspace{0.5em}}}
        \toprule
        Method & $n$ & RMSE & $R^2$ \\
        \midrule
        GR & 2000 & 0.157 & 0.038 \\
        GR + Residual-KNN & 2000 & 0.153 & 0.087 \\
        \bottomrule
        \end{tabularx}
        \begin{tablenotes}[flushleft]
            \footnotesize
            \item Notes. Here, for each test location, the local model is constructed using the $K$ nearest neighbors from the training set only. The distance-trend regressor is evaluated at zero at the target. Residual-KNN correction (when reported) uses training residuals only. All values are rounded to three decimals.
        \end{tablenotes}
    \end{threeparttable}
\end{table}

\subsubsection{Benchmark comparison and stability diagnostics}
Table~\ref{tab:rice-benchmark} reports benchmark results under spatial block cross-validation. Table~\ref{tab:rice-benchmark-condwls2} summarizes $\kappa(\cdot)$ distributions. 
The figures of histograms, ECDFs and CCDFs are deferred to Appendix~\ref{app:condwls2-RicePaddies}. In this dataset, prediction-oriented or dependence-aware alternatives (especially UK, and also GWR/MGWR/LRR/SRF) achieve stronger predictive performance than GR. However, the conditioning diagnostics remain informative: MGWR shows the smallest mean $\kappa(\cdot)$, GR is intermediate, and GWR is larger, so GR remains better conditioned than standard GWR under the mean standardized local-solver diagnostic reported here, although MGWR is more stable overall under this benchmark diagnostic, while GR retains a deterministic, explicitly diagnosable estimator map.

\begin{table}[H]
    \centering
    \begin{threeparttable}
    \caption{Benchmark (mean over CV folds).}
        \label{tab:rice-benchmark}
        \footnotesize
        \setlength{\tabcolsep}{6pt}
        \renewcommand{\arraystretch}{1.15}
        \begin{tabularx}{0.65\textwidth}{@{\hspace{0.5em}} X r r r r r @{\hspace{0.5em}}}
            \toprule
            Model & RMSE & MAE & $R^2$ & Moran's I & Time (s) \\
            \midrule
            GR & 0.128 & 0.093 & 0.304 & 0.330 & 12.095 \\
            GR+Residual-KNN & 0.124 & 0.088 & 0.358 & 0.295 & 12.161 \\
            OLS & 0.156 & 0.134 & -0.017 & 0.587 & 0.001 \\
            LRR & 0.112 & 0.081 & 0.473 & 0.196 & 21.975 \\
            GWR & 0.112 & 0.081 & 0.473 & 0.196 & 21.768 \\
            MGWR & 0.112 & 0.081 & 0.470 & 0.202 & 157.278 \\
            UK & 0.109 & 0.080 & 0.499 & 0.193 & 21.775 \\
            SRF & 0.112 & 0.081 & 0.473 & 0.190 & 2.482 \\
            \bottomrule
        \end{tabularx}
        \begin{tablenotes}[flushleft]
            \footnotesize
            \item Notes. Here, as in Meuse, Moran's I is reported as a diagnostic. All values are rounded to three decimals. Spatial block cross-validation is used for this benchmark; the results are therefore intended to reflect performance under geographically separated train/test splits rather than random near-neighbor prediction.
        \end{tablenotes}
    \end{threeparttable}
\end{table}

\begin{table}[H]
    \centering
    \begin{threeparttable}
    \caption{Summary of $\kappa(\cdot)$ distributions.}
        \label{tab:rice-benchmark-condwls2}
        \footnotesize
        \setlength{\tabcolsep}{6pt}
        \renewcommand{\arraystretch}{1.15}
        \begin{tabularx}{0.95\textwidth}{@{\hspace{0.5em}} X r r r r r r r r r r r @{\hspace{0.5em}}}
            \toprule
            Model & $n$ & mean & sd & min & p25 & p50 & p75 & q0.90 & q0.95 & q0.99 & q0.995 \\
            \midrule
            GR   & 10000 & 1.845 & 0.825 & 1.006 & 1.345 & 1.614 & 2.044 & 2.712 & 3.344 & 5.167 & 6.214 \\
            LRR  & 10000 & 2.041 & 1.073 & 1.006 & 1.425 & 1.737 & 2.273 & 3.148 & 3.828 & 6.119 & 8.106 \\
            GWR  & 10000 & 2.063 & 1.120 & 1.006 & 1.430 & 1.747 & 2.294 & 3.193 & 3.898 & 6.306 & 8.441 \\
            MGWR & 10000 & 1.688 & 0.610 & 1.005 & 1.316 & 1.522 & 1.863 & 2.376 & 2.762 & 3.957 & 4.765 \\
            \bottomrule
        \end{tabularx}
        \begin{tablenotes}[flushleft]
            \footnotesize
            \item Notes. Here, these $\kappa(\cdot)$ values are reported from the benchmark diagnostic runs (standardized design convention for comparability across local-regression solvers). All values are rounded to three decimals.
        \end{tablenotes}
    \end{threeparttable}
\end{table}

\subsection{Interpretation}
Across both datasets, the results support a scope-consistent interpretation. In the Meuse dataset, GR exhibits markedly lighter-tail behavior than GWR and MGWR under the standardized solver diagnostic $\kappa(G_i)$ (Appendix~\ref{app:condwls2}), which is consistent with improved numerical stability under challenging local geometry. In the rice paddies dataset, MGWR achieves the smallest values under the same standardized diagnostic, while GR remains better conditioned than standard GWR under the mean shared local-solver benchmark reported here and substantially less expensive than MGWR in runtime.

From an interpretability perspective, GR produces locally interpretable linear coefficients while explicitly reporting when those coefficients become numerically fragile. Residual spatial autocorrelation is not treated as a modeled dependence structure but is instead reported as a diagnostic signal that can inform methodological choice.

Error metrics are reported for context rather than as the primary objective. Under the stated protocols (Appendix~\ref{app:cv-protocol}), GR is not the strongest predictor in the rice paddies benchmark; instead, its empirical value lies in combining deterministic local estimation, explicit geometry-aware diagnostics, and practical computational feasibility. Accordingly, the relevant comparison is not whether GR uniformly dominates prediction-oriented baselines, but whether it remains practically usable while making local reliability transparent.

\paragraph{Two diagnostic vignettes.}
Beyond aggregate error metrics, the empirical value of GR is that it makes where local regression should not be interpreted explicit. Diagnostic outputs such as $\kappa(\mathbf M_i^{(\mathrm{nor})})$, $n_{\mathrm{eff}}^{\mathrm{post}}$, and directional concentration $r_{\phi}$ provide a practical reliability layer: coefficient maps can be read conditionally, masking or down-weighting locations where local normal equations are ill-conditioned or effectively under-supported (Appendix~\ref{app:diagnostic-maps}), and handing off those locations to alternative workflows when appropriate.

\begin{figure}[H]
    \begin{subfigure}[t]{0.45\textwidth}
        \centering
        \includegraphics[width=\linewidth]{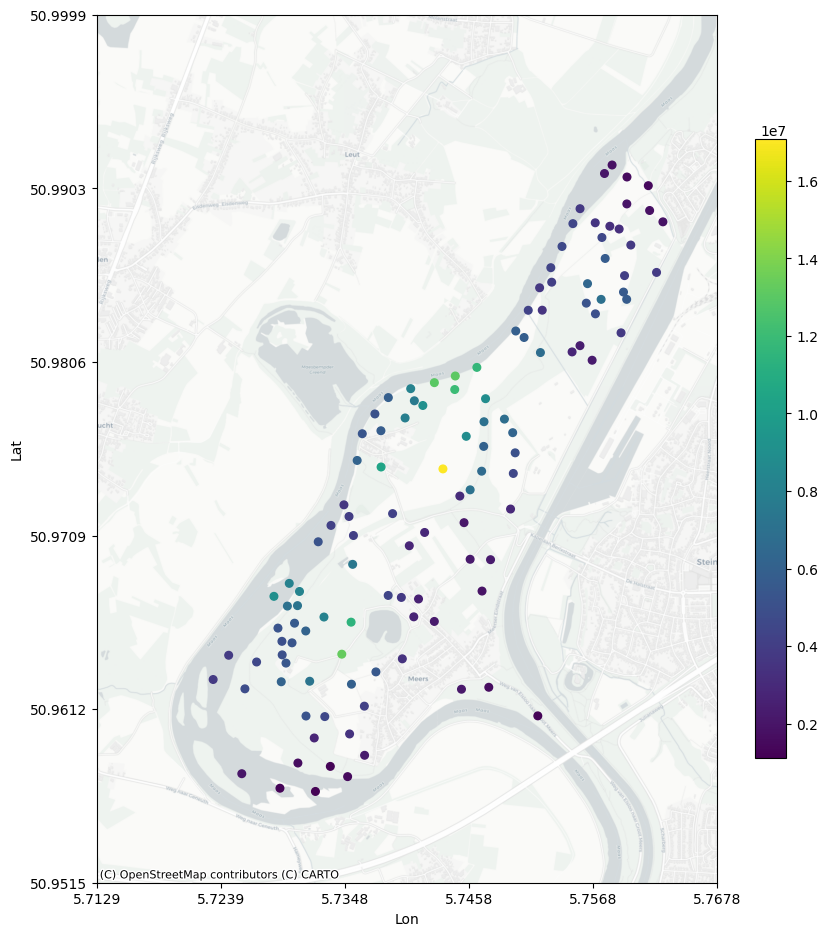}
        \caption{$\kappa(\mathbf M_i^{(\mathrm{nor})})$}
        \label{fig:meuse-diag-condMnor}
    \end{subfigure}\hfill
    \begin{subfigure}[t]{0.45\textwidth}
        \centering
        \includegraphics[width=\linewidth]{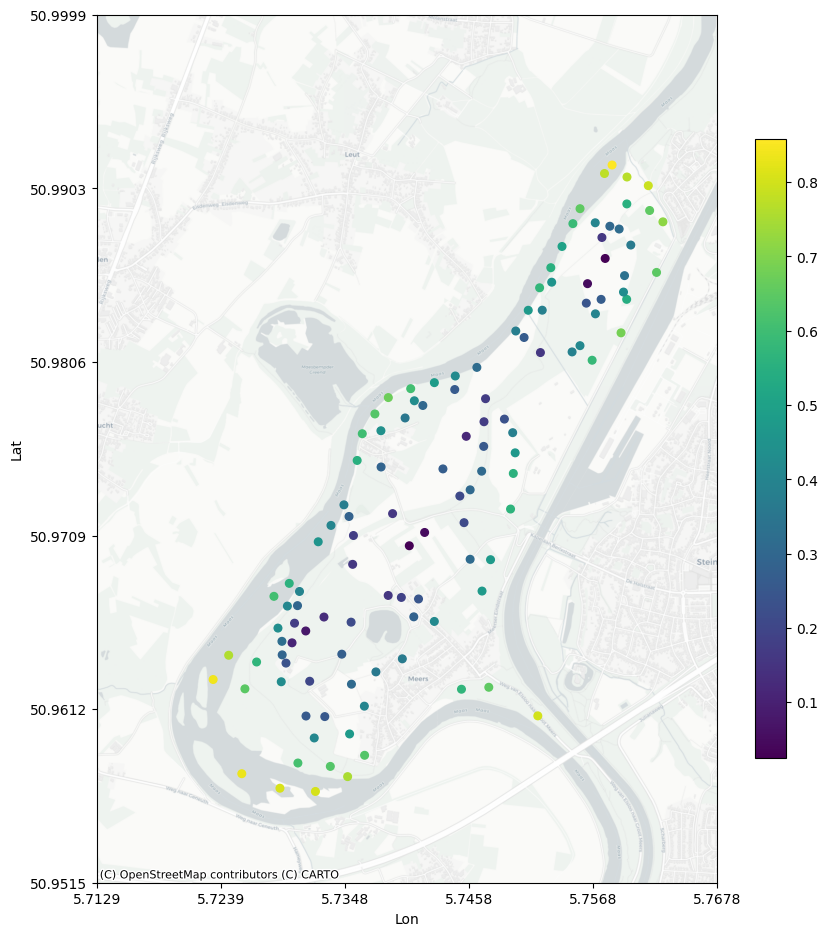}
        \caption{$r_{\phi}$}
        \label{fig:meuse-diag-rphi}
    \end{subfigure}
    \caption{Meuse: spatial distribution of solver and geometry diagnostics.}
    \label{fig:meuse-diag}
\end{figure}

\paragraph{Meuse.}
In Meuse, a risk of local-regression coefficient maps is that visually striking patterns may be artifacts of an unstable local solve. When the realized weighted normal matrix $\mathbf M_i^{(\mathrm{nor})}$ is ill-conditioned, small perturbations in the local neighborhood can induce disproportionately large swings in the estimated coefficients. GR makes this concrete by explicitly reporting solver and geometry diagnostics. Figure~\ref{fig:meuse-diag-condMnor} maps $\kappa(\mathbf M_i^{(\mathrm{nor})})$, and Figure~\ref{fig:meuse-diag-rphi} maps the directional concentration diagnostic $r_{\phi}$. Locations in the heavy tail of $\kappa(\mathbf M_i^{(\mathrm{nor})})$---and/or exhibiting strongly one-sided directional support as indicated by large $r_{\phi}$---are treated as \emph{numerically fragile}. In these regions, sharp spikes or sign reversals are not interpreted as substantive local effects under the chosen neighborhood geometry. Instead, they are labeled as \emph{diagnostically unsupported} and may be (i) masked in coefficient interpretation, or (ii) handed off to dependence-modeling workflows when spatial structure explanation is the primary objective. For completeness, Appendix~\ref{app:diagnostic-maps-meuse} reports the full set of diagnostic maps (including $n_{\mathrm{eff}}^{\mathrm{post}}$) and example reliability masking.

\begin{figure}[H]
    \begin{subfigure}[t]{0.45\textwidth}
        \centering
        \includegraphics[width=\linewidth]{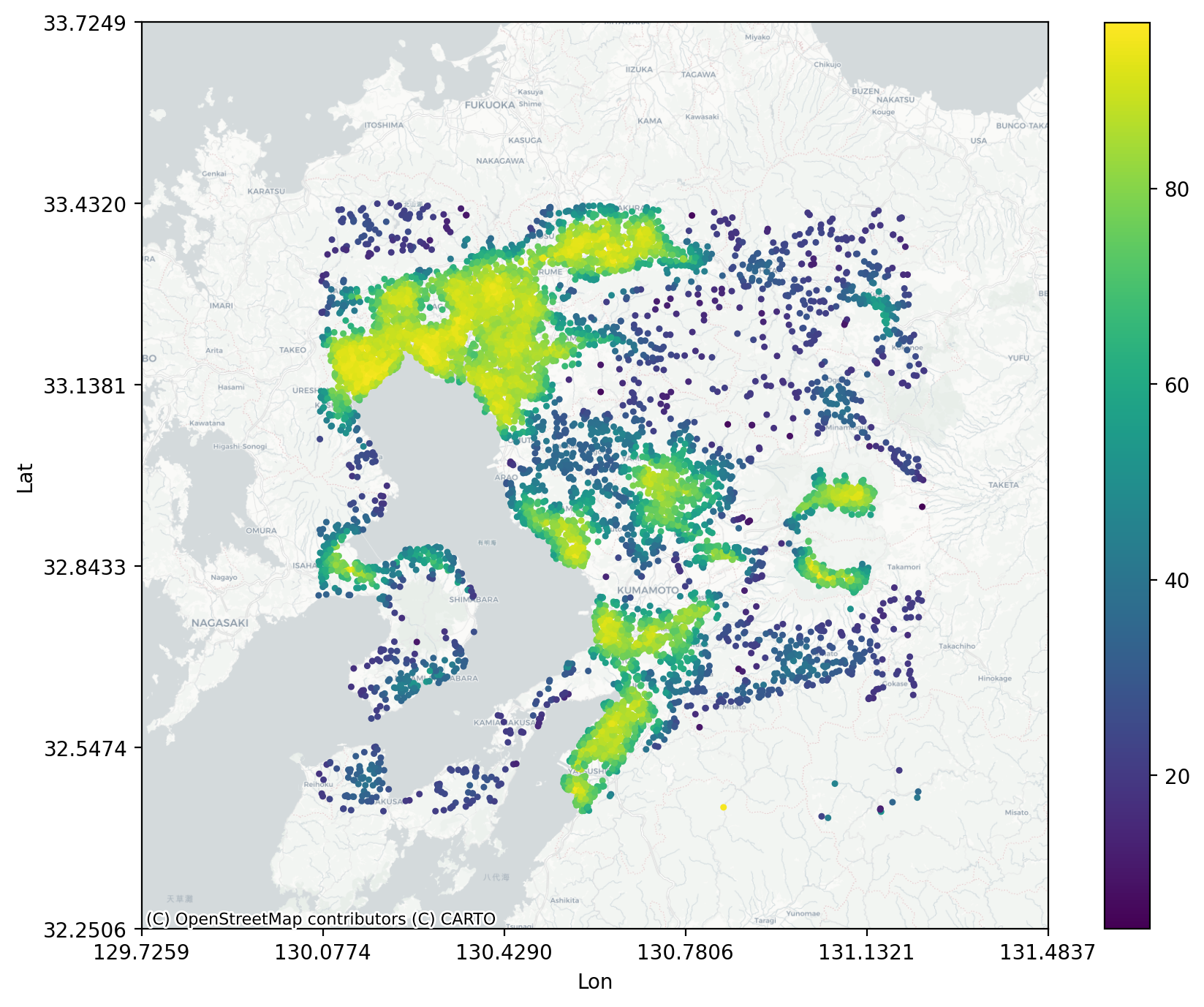}
        \caption{ $n_{\mathrm{eff}}^{\mathrm{post}}$}
        \label{fig:RicePaddies-diag-neffpost}
    \end{subfigure}\hfill
    \begin{subfigure}[t]{0.45\textwidth}
        \centering
        \includegraphics[width=\linewidth]{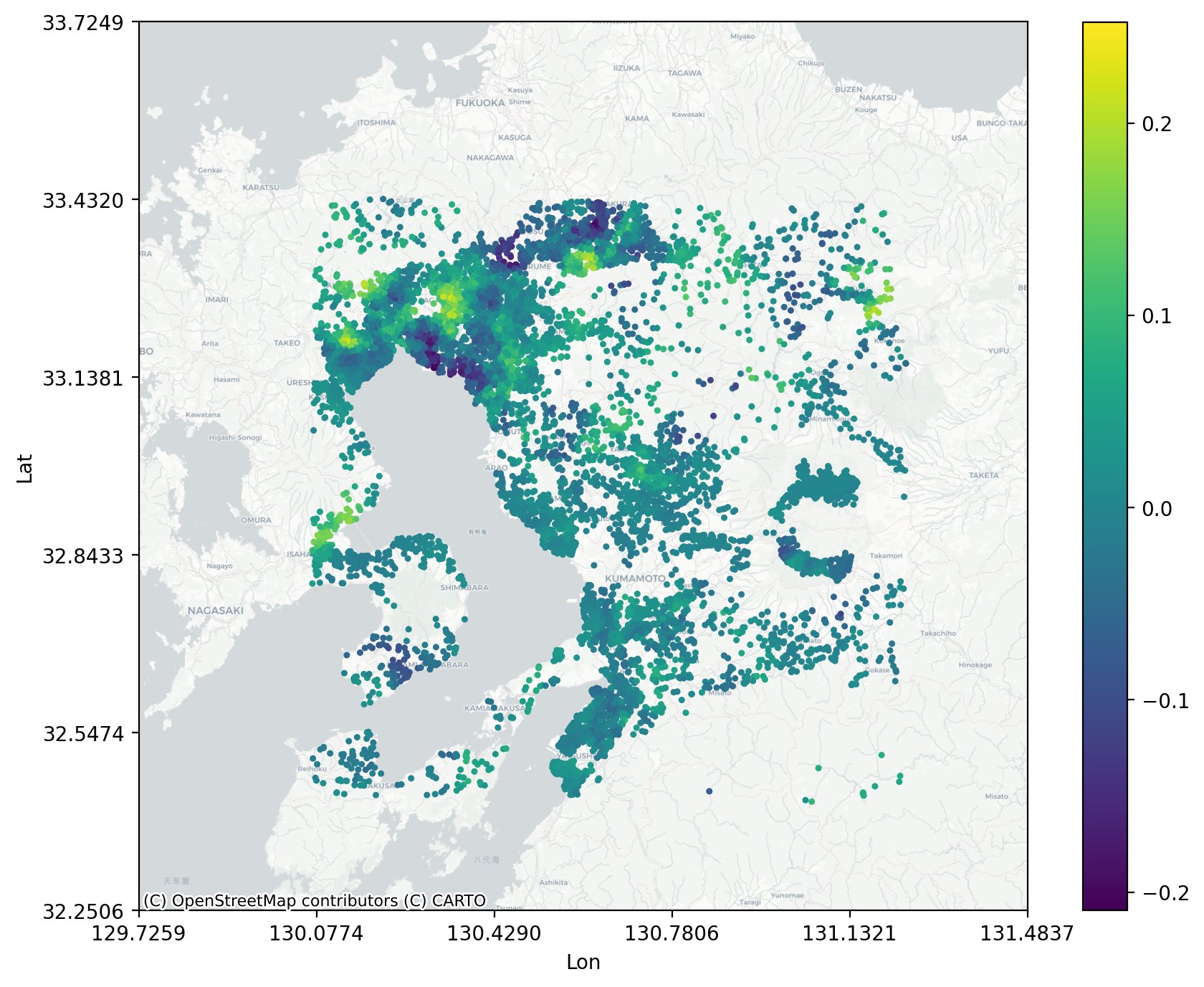}
        \caption{$\hat\beta_1(s)$}
        \label{fig:RicePaddies-diag-beta1}
    \end{subfigure}
    \caption{Rice paddies: spatial distribution of effective support and the local slope surface.}
    \label{fig:diagnostic-maps-RicePaddies}
\end{figure}

\paragraph{Rice paddies.}
In rice paddies, the main practical difficulty is uneven local support induced by the spatial configuration (irregular density and geometry), together with the fact that prediction-oriented alternatives can outperform GR in aggregate error metrics. GR is therefore not presented here as the best predictive model. Its role is instead diagnostic and interpretive: it makes the effective local support explicit through $n_{\mathrm{eff}}^{\mathrm{post}}$ (Figure~\ref{fig:RicePaddies-diag-neffpost}) and yields a local slope surface $\hat\beta_1(s)$ (Figure~\ref{fig:RicePaddies-diag-beta1}) that is intended to be read conditionally on that support. When $n_{\mathrm{eff}}^{\mathrm{post}}$ is small, the realized fit is effectively supported by only a small fraction of the nominal $K$ neighbors after deterministic weighting and ESS enforcement, so the corresponding coefficients are best read as high-variance local summaries rather than as robust substantive signals. Additional diagnostic maps (including $\kappa(\mathbf M_i^{(\mathrm{nor})})$ and $r_{\phi}$) and example reliability masking are provided in Appendix~\ref{app:diagnostic-maps-RicePaddies}.

\subsection{Summary}
This chapter demonstrates that GR can be applied to datasets with markedly different spatial structures while maintaining stable and transparent local estimation diagnostics and computational feasibility from small to large scale. The $\kappa(\cdot)$ diagnostics show that stability advantages are configuration- and dataset-dependent: GR has a pronounced tail-stability advantage over GWR/MGWR in Meuse, while MGWR is most stable in rice paddies under the standardized diagnostic convention (Appendix~\ref{app:condwls2}). In the larger rice paddies benchmark, prediction-oriented or dependence-aware alternatives achieve stronger predictive performance, but GR remains practically usable and continues to provide an explicit diagnostic layer for local reliability, including effective support and realized conditioning. These findings align with the Chapters~4--7 viewpoint that GR is a deterministic, geometry-aware local regression framework intended primarily as a diagnostic and interpretive tool for stable local estimation, rather than as a universally prediction-optimal replacement for dependence-modeling or nonlinear learners.
\section{Discussion}
This chapter consolidates the contribution of Gimbal Regression (GR) and clarifies its intended scope in relation to the estimator construction (Chapters~4--6), simulations (Chapter~7), and the empirical results (Chapter~8). 

The key novelties are to treat orientation as a primitive reference-frame object, to separate reference-frame choice (rotation/orientation) from influence redistribution (directional weighting), and to formalize the full weighting-and-safeguard procedure as a realized piecewise estimator map. Accordingly---prioritizing transparency of estimator assumptions, numerical behavior, and diagnostic outputs over universal predictive dominance---GR is best understood as an interpretable local modeling and diagnostic system that remains computationally predictable at scale.

\subsection{Summary of contribution}
GR makes three design commitments explicit. First, orientation is treated as a primitive object that fixes the reference frame in which directional similarity is evaluated. It is not introduced as an additional regression parameter, nor is it used to rotate the local design matrix. Second, GR separates two distinct mechanisms: the choice of reference frame for weighting, which is determined by orientation through the metric used to compute weights, and the redistribution of influence, which is implemented through directional weighting and acts only via the diagonal weight matrix. This separation clarifies which component changes the reference frame for weighting and which component changes leverage, while preserving the original local design matrix. Third, the full weight construction---including numerical safeguards and fallback branches---is fixed as a single realized piecewise mapping from data to weights and local estimates. Stability is therefore stated conditionally on the realized weights produced by this map, so that the theoretical statements correspond directly to the computational object that is actually executed.

Under these commitments, GR is a deterministic and auditable local linear framework: neighborhoods and weights are explicit per location, solves are closed-form, and diagnostics expose when local estimation is ill-posed due to anisotropic sampling or effective-information loss. Chapter~7 shows that the weight map activates under the intended stress regimes, and Chapter~8 shows that the same mechanisms remain numerically stable and computationally feasible on contrasting real datasets.

\subsection{Scope of applicability}
GR is most appropriate when spatial variation in relationships is expected to be locally smooth and substantively meaningful, and when the analyst values direct interpretability of local coefficients and diagnostics. Typical examples include environmental processes, agricultural systems, and regional socioeconomic settings where local conditions plausibly modulate covariate effects.

All estimation in GR is local. Any ``global'' summaries reported are post-hoc aggregations of local outputs (e.g., averaged errors or mapped diagnostics) and should not be interpreted as defining a separate global estimator. The value of the method lies in the spatial distribution of local coefficient estimates and diagnostics, not in a single global parameter vector.

\subsection{Design scope and interpretation}
The framework is intentionally specified as a deterministic local estimator with explicit diagnostics. Accordingly, several scope commitments guide how its outputs should be interpreted.

First, GR adopts deterministic weighting and does not specify a stochastic dependence model. The method does not estimate a covariance structure, dependence parameters, or a generative spatial process. Directional weighting is therefore a deterministic rule for allocating influence across directions and should not be interpreted as specifying stochastic spatial dependence.

Second, GR operates as a local estimation framework and does not pursue a global pooling objective. The estimator does not optimize a single global criterion through spatial pooling, hierarchical shrinkage, or global smoothing. Any ``global'' summaries reported in the empirical sections should therefore be understood as post-hoc aggregations of location-wise outputs rather than as quantities optimized by the estimation procedure itself.

Third, the framework prioritizes interpretability together with auditable numerical behavior. The benchmark comparisons in Chapter~8 include methods designed primarily for predictive performance under model-based assumptions (for example, covariance-based approaches) or for flexible function approximation. GR is not intended to replace those approaches. Instead, its role is to provide stable and interpretable local coefficient surfaces while reporting diagnostics that make the numerical behavior of local estimation transparent, including situations in which the local solve may be unreliable.

These scope statements are deliberate design choices: they support a single-pass local estimator whose assumptions, numerical behavior, and diagnostic outputs are reproducible and auditable at scale.

\subsection{Diagnostic-first regimes}
GR is designed so that its diagnostics remain informative even when local estimation becomes ill-posed or is repeatedly driven by safeguard branches.

When the realized neighborhood contains insufficient effective local information---for example due to severe local rank deficiency, extreme anisotropy, or systematic ESS-triggered fallback---the framework is still able to make the failure mode explicit through conditioning and ESS diagnostics. In such regions, the resulting coefficient surfaces should be interpreted as unstable local summaries. The appropriate use in these cases is therefore to map and report the diagnostic fields, such as conditioning measures and ESS or fallback frequencies, and to treat substantive interpretation of the coefficients as secondary.

A second situation arises near sharp regime boundaries or discontinuities. GR can help detect boundary-like structure through patterns of instability, residual diagnostics, or abrupt changes in local diagnostic fields. However, any coefficient surface produced by local smoothing may blur genuine structural breaks when neighborhoods straddle a boundary. In such boundary-candidate regions, interpretation should therefore be guided primarily by the diagnostics themselves, for example through boundary-aware neighborhood rules, explicit reporting of boundary proximity, or stratified fits estimated separately on each side of the suspected boundary.

In both regimes, the intended behavior is explicit: the realized branch (including any fallback) and the numerical diagnostics identify where local estimation is well-posed. Substantive claims about coefficient heterogeneity should be restricted to such well-posed regions, while elsewhere GR should be used primarily to map and report diagnostic fields.

\subsection{Relation to GWR/MGWR and the role of numerical diagnostics}

The inclusion of GWR and MGWR in Chapter~8 is intentional and should be interpreted carefully. These methods represent widely used local regression baselines, but their practical performance can be sensitive to local rank deficiency and geometry-induced ill-conditioning under fixed neighborhood specifications.

Chapter~8 highlights two complementary empirical observations. First, numerical behavior can differ substantially across local methods even when identical neighborhood rules are applied. Second, conditioning diagnostics (e.g., \texttt{condWLS2}) provide concrete evidence about the stability of local solves and help reveal potential failure modes. Together, these results illustrate that numerical behavior is itself an empirical property of the estimator implementation and that explicit conditioning diagnostics provide a practical way to assess it.

The purpose of these comparisons is not to argue that GWR/MGWR are ``inferior'', but to empirically contextualize the stability goal of GR: when local design matrices become difficult to invert or near-singular, stability is a first-order requirement for interpretability and reproducibility. In this sense, conditioning diagnostics are not merely implementation details; they are central to making local regression scientifically auditable.

\subsection{Relation to kriging and machine learning}

GR occupies a complementary role relative to model-based geostatistics and modern machine learning.

\paragraph{Kriging and covariance models.}
Kriging-based approaches are designed around a stochastic dependence structure and can achieve strong prediction when the covariance specification is adequate. Their explanatory outputs typically emphasize trend--covariance decomposition and dependence parameters. GR does not compete on that axis; instead, it provides direct access to spatially varying local covariate effect surfaces and per-location numerical diagnostics.

\paragraph{Machine learning models.}
Flexible learners (e.g., spatial random forests) can achieve strong prediction, particularly with large training sets and nonlinear interactions. However, spatial structure is often absorbed implicitly within complex model architectures, making it difficult to diagnose why a model succeeds or fails locally. GR prioritizes local interpretability: coefficients retain conventional meanings, and the method exposes when local estimation is ill-posed.

Accordingly, GR should be understood as an interpretable local modeling and diagnostic baseline that can inform when more complex predictive approaches (kriging or nonlinear learners) are warranted.

\subsection{Interpretability and diagnostic value}
Interpretability in GR follows from its construction: each local estimate is associated with an explicit neighborhood, an explicit directional structure, and a closed-form local regression problem. This enables coefficient surfaces to be interpreted as evidence of spatial heterogeneity conditional on the chosen locality rule rather than as artifacts of hidden global optimization.

Residual spatial diagnostics are reported explicitly in Chapter~8 and computed using a fixed adjacency structure independent of regression weights. This separation is important: it ensures that residual diagnostics reflect remaining spatial structure rather than a mechanical consequence of the weighting scheme. When residual spatial autocorrelation remains non-negligible, it should be interpreted as a signal that (i) the local linear specification is incomplete, (ii) unmodeled dependence or omitted covariates remain, or (iii) a dependence-modeling approach (e.g., kriging) may be more appropriate for the analysis goal.

Crucially, GR is designed to make failure modes visible. Numerical safeguards and conditioning diagnostics allow the analyst to distinguish substantive heterogeneity from artifacts of local rank deficiency, anisotropic sampling, or insufficient effective information.

\subsection{Implications for spatial modeling practice}
GR should be understood as a specialized framework for studying spatial heterogeneity under controlled and explicitly stated assumptions. Its contribution lies not in universal applicability, but in offering a disciplined alternative to both deterministic local smoothing approaches that can silently fail under ill-conditioned neighborhoods and opaque predictive models that may obscure local failure modes and reduce interpretability.

When used within its intended scope, GR provides a transparent and computationally predictable approach to local spatial modeling, and it supplies diagnostics that support model choice: it can justify when local linear modeling is adequate and when dependence modeling or nonlinear prediction is warranted.

\subsection{Scalability and computational predictability}
A practical advantage of GR is that its estimator is a deterministic single-pass map with predictable per-location cost: for each target location, the computation consists of (i) neighborhood construction under a fixed rule (e.g., KNN), (ii) diagonal weight evaluation under an oriented metric, and (iii) a closed-form local WLS solve. There is no estimator-defined iterative optimization, so runtime scales linearly in the number of target locations up to the cost of neighborhood search and small linear-algebra kernels (typically dominated by a $p\times p$ solve). This structure makes large-scale runs straightforward to parallelize across CPU cores or GPUs and keeps failure modes auditable: when local configurations are ill-posed, the realized branch (ESS fallback / uniform weighting) is explicit and reproducible, enabling downstream analyses to condition on—and report—the exact computational regime under which each local estimate was produced.

\subsection{Extensions and future work}
The framework is intentionally deterministic and modular, which creates clear interfaces for extensions without changing the core estimator structure.

\subsubsection{Incorporating prior knowledge via covariate construction}
Prior information can be injected through feature construction and basis expansion. For example, when stage-like metadata are available (e.g., phenology), local regressions can include phase indicators, interactions, or smooth basis functions. Estimation remains linear in parameters at each location, retaining closed-form solves while enabling nonlinear response shapes through transformed features.

\subsubsection{Informed neighborhoods and kernels}
Prior knowledge can also enter through neighborhood selection and weighting. Beyond geographic proximity, neighborhoods may incorporate similarity in auxiliary state variables. Directional weighting can likewise be centered using externally supported orientation cues. These are deterministic modifications that preserve reproducibility while refining the notion of local relevance.

\subsubsection{Locally adaptive nonlinearity with reproducible selection rules}
A further extension is locally adaptive functional flexibility using a finite candidate set of basis families (e.g., linear, saturating transforms, spline bases) selected by a deterministic rule (e.g., held-out error, information criteria, or a pre-specified domain rule). This introduces limited nonlinearity while retaining reproducibility and interpretability.

In summary, Gimbal Regression is positioned as a deterministic, interpretable, and scalable framework for local modeling when anisotropic geometry and local ill-conditioning are practical concerns. Its contribution is to make numerical stability and residual spatial diagnostics explicit outputs of the analysis, allowing practitioners to identify where local regression is trustworthy and where alternative spatial models are warranted. The empirical and simulation results support this scope. The method does not claim universal predictive dominance; rather, it provides a reproducible local estimator whose stability behavior is visible, controllable, and computationally predictable.

\section{Conclusion}
This paper proposed Gimbal Regression (GR), a deterministic, geometry-aware local linear regression framework aimed at making local estimation under spatial heterogeneity stable and auditable when neighborhood geometry is anisotropic or effectively low-dimensional. The core novelties are to treat orientation as a primitive reference-frame object, to separate reference-frame choice (rotation/orientation) from influence redistribution (directional weighting), and to specify the full weighting-and-safeguard procedure as a realized piecewise estimator map.

The theoretical and empirical results support the intended scope: GR provides closed-form local solves with explicit conditioning and ESS diagnostics, so well-posed regions can be identified and interpreted while ill-posed regions are handled transparently via realized safeguard branches. GR is therefore positioned as an interpretable local modeling and diagnostic system with predictable computation at scale, complementary to dependence-modeling approaches (e.g., kriging) and prediction-optimized nonlinear learners.

\section*{Author Contributions and Use of Generative AI}
The author used generative AI tools for language editing and formatting assistance. All methodological development, theoretical derivations, simulations, and empirical analyses were conceived, implemented, and verified by the author. The author takes full responsibility for the content of the manuscript.

\appendix
\section*{Appendix}

\setcounter{section}{0}

This appendix collects the explicit computational specification, supplementary experiment diagnostics, and implementation details referenced in the main text. No extensions, variants, or additional methodological claims are introduced.

\section{Computationally Realized Estimator Map (Piecewise Definition)}
\label{app:A}
This appendix provides a fully explicit, reproducible, piecewise definition of the computationally realized Gimbal Regression (GR) estimator map referenced in Chapters~4--5. The purpose is to define the realized orientation and weight construction deterministically, to formalize the one-shot ESS correction and uniform fallback, and to make precise what is conditioned on in the conditional stability statements of Chapter~5.

\subsection{Inputs, neighborhood rule, and deterministic constants}
\label{app:inputs-neighborhood-rule-and-deterministic-constants}
Fix a target location $s_i$ and a deterministic neighborhood rule producing $\mathcal N_i\subset\{1,\dots,n\}$ with cardinality $n_i=|\mathcal N_i|$. The realized estimator map is defined pointwise for each $i$.

Inputs per neighborhood: $\{(s_j,x_j,y_j)\}_{j\in\mathcal N_i}$, where $s_j\in\mathbb R^2$ are locations, $x_j\in\mathbb R^p$ are covariates (rows of the local design matrix), and $y_j\in\mathbb R$ are responses.

Deterministic tuning parameters: 
\begin{itemize}
    \item bandwidth $h>0$, 
    \item modulation parameter $\gamma\ge 0$, 
    \item ESS target $n_0>0$, 
    \item minimum ESS threshold $n_{\min}>0$, and 
    \item maximum anisotropy ratio $\eta_{\max}\ge 1$.
\end{itemize}

Numerical safeguard thresholds:
\begin{itemize}
    \item $\varepsilon_\phi>0$ (bearing isotropy threshold), 
    \item $\varepsilon_\theta>0$ (value-orientation identifiability threshold), and 
    \item $\varepsilon_\eta>0$ (geometry eigenvalue floor). 
\end{itemize}

These are numerical stability constants that make the map single-valued and reproducible; they are not stochastic model parameters.

Distances use a fixed metric $d_{ij}=d(s_i,s_j)$. Bearings and displacement vectors are computed in a fixed local convention (e.g., tangent-plane East–North), consistent across all $i$.

\subsection{Tangent-plane displacement and bearing angles}
\label{app:tangent-plane-displacement-and-bearing-angles}
Let $\mathbf\Delta_{ij}\in\mathbb R^2$ denote the tangent-plane displacement from $s_i$ to $s_j$, and let $\theta_{ij}\in(-\pi,\pi]$ denote the corresponding bearing angle (consistent with the chosen displacement convention). Let $z_{ij}$ denote the (optionally normalized) distance regressor; in the notation of Chapter 4, this may be taken as $z_{ij}=d_{ij}/u$ for a fixed scale $u>0$.

\subsection{Bearing resultant, isotropy detection, and $\phi_i$}
\label{app:bearing-resultant-isotropy-detection}
Define distance-decay weights for bearing aggregation:
\begin{equation}
    \omega_{ij}=\exp\left(-\frac{d_{ij}^2}{h^2}\right).
\end{equation}

Define the weighted cosine–sine sums:
\begin{equation}
    c_i=\sum_{j\in\mathcal N_i}\omega_{ij}\cos\theta_{ij},
\end{equation}
\begin{equation}
    s_i^{(\mathrm{sin})}=\sum_{j\in\mathcal N_i}\omega_{ij}\sin\theta_{ij}.
\end{equation}

Define the normalized resultant length:
\begin{equation}
    r_i=\frac{\sqrt{c_i^2+(s_i^{(\mathrm{sin})})^2}}{\sum_{j\in\mathcal N_i}\omega_{ij}}.
\end{equation}

Define the bearing-based dominant direction (with isotropy deactivation):
\begin{equation}
    \phi_i=
    \begin{cases}
        \mathrm{atan2}(s_i^{(\mathrm{sin})},c_i), & r_i>\varepsilon_\phi,\\
        0, & r_i\le \varepsilon_\phi.
    \end{cases}
\end{equation}

Operational meaning: if $r_i\le \varepsilon_\phi$, the local bearing field is treated as angularly isotropic, the bearing component is deactivated by taking $\phi_i=0$, and in subsequent steps the rotation factor $R(\phi_i)$ equals the identity.

\subsection{Value-based orientation $\theta_{z,i}^\ast$ with identifiability rule (using $\mathrm{atan2}$)}
\label{app:value-based-orientation}
Define neighborhood means:
\begin{equation}
    \bar z_i=\frac{1}{n_i}\sum_{j\in\mathcal N_i} z_{ij},
\end{equation}

\begin{equation}
    \bar y_i=\frac{1}{n_i}\sum_{j\in\mathcal N_i} y_j.
\end{equation}

Define second-moment summaries:
\begin{equation}
    \operatorname{Var}_i(z)=\frac{1}{n_i}\sum_{j\in\mathcal N_i}(z_{ij}-\bar z_i)^2,
\end{equation}

\begin{equation}
    \operatorname{Var}_i(y)=\frac{1}{n_i}\sum_{j\in\mathcal N_i}(y_j-\bar y_i)^2,
\end{equation}

\begin{equation}
    \operatorname{Cov}_i(z,y)=\frac{1}{n_i}\sum_{j\in\mathcal N_i}(z_{ij}-\bar z_i)(y_j-\bar y_i).
\end{equation}

Define the identifiability score:
\begin{equation}
    g_{\mathrm{ident},i}=|\operatorname{Var}_i(y)-\operatorname{Var}_i(z)|+|2\operatorname{Cov}_i(z,y)|.
\end{equation}

Define the two-argument inverse tangent $\mathrm{atan2}(y,x)$ as the quadrant-identifying inverse tangent used in implementation, and then the value-based orientation (piecewise):
\begin{equation}
    \theta_{z,i}^\ast=
    \begin{cases}
        \frac{1}{2}\mathrm{atan2}\left(\operatorname{Var}_i(y)-\operatorname{Var}_i(z),2\operatorname{Cov}_i(z,y)\right), & g_{\mathrm{ident},i}>\varepsilon_\theta,\\
        0, & g_{\mathrm{ident},i}\le \varepsilon_\theta.
    \end{cases}
\end{equation}

Implementation note (non-identifiability): when both arguments passed to $\mathrm{atan2}$ are numerically small, the rule above deterministically selects the non-identifiable branch and sets $\theta_{z,i}^\ast=0$.

\subsection{Geometry-based anisotropy ratio $\eta_i$ (unique and reproducible)}
\label{app:geometry-based-anisotropy-ratio}
Define normalized weights for geometry aggregation:
\begin{equation}
    \tilde\omega_{ij}=\frac{\omega_{ij}}{\sum_{k\in\mathcal N_i}\omega_{ik}}.
\end{equation}

Define the weighted second-moment matrix of the displacement field:
\begin{equation}
    \mathbf S_i=\sum_{j\in\mathcal N_i}\tilde\omega_{ij}\mathbf\Delta_{ij}\mathbf\Delta_{ij}^\top.
\end{equation}

Let the eigenvalues of $\mathbf S_i$ be $\lambda_{\max}(\mathbf S_i)\ge \lambda_{\min}(\mathbf S_i)\ge 0$. Define the floored minimum eigenvalue:
\begin{equation}
    \lambda_{\min}^\dagger(\mathbf S_i)=\max\{\lambda_{\min}(\mathbf S_i),\varepsilon_\eta\}.
\end{equation}        

Define the raw anisotropy ratio:
\begin{equation}
    \eta_i^{raw}=\sqrt{\frac{\lambda_{\max}(\mathbf S_i)}{\lambda_{\min}^\dagger(\mathbf S_i)}}.
\end{equation}        

Define the clipped anisotropy ratio:
\begin{equation}
    \eta_i=\min\{\max\{\eta_i^{raw},1\},\eta_{\max}\}.
\end{equation}

Interpretation: $\varepsilon_\eta$ and $\eta_{\max}$ are deterministic numerical constants ensuring stability and preventing extreme anisotropy; this makes $\eta_i$ uniquely reproducible from neighborhood geometry.

\subsection{Metric construction and raw directional weights}
\label{app:metric-construction-and-raw-directional-weights}
Let $R(\alpha)\in\mathbb R^{2\times 2}$ denote the standard planar rotation matrix. Define the metric basis:
\begin{equation}   
    \mathbf Q_i=R(\phi_i)R(\theta_{z,i}^\ast).
\end{equation}

Define the anisotropy scaling matrix:
\begin{equation}
    \mathbf\Lambda_i=h^{-2}\operatorname{diag}(1,\eta_i^{-2}).
\end{equation}

Define the induced (weight-evaluation) metric:
\begin{equation}
    \mathbf M_i^{(\mathrm{met})}=\mathbf Q_i\mathbf\Lambda_i\mathbf Q_i^\top.
\end{equation}

Define raw directional weights:
\begin{equation}
    w_{ij}^{raw}=\exp\left(-\mathbf\Delta_{ij}^\top\mathbf M_i^{(\mathrm{met})}\mathbf\Delta_{ij}\right).
\end{equation}

\subsection{One-shot ESS correction and uniform fallback (piecewise)}
\label{app:one-shot-ESS-correction-and-uniform-fallback}
Define the raw normalization constant:
\begin{equation}
    S_i^{raw}=\sum_{k\in\mathcal N_i}w_{ik}^{raw}.
\end{equation}

Define normalized raw weights:
\begin{equation}
    \tilde w_{ij}^{raw}=\frac{w_{ij}^{raw}}{S_i^{raw}}.
\end{equation}

Define effective sample size (ESS) of the normalized raw weights:
\begin{equation}
    n_{\mathrm{eff}}(w_i^{raw})=\frac{1}{\sum_{j\in\mathcal N_i}(\tilde w_{ij}^{raw})^2}.
\end{equation}

Define the one-shot effective bandwidth:
\begin{equation}
    h_{\mathrm{eff},i}=h\sqrt{\frac{n_0}{n_{\mathrm{eff}}(w_i^{raw})}}.
\end{equation}

Recompute the metric using $h=h_{\mathrm{eff},i}$:
\begin{equation}
    \mathbf\Lambda_i^{(\mathrm{eff})}=h_{\mathrm{eff},i}^{-2}\operatorname{diag}(1,\eta_i^{-2}),
\end{equation}

\begin{equation}
    \mathbf M_{i,\mathrm{eff}}^{(\mathrm{met})}=\mathbf Q_i\mathbf\Lambda_i^{(\mathrm{eff})}\mathbf Q_i^\top.
\end{equation}

Define recomputed weights:
\begin{equation}
    w_{ij}^{(1)}=\exp\left(-\mathbf\Delta_{ij}^\top\mathbf M_{i,\mathrm{eff}}^{(\mathrm{met})}\mathbf\Delta_{ij}\right).
\end{equation}

Normalize recomputed weights:
\begin{equation}
    S_i^{(1)}=\sum_{k\in\mathcal N_i}w_{ik}^{(1)},
\end{equation}

\begin{equation}
    \tilde w_{ij}^{(1)}=\frac{w_{ij}^{(1)}}{S_i^{(1)}}.
\end{equation}

Compute ESS after the one-shot correction:
\begin{equation}
    n_{\mathrm{eff}}(w_i^{(1)})=\frac{1}{\sum_{j\in\mathcal N_i}(\tilde w_{ij}^{(1)})^2}.
\end{equation}

Define the final weights by the uniform fallback rule:
\begin{equation}
    w_{ij}=
    \begin{cases}
        \frac{1}{n_i}, & n_{\mathrm{eff}}(w_i^{(1)})<n_{\min},\\
        \tilde w_{ij}^{(1)}, & n_{\mathrm{eff}}(w_i^{(1)})\ge n_{\min}.\end{cases}
\end{equation}

Finally define:
\begin{equation}
    \mathbf W_i=\mathrm{diag}(w_{ij}).
\end{equation}

This makes the realized weight map explicitly piecewise: the branch is selected deterministically by the post-correction ESS threshold test.

\subsection{Realized weight map (summary definition)}
\label{app:realized-weight-map}
For fixed $i$ and neighborhood $\mathcal N_i$, the realized weight map is the deterministic composition:
\begin{equation}
    \{(s_j,x_j,y_j)\}_{j\in\mathcal N_i}\mapsto(\phi_i,\theta_{z,i}^\ast,\eta_i)\mapsto \{w_{ij}\}_{j\in\mathcal N_i}\mapsto \mathbf W_i.
\end{equation}

All conditional statements in Chapter~5 are to be interpreted as conditioning on the realized neighborhood configuration and the realized output of this piecewise map, including the realized branch selection and the realized weight matrix $\mathbf W_i$.

\subsection{Realized local estimator map and well-posedness convention}
\label{app:realized-local-estimator-and-well-posedness-convention}
Given the realized $\mathbf W_i$, define the local design matrix $\mathbf X_i\in\mathbb R^{n_i\times p}$ and response vector $\mathbf y_i\in\mathbb R^{n_i}$ over $\mathcal N_i$. The realized GR estimator (Section~4.6) is:
\begin{equation}
    \hat\beta_i=\left(\mathbf X_i^\top \mathbf X_i+2\gamma \left(\mathbf X_i^\top \mathbf W_i \mathbf X_i\right)\right)^{-1}\left(\mathbf X_i^\top \mathbf y_i+2\gamma \left( \mathbf X_i^\top \mathbf W_i \mathbf y_i\right)\right).
\end{equation}

Well-posedness convention (deterministic): the estimator is reported only when the normal matrix is nonsingular. Operationally, pathological cases are handled deterministically as follows.

If the uniform-fallback branch is triggered in Appendix~\ref{app:one-shot-ESS-correction-and-uniform-fallback}, then $\mathbf W_i$ is set to the uniform weights over $\mathcal N_i$.

If, after this realized weighting, the normal matrix remains singular (e.g., due to rank deficiency of $\mathbf X_i$ on $\mathcal N_i$), the local estimate at $i$ is declared undefined and the location is flagged/excluded in downstream mapping. This exclusion is part of the realized estimator map and is deterministic under the fixed neighborhood rule and fixed design specification.

This convention matches the role of Appendix~A as the explicit computational specification referenced in Chapter 4 (safeguards ensuring well-posedness of the realized map) and Chapter 5 (conditional stability given the realized output and branch).

\section{Additional diagnostic maps for Experiment 7.1}
\label{app:B}
\begin{figure}[H]
    \centering
    \small
    \begin{subfigure}[t]{0.45\textwidth}
        \centering
        \includegraphics[width=\linewidth]{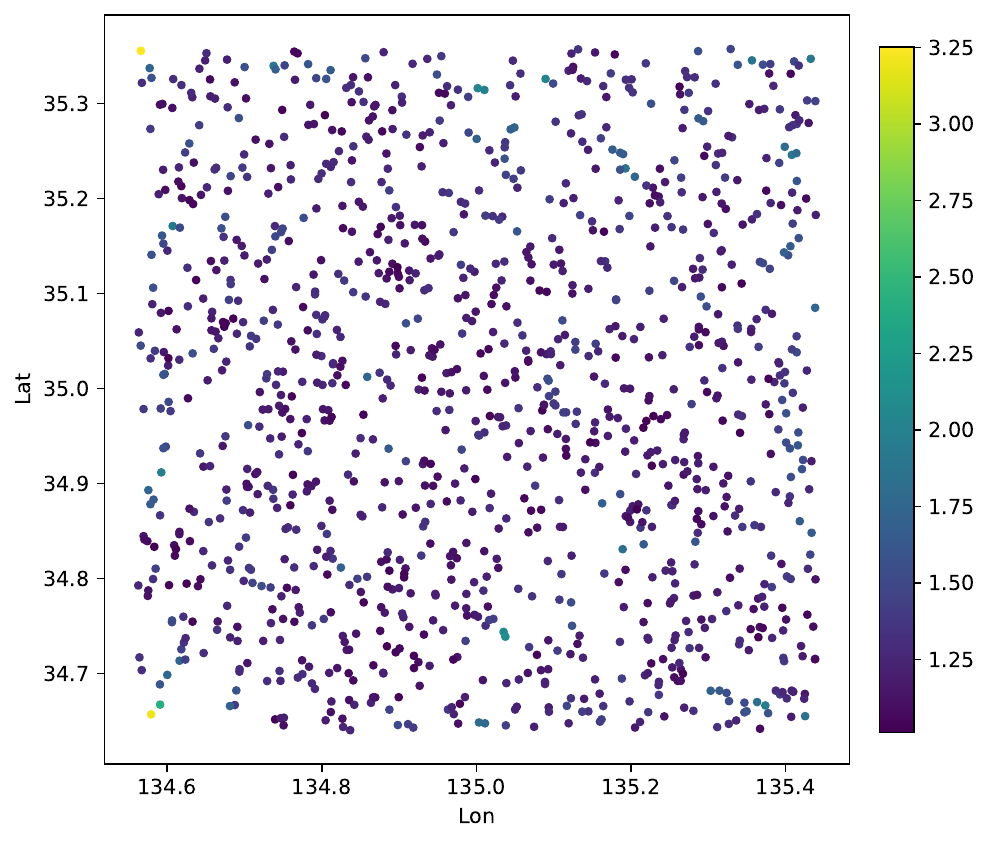}
        
        {(g) $\eta$}\vspace{4pt}
    \end{subfigure}\hfill
    \begin{subfigure}[t]{0.45\textwidth}
        \centering
        \includegraphics[width=\linewidth]{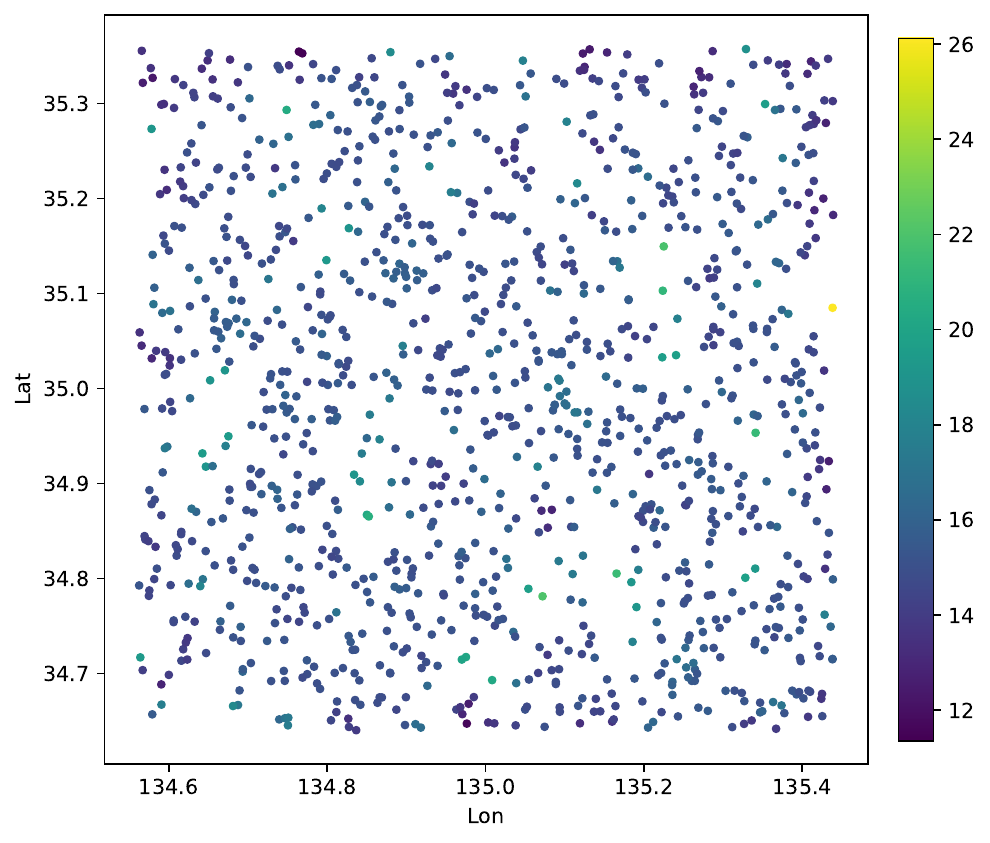}
        
        {(h) $n_{\mathrm{eff}}^{\mathrm{post}}$}\vspace{4pt}
    \end{subfigure}
    \begin{subfigure}[t]{0.45\textwidth}
        \centering
        \includegraphics[width=\linewidth]{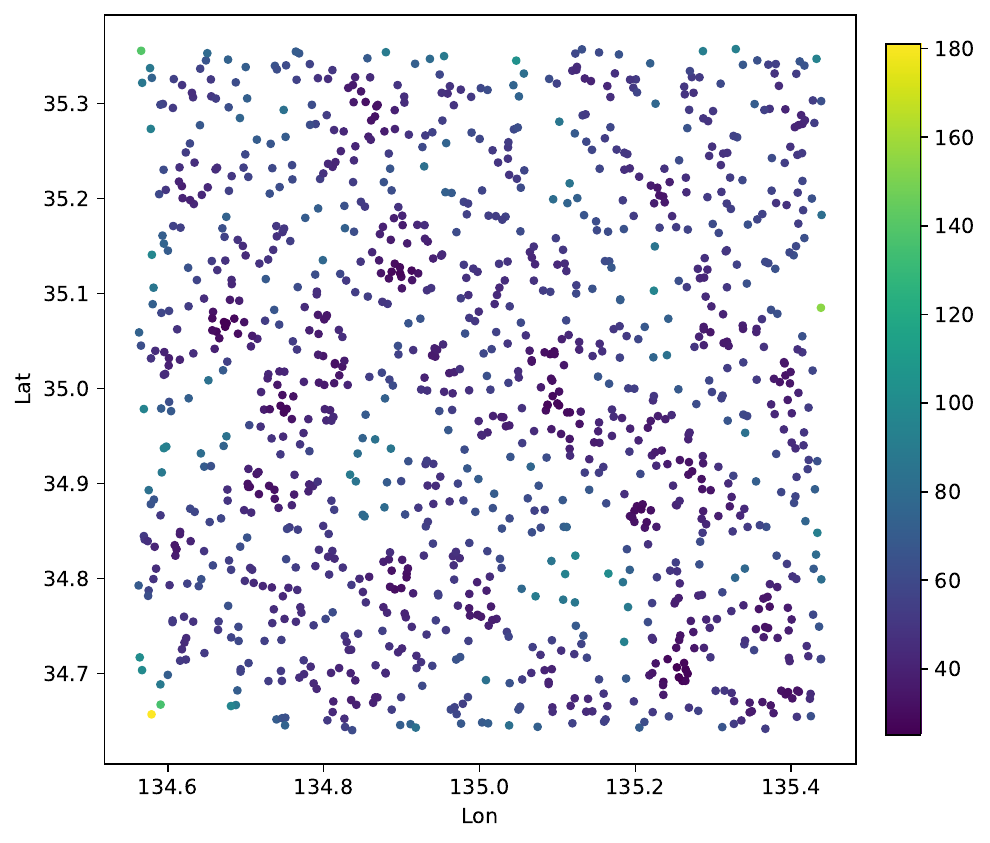}
        
        {(i) $\kappa(\mathbf M_i^{(\mathrm{nor})})$}\vspace{4pt}
    \end{subfigure}\hfill
    \begin{subfigure}[t]{0.45\textwidth}
        \centering
        \includegraphics[width=\linewidth]{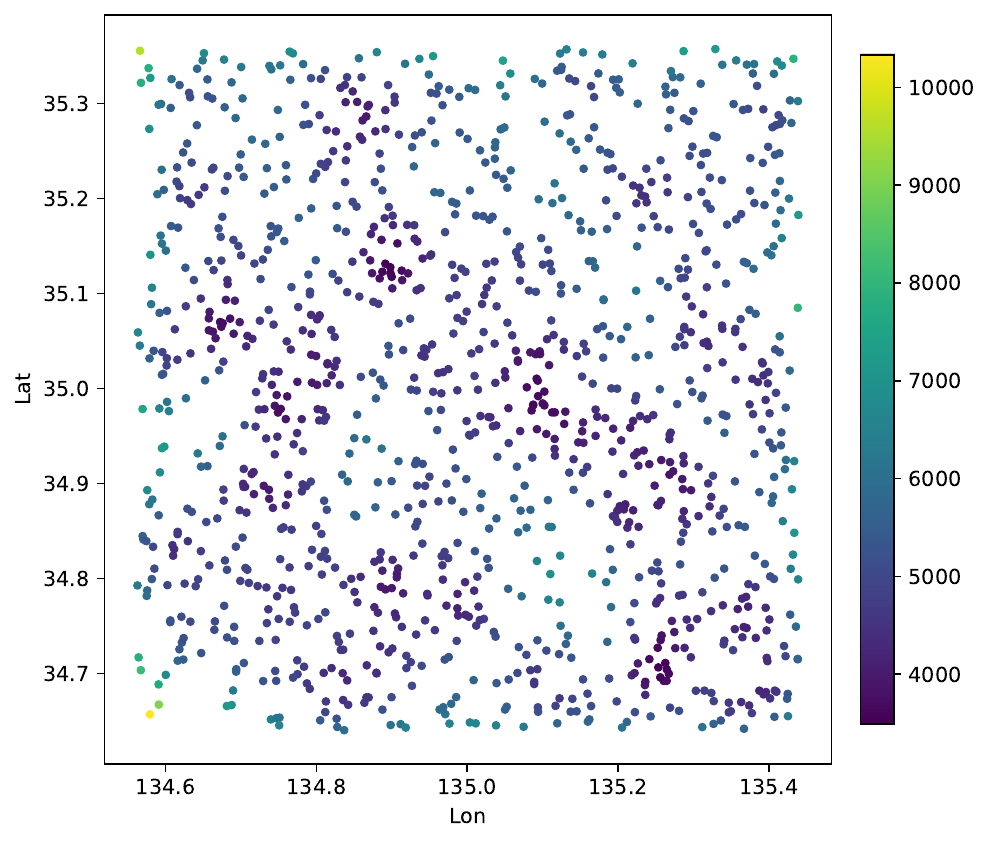}
        
        {(j) KNN mean distance ($K{=}30$)}\vspace{4pt}
        
    \end{subfigure}
    \caption{Spatial distributions}
    \label{fig:ex71}
\end{figure}

This appendix provides the full per-metric log (Table~\ref{tab:exp71-full-log}) and additional figures (Figure~\ref{fig:ex71}) for Exp 7.1. The figures visualize the spatial distributions of the variables ($X$, $y$) and the diagnostic quantities computed under the GR(full) specification.
\subsection{Full map-level log}
\begin{table}[H]
    \centering
    \begin{threeparttable}    
    \caption{Full map-level log.}
    \label{tab:exp71-full-log}
    \footnotesize    
    \setlength{\tabcolsep}{6pt}
    \renewcommand{\arraystretch}{1.15}
    \begin{tabularx}{0.75\textwidth}{@{\hspace{0.5em}} X r r r r @{\hspace{0.5em}}}
            \toprule
            Quantity
            & Isotropic-proxy
            & GR ($\theta_z$ OFF)
            & GR (full)
            & GR (full; strict $\varepsilon_\phi$) \\
            \midrule
            \multicolumn{5}{@{}l}{\textit{Fit summaries (context only)}}\\
                $N$ & 1200 & 1200 & 1200 & 1200 \\
                $\mu(\mathrm{RMSE})$ & 0.960 & 0.960 & 0.960 & 0.960 \\
                $\sigma(\mathrm{RMSE})$ & 0.096 & 0.096 & 0.096 & 0.096 \\
                $\mu(R^2)$ & 0.493 & 0.493 & 0.493 & 0.493 \\
                $\sigma(R^2)$ & 0.174 & 0.174 & 0.174 & 0.174 \\
            \addlinespace
            \multicolumn{5}{@{}l}{\textit{Conditioning diagnostics}}\\
                $\mu(\kappa(\mathbf M_i^{(\mathrm{nor})}))$ & 51.235 & 51.247 & 51.257 & 51.260 \\
                $\sigma(\kappa(\mathbf M_i^{(\mathrm{nor})}))$ & 15.211 & 15.088 & 15.156 & 15.165 \\
                $p_{50}(\kappa(\mathbf M_i^{(\mathrm{nor})}))$ & 48.626 & 48.810 & 48.773 & 48.794 \\
                $p_{95}(\kappa(\mathbf M_i^{(\mathrm{nor})}))$ & 79.737 & 79.508 & 79.149 & 79.188 \\
                $p_{99}(\kappa(\mathbf M_i^{(\mathrm{nor})}))$ & 95.524 & 94.819 & 94.870 & 94.870 \\
            \addlinespace
                
            \multicolumn{5}{@{}l}{\textit{ESS-related quantities}}\\
                $\mu(n_{\mathrm{eff}}^{\mathrm{raw}})$ & 10.565 & 12.883 & 12.900 & 12.940 \\
                $\sigma(n_{\mathrm{eff}}^{\mathrm{raw}})$ & 2.953 & 3.274 & 3.257 & 3.309 \\
                $\mu(n_{\mathrm{eff}}^{\mathrm{post}})$ & 15.224 & 15.201 & 15.206 & 15.211 \\
                $\sigma(n_{\mathrm{eff}}^{\mathrm{post}})$ & 1.914 & 1.222 & 1.201 & 1.208 \\
            \addlinespace
                
            \multicolumn{5}{@{}l}{\textit{Geometry / orientation diagnostics}}\\
                $\mu(\eta_i)$ & 1.000 & 1.263 & 1.263 & 1.263 \\
                $\sigma(\eta_i)$ & 0.000 & 0.193 & 0.193 & 0.193 \\
                $\mu(r_{\phi,i})$ & 0.301 & 0.301 & 0.301 & 0.301 \\
                $\sigma(r_{\phi,i})$ & 0.158 & 0.158 & 0.158 & 0.158 \\
                $\mu(g_{\mathrm{ident},i})$ & 1.608 & 1.608 & 1.608 & 1.608 \\
                $\sigma(g_{\mathrm{ident},i})$ & 0.804 & 0.804 & 0.804 & 0.804 \\
                $\mu(\theta_{z,i}^{\ast})$ & 0.000 & 0.000 & 0.748 & 0.748 \\
                $\sigma(\theta_{z,i}^{\ast})$ & 0.000 & 0.000 & 0.301 & 0.301 \\
            \addlinespace
                
            \multicolumn{5}{@{}l}{\textit{Branch rates / fallback proxies}}\\
                $\Pr(\phi_i=0)$ & 1.000 & 0.000 & 0.000 & 0.550 \\
                $\Pr(\theta_{z,i}^{\ast}=0)$ & 1.000 & 1.000 & 0.000 & 0.000 \\    
                $\Pr(\mathrm{uniform})$ & 0.000 & 0.000 & 0.000 & 0.000 \\
            \bottomrule
        \end{tabularx}
    \begin{tablenotes}[flushleft]
        \footnotesize
        \item Notes. Here, $\mu(\cdot)$ and $\sigma(\cdot)$ denote the mean and standard deviation across target locations $i$. $p_{50}(\cdot)$ denotes median across targets, and $p_{95}(\cdot)$ and $p_{99}(\cdot)$ denote the empirical 95th and 99th percentiles. The “strict $\varepsilon_\phi$” column corresponds to the diagnostic-only rerun that increases the frequency of the $\phi_i=0$ identifiability declaration. All values are rounded to three decimals.      
    \end{tablenotes}    
    \end{threeparttable}
\end{table}

\subsection{Spatial distributions of the variables ($X$, $y$) and diagnostic quantities under GR(full)}
\begin{figure}[H]
    \small
    \centering
    \begin{subfigure}[t]{0.45\textwidth}
        \centering
        \includegraphics[width=\linewidth]{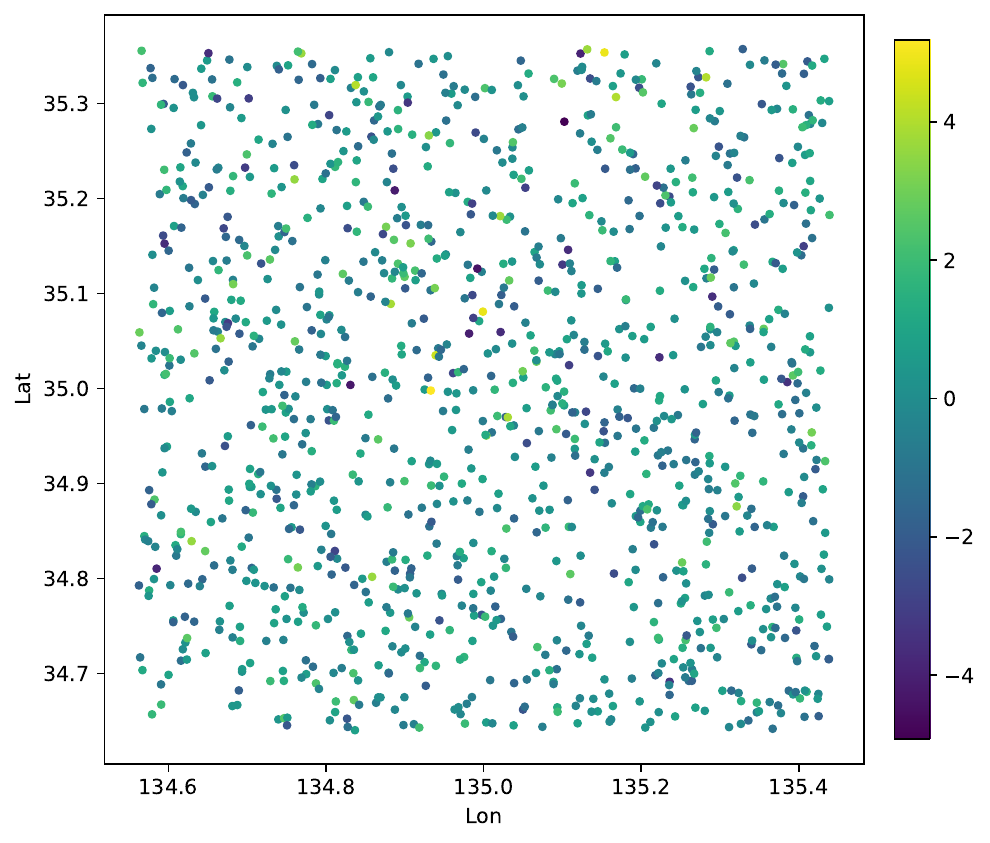}
        
        {(a) $y$}\vspace{4pt}
    \end{subfigure}\hfill
    \begin{subfigure}[t]{0.45\textwidth}
        \centering
        \includegraphics[width=\linewidth]{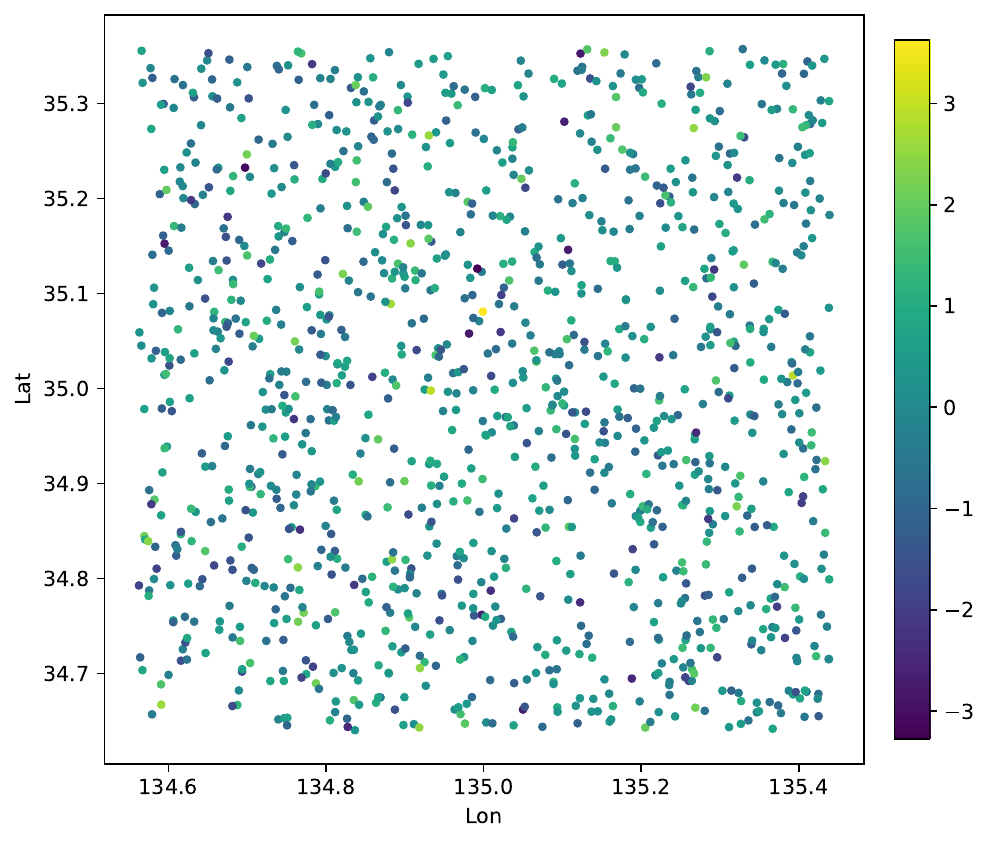}
        
        {(b) $X$}\vspace{4pt}
    \end{subfigure}
    \begin{subfigure}[t]{0.45\textwidth}
        \centering
        \includegraphics[width=\linewidth]{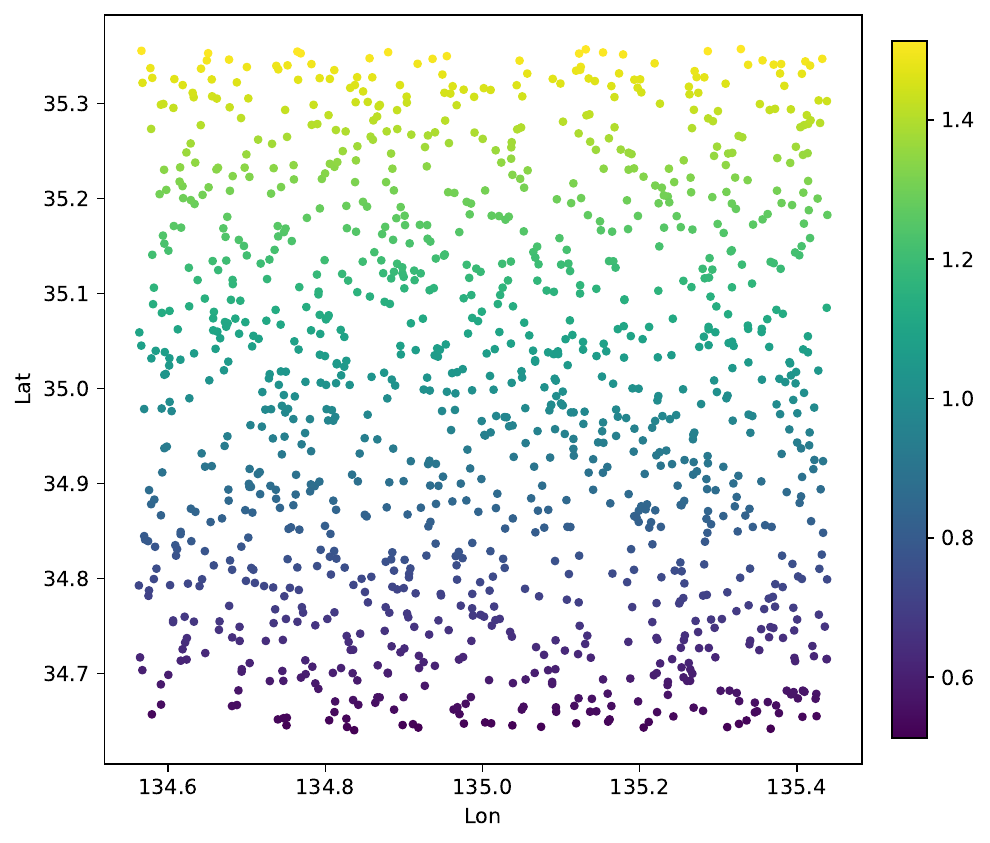}
        
        {(c) $\hat{\beta}_1(s)$}\vspace{4pt}
    \end{subfigure}\hfill
    \begin{subfigure}[t]{0.45\textwidth}
        \centering
        \includegraphics[width=\linewidth]{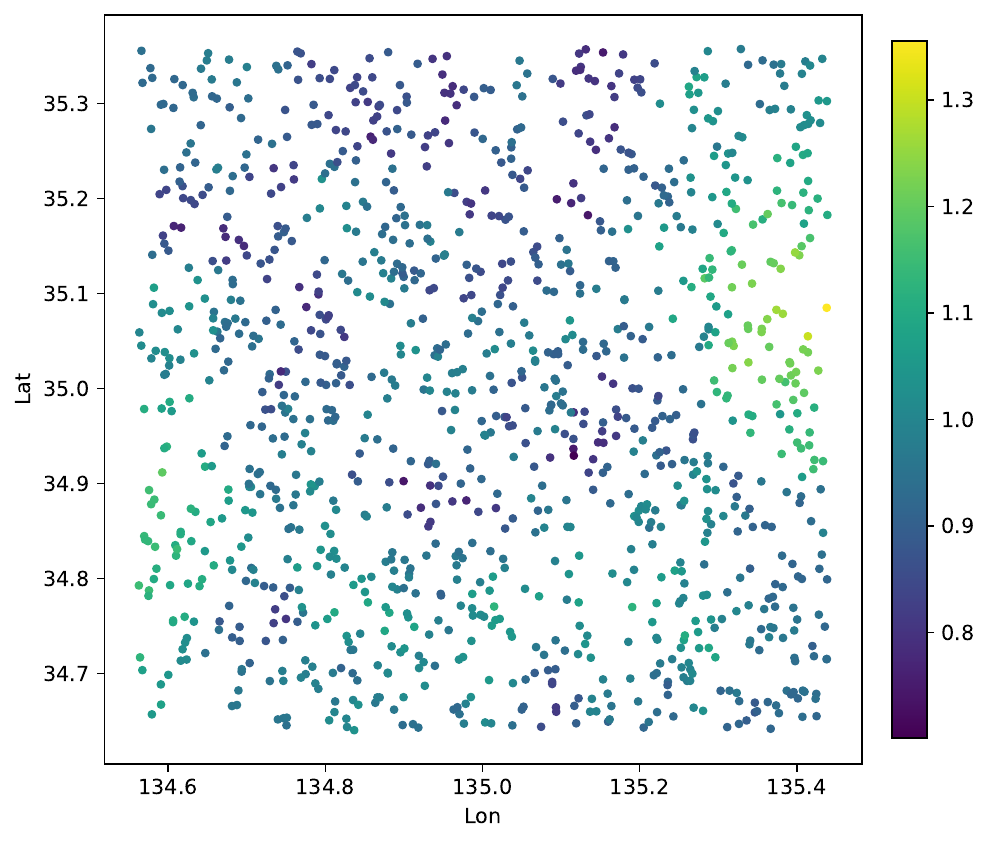}
        
        {(d) RMSE}\vspace{4pt}
    \end{subfigure}
    \begin{subfigure}[t]{0.45\textwidth}
        \centering
        \includegraphics[width=\linewidth]{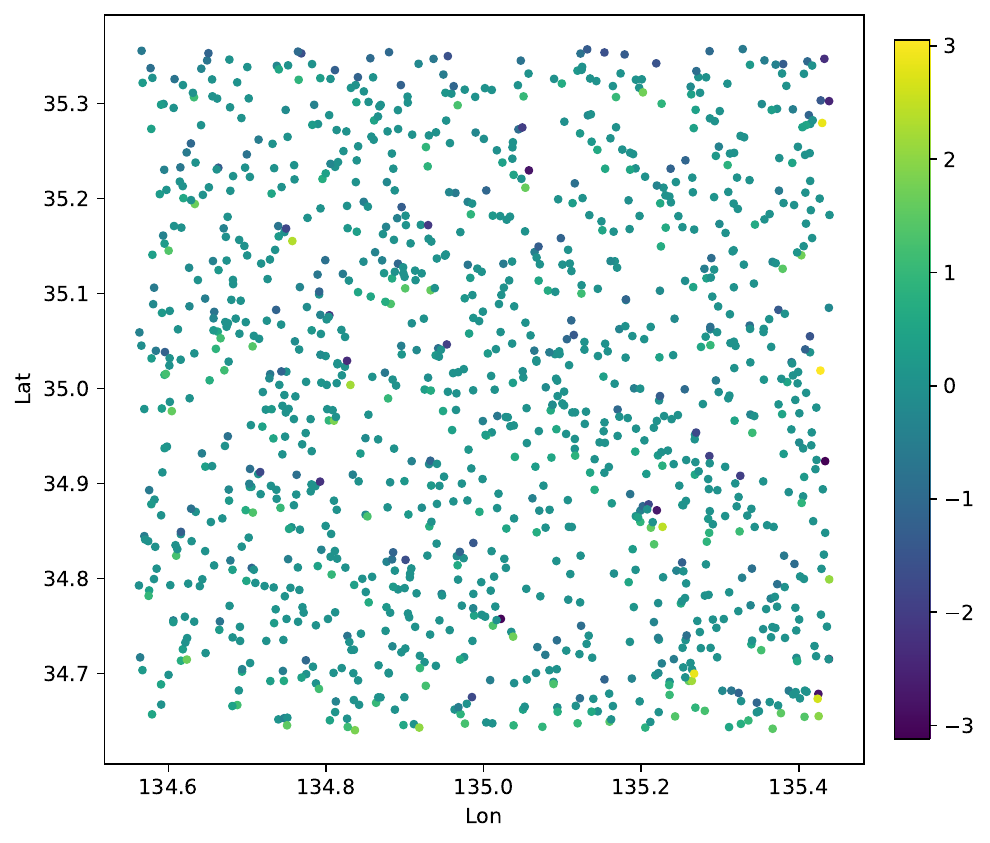}
        
        {(e) $\phi$}\vspace{4pt}
    \end{subfigure}\hfill
    \begin{subfigure}[t]{0.45\textwidth}
        \centering
        \includegraphics[width=\linewidth]{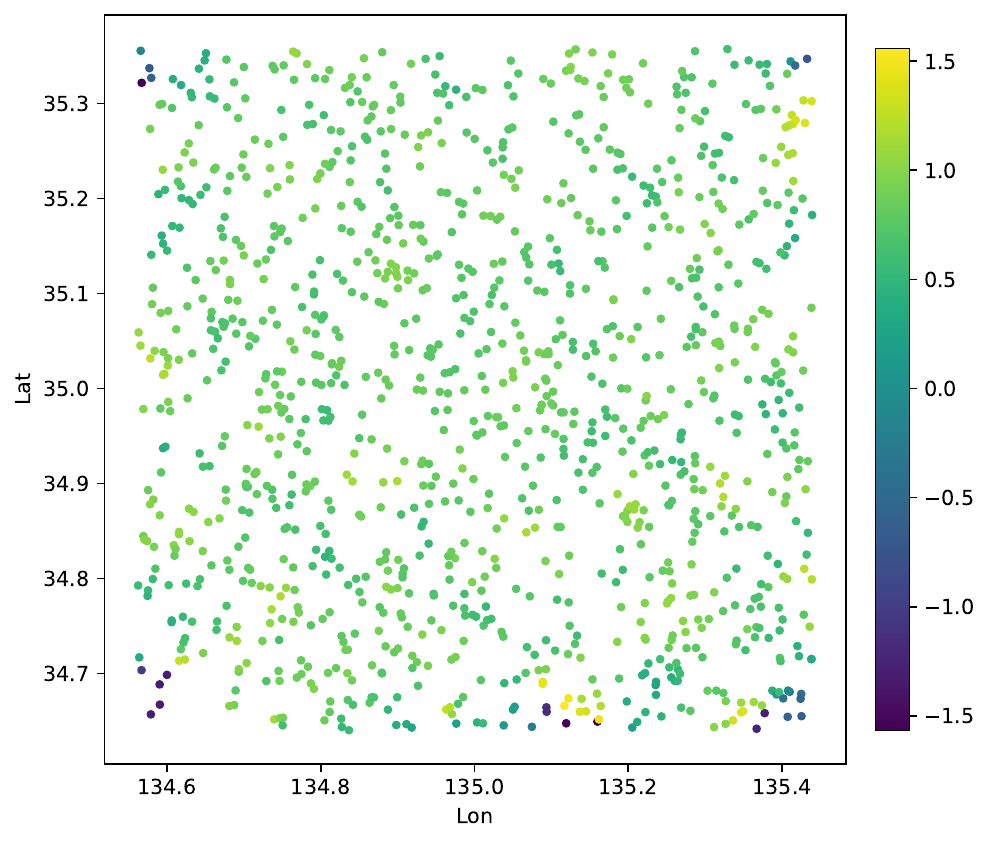}
        
        {(f) $\theta_{z}^\ast$}\vspace{4pt}
    \end{subfigure}
\end{figure}
\clearpage
\section{Additional diagnostic maps for Experiment 7.2}
\label{app:C}
\begin{figure}[H]
    \centering
    \begin{subfigure}[t]{0.45\textwidth}
        \centering
        \includegraphics[width=\linewidth]{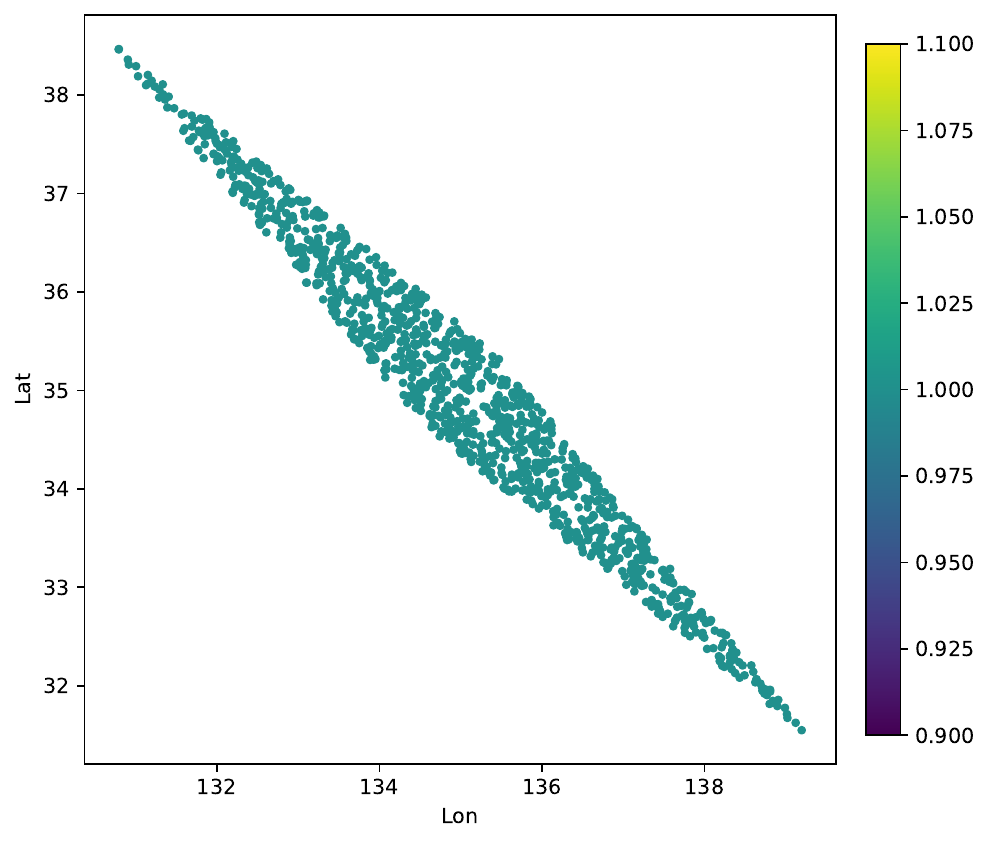}
        
        {(g) $\eta$}\vspace{4pt}
    \end{subfigure}\hfill
    \begin{subfigure}[t]{0.45\textwidth}
        \centering
        \includegraphics[width=\linewidth]{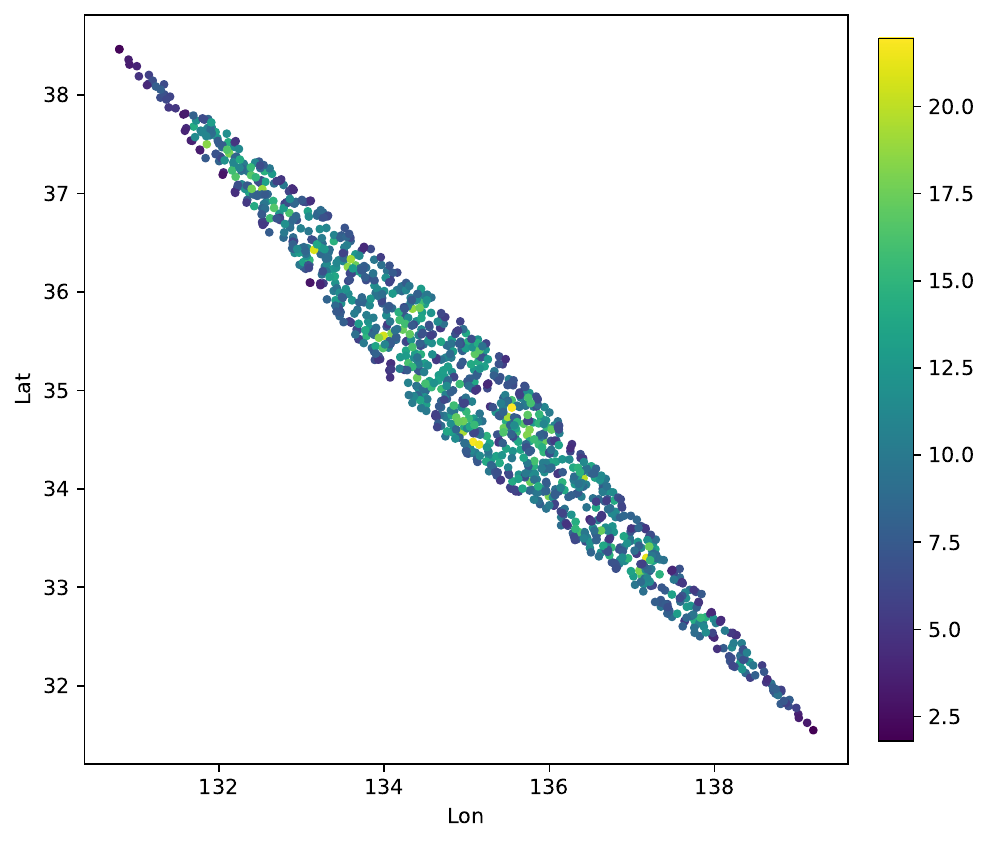}
        
        {(h) $n_{\mathrm{eff}}^{\mathrm{post}}$}\vspace{4pt}
    \end{subfigure}
    \begin{subfigure}[t]{0.45\textwidth}
        \centering
        \includegraphics[width=\linewidth]{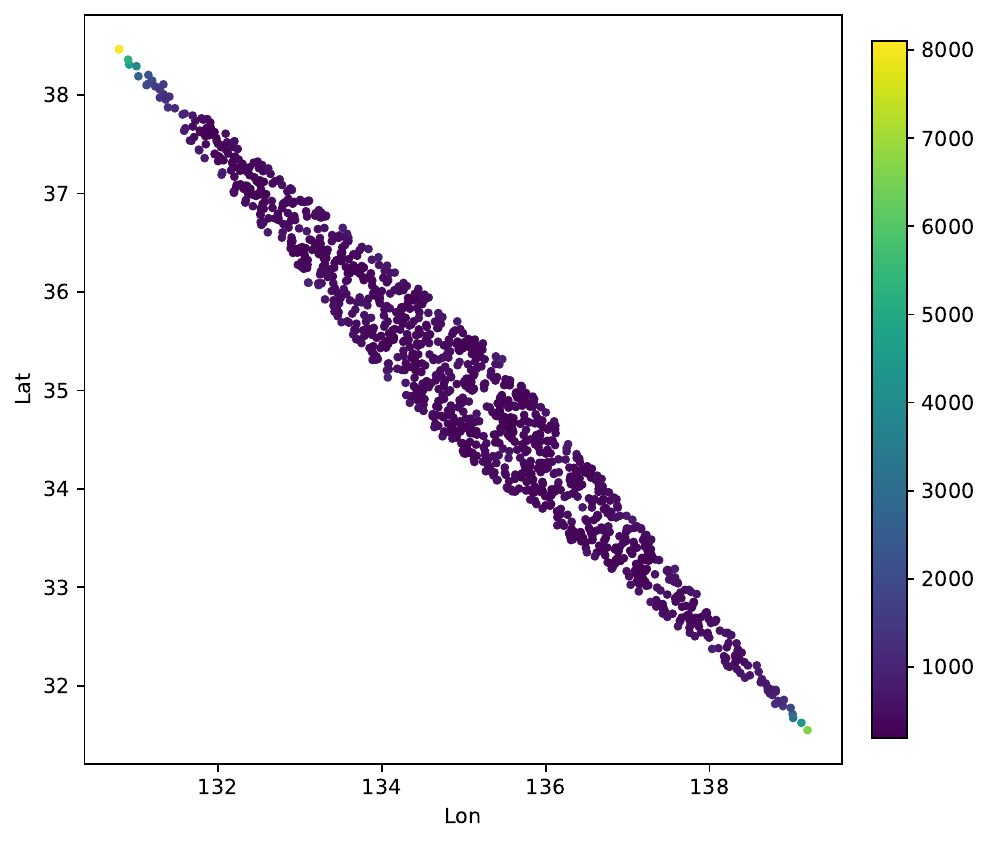}
        
        {(i) $\kappa(\mathbf M_i^{(\mathrm{nor})})$}\vspace{4pt}
    \end{subfigure}\hfill
    \begin{subfigure}[t]{0.45\textwidth}
        \centering
        \includegraphics[width=\linewidth]{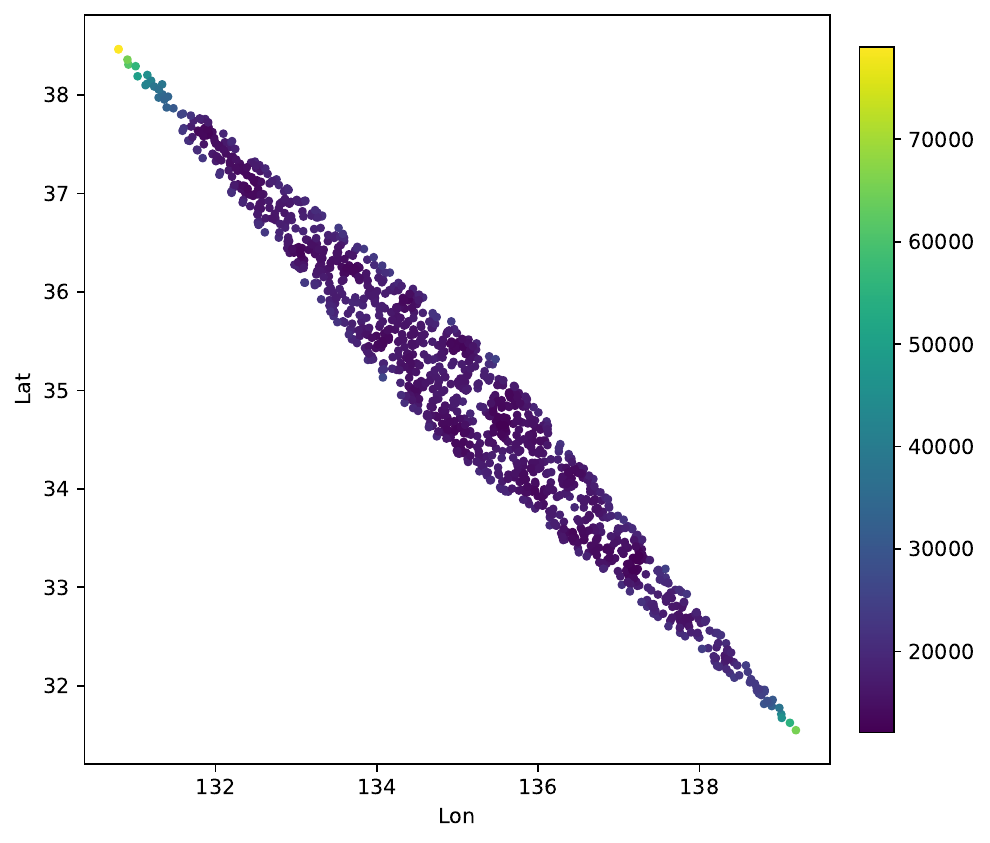}
        
        {(j) KNN mean distance ($K{=}30$)}\vspace{4pt}
    \end{subfigure}
    \caption{Spatial distributions: Isotropic-proxy ($\rho=10.0$).}
    \label{fig:ex72-ISO}
\end{figure}

\begin{figure}[H]
    \begin{subfigure}[t]{0.45\textwidth}
        \centering
        \includegraphics[width=\linewidth]{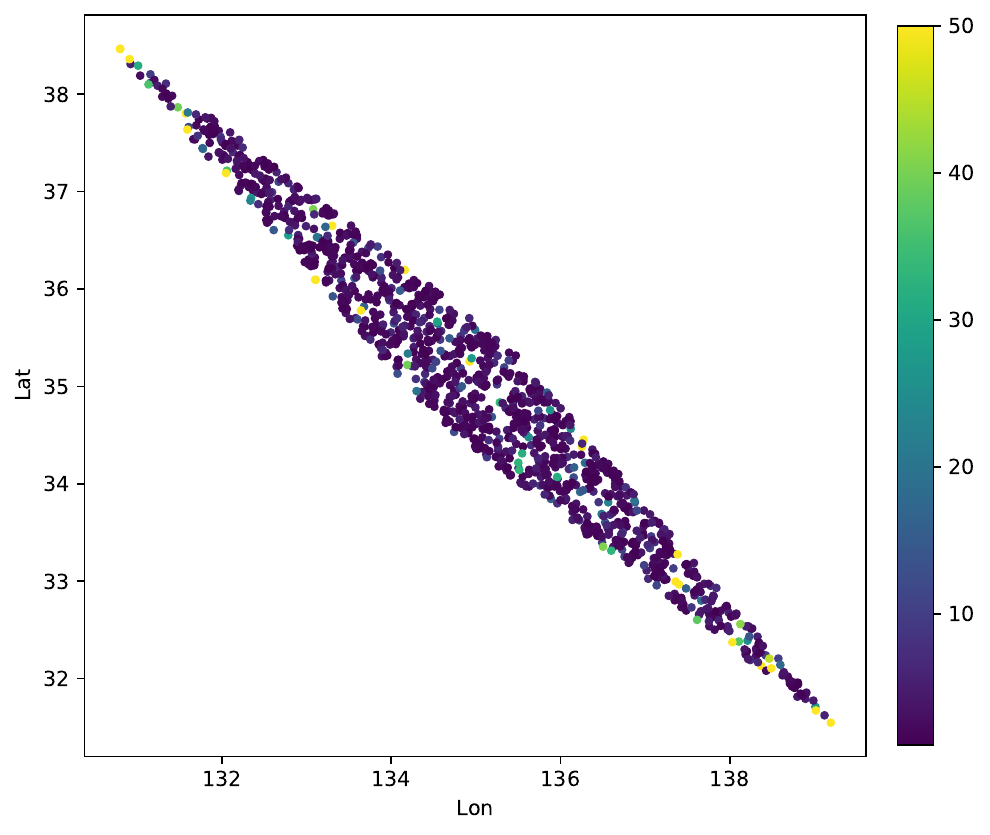}
        
        {(g) $\eta$}\vspace{4pt}
    \end{subfigure}\hfill
    \begin{subfigure}[t]{0.45\textwidth}
        \centering
        \includegraphics[width=\linewidth]{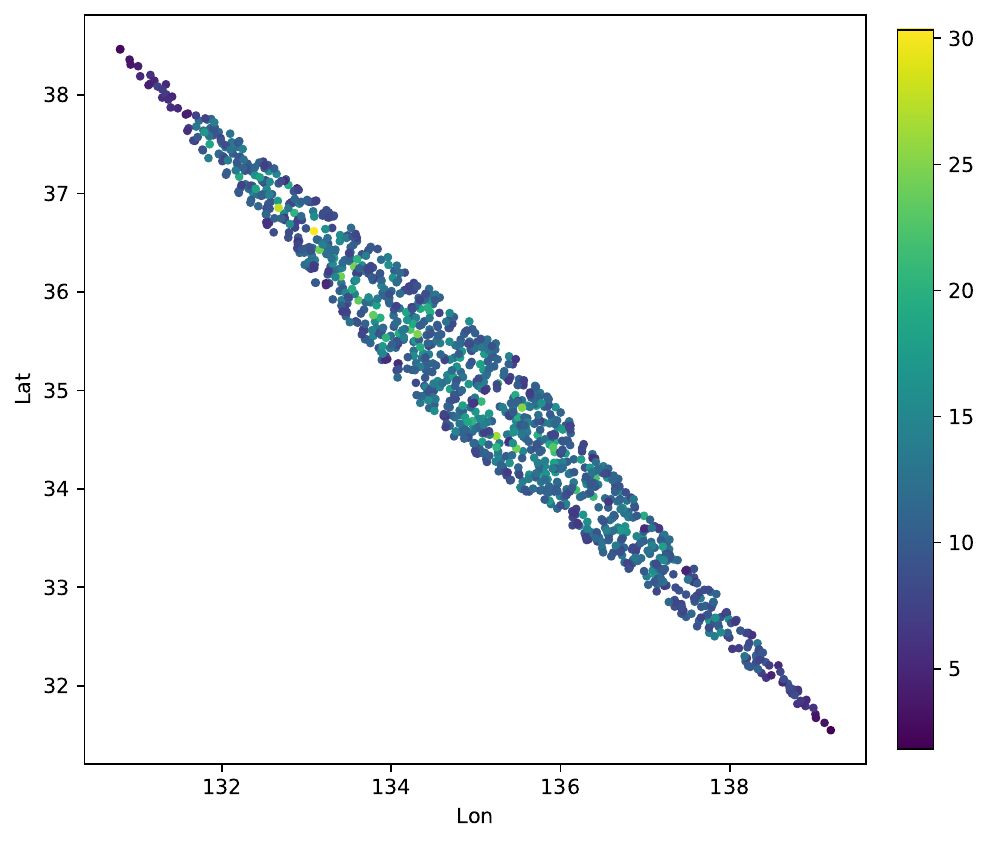}
        
        {(h) $n_{\mathrm{eff}}^{\mathrm{post}}$}\vspace{4pt}
    \end{subfigure}
    \begin{subfigure}[t]{0.45\textwidth}
        \centering
        \includegraphics[width=\linewidth]{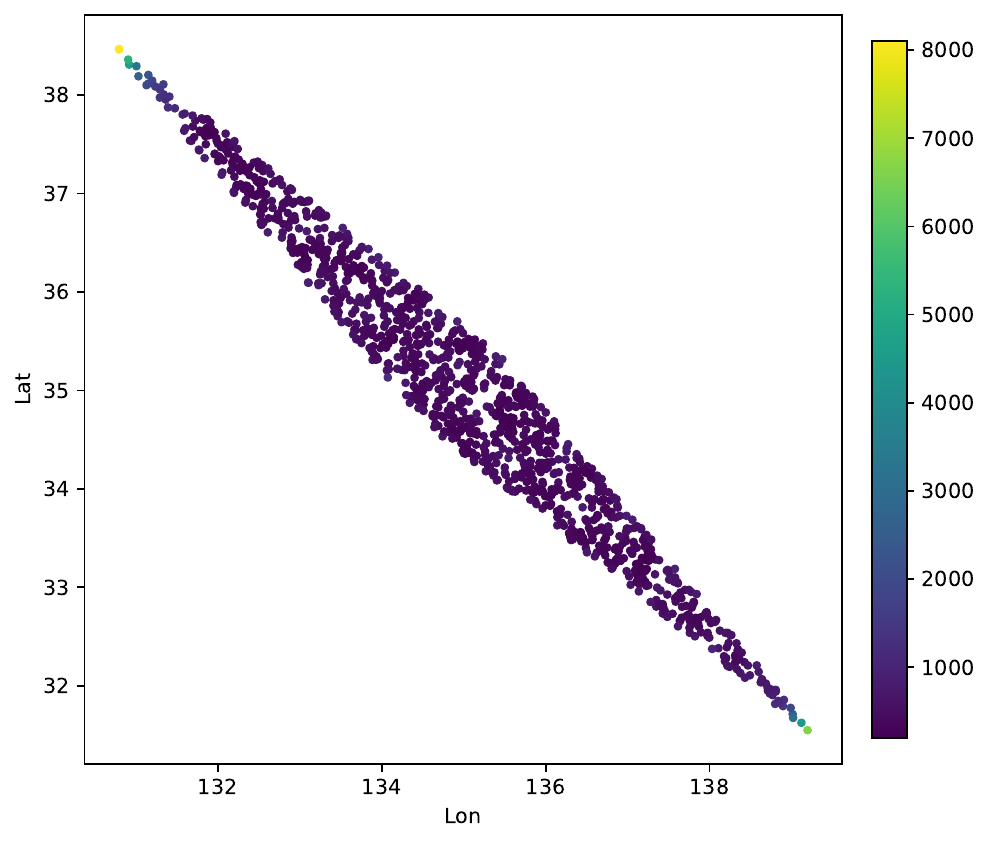}
        
        {(i) $\kappa(\mathbf M_i^{(\mathrm{nor})})$}\vspace{4pt}
    \end{subfigure}\hfill
    \begin{subfigure}[t]{0.45\textwidth}
        \centering
        \includegraphics[width=\linewidth]{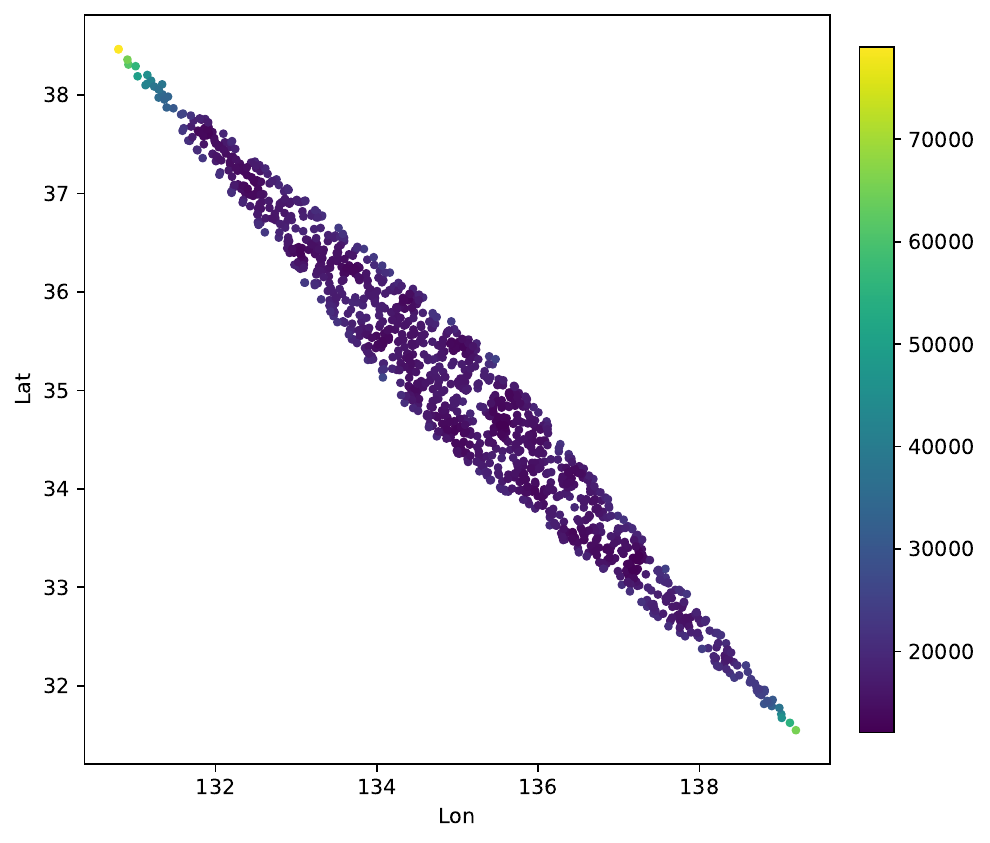}
        
        {(j) KNN mean distance ($K{=}30$)}
    \end{subfigure}
    \caption{Spatial distributions: GR ($\rho$=10.0).}
    \label{fig:ex72-GR}
\end{figure}

This appendix provides the full per-metric log (Table~\ref{tab:exp72-full-log}) and additional figures (Figure~\ref{fig:ex72-ISO} and Figure~\ref{fig:ex72-GR}) for Exp 7.2. The figures visualize spatial distributions of variables and diagnostic quantities.
\subsection{Full per-metric log}
\begin{table}[H]
    \centering
    \begin{threeparttable}
    \caption{Full per-metric log: $\rho$  = 10.0.}
    \label{tab:exp72-full-log}
    \footnotesize
    \setlength{\tabcolsep}{6pt}
    \renewcommand{\arraystretch}{1.15}
    \begin{tabularx}{1.0\textwidth}{@{\hspace{0.5em}} X r r @{\hspace{0.5em}}}
    \toprule
    Quantity & Isotropic-proxy ($\rho$=10.0) & GR ($\rho$=10.0) \\
    \midrule
    \multicolumn{3}{@{}l}{\textit{Fit summaries (context only)}}\\
    $N$ & 1200.000 & 1200.000 \\
    $\mu(\mathrm{RMSE})$ & 0.965 & 0.965 \\
    $\sigma(\mathrm{RMSE})$ & 0.085 & 0.085 \\
    $\mu(R^2)$ & 0.499 & 0.499 \\
    $\sigma(R^2)$ & 0.141 & 0.141 \\
    
    \addlinespace
    \multicolumn{3}{@{}l}{\textit{Conditioning diagnostics}}\\
    $\mu(\kappa(\mathbf M_i^{(\mathrm{nor})}))$ & 449.954 & 460.314 \\
    $\sigma(\kappa(\mathbf M_i^{(\mathrm{nor})}))$ & 487.146 & 484.666 \\
    $p_{50}(\kappa(\mathbf M_i^{(\mathrm{nor})}))$ & 374.853 & 384.615 \\
    $p_{95}(\kappa(\mathbf M_i^{(\mathrm{nor})}))$ & 738.580 & 761.124 \\
    $p_{99}(\kappa(\mathbf M_i^{(\mathrm{nor})}))$ & 1904.084 & 1906.961 \\
    \addlinespace
    \multicolumn{3}{@{}l}{\textit{ESS-related quantities}}\\
    $\mu(n_{\mathrm{eff}}^{\mathrm{raw}})$  & 1.873 & 3.542 \\
    $\sigma(n_{\mathrm{eff}}^{\mathrm{raw}})$ & 0.794 & 1.733 \\
    $\mu(n_{\mathrm{eff}}^{\mathrm{post}})$ & 9.281 & 10.885 \\
    $\sigma(n_{\mathrm{eff}}^{\mathrm{post}})$ & 3.376 & 3.506 \\
    \addlinespace
    \multicolumn{3}{@{}l}{\textit{Geometry / orientation diagnostics}}\\
    $\mu(\eta_i)$ & 1.000 & 4.962 \\
    $\sigma(\eta_i)$ & 0.000 & 8.225 \\
    $\mu(r_{\phi,i})$ & 0.754 & 0.754 \\
    $\sigma(r_{\phi,i})$ & 0.233 & 0.233 \\
    $\mu(g_{\mathrm{ident},i})$ & 10.380 & 10.380 \\
    $\sigma(g_{\mathrm{ident},i})$ & 17.678 & 17.678 \\
    $\mu(\theta_{z,i}^{\ast})$ & 0.000 & -0.785 \\
    $\sigma(\theta_{z,i}^{\ast})$ & 0.000 & 0.109 \\
    \addlinespace
    \multicolumn{3}{@{}l}{\textit{Branch rates / fallback proxies}}\\
    $\Pr(\phi_i=0)$ & 1.000 & 0.000 \\
    $\Pr(\theta_{z,i}^{\ast}=0)$ & 1.000 & 0.000 \\
    $\Pr(\mathrm{uniform})$ & 0.018 & 0.008 \\
    $N_{\mathrm{uniform}}$ & 22.000 & 9.000 \\
    \bottomrule
    \end{tabularx}
    \begin{tablenotes}[flushleft]
        \footnotesize
        \item Notes. Here, $\mu(\cdot)$ and $\sigma(\cdot)$ denote mean and standard deviation across target locations $i$. $p_{50}(\cdot)$ denotes median across targets, and percentiles $p_{95}(\cdot)$ and $p_{99}(\cdot)$ are computed across targets. $\Pr(\mathrm{uniform})$ and $N_{\mathrm{uniform}}$ are the proxy fallback rate and count induced by the one-shot ESS safeguard (Appendix~\ref{app:one-shot-ESS-correction-and-uniform-fallback}). All values are rounded to three decimals.
    \end{tablenotes}
    \end{threeparttable}
\end{table}

\subsection{Spatial distributions of variables and diagnostic quantities}
\subsubsection{Isotropic-proxy ($\rho=10.0$)}
\begin{figure}[H]
    \centering
    \small
    \begin{subfigure}[t]{0.45\textwidth}
        \centering
        \includegraphics[width=\linewidth]{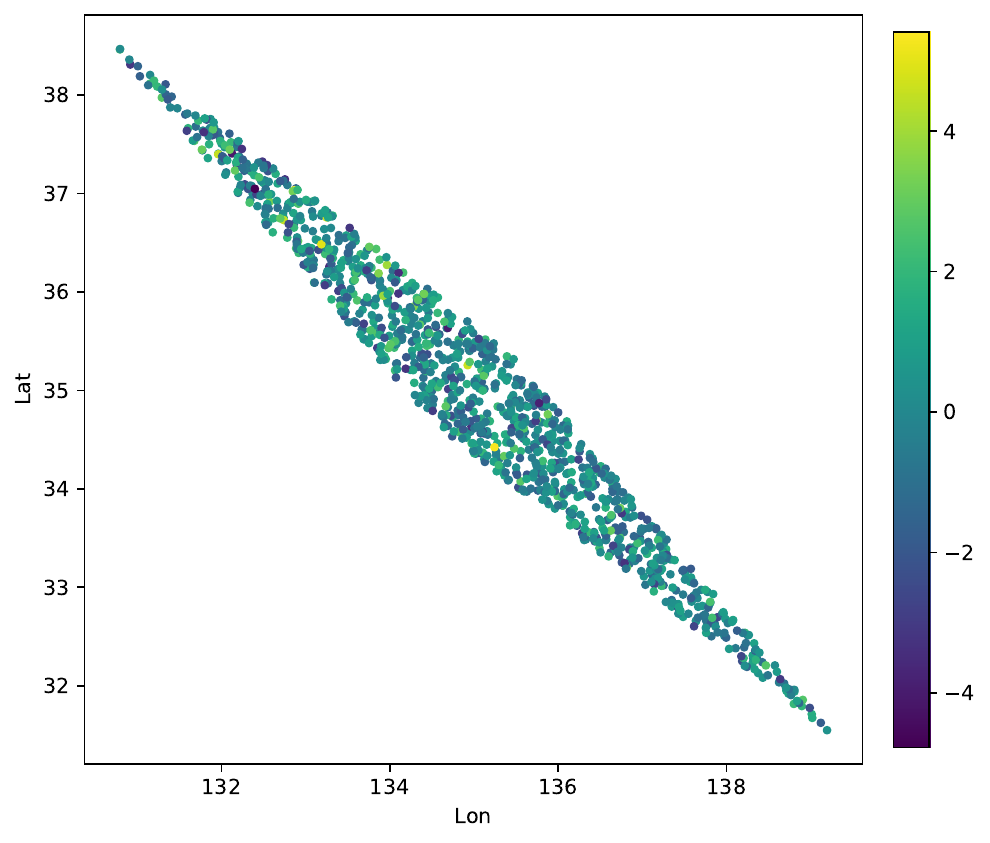}
        
        {(a) $y$}\vspace{4pt}
    \end{subfigure}\hfill
    \begin{subfigure}[t]{0.45\textwidth}
        \centering
        \includegraphics[width=\linewidth]{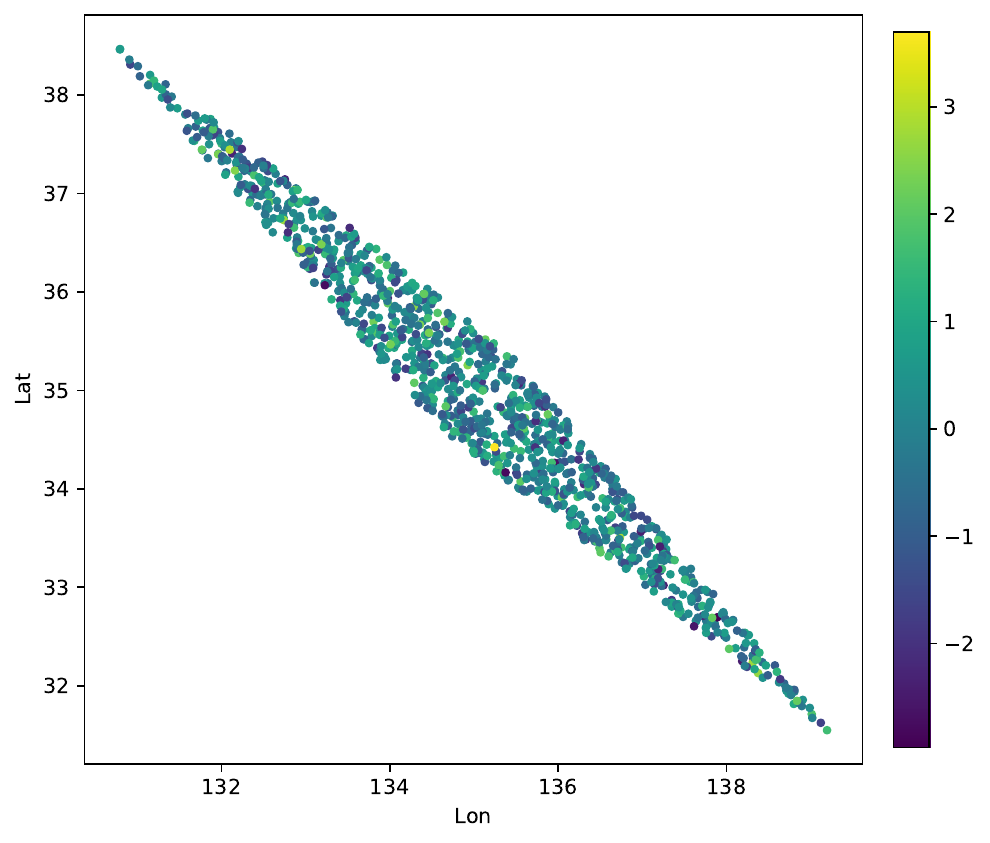}
        
        {(b) $X$}\vspace{4pt}
    \end{subfigure}
    \begin{subfigure}[t]{0.45\textwidth}
        \centering
        \includegraphics[width=\linewidth]{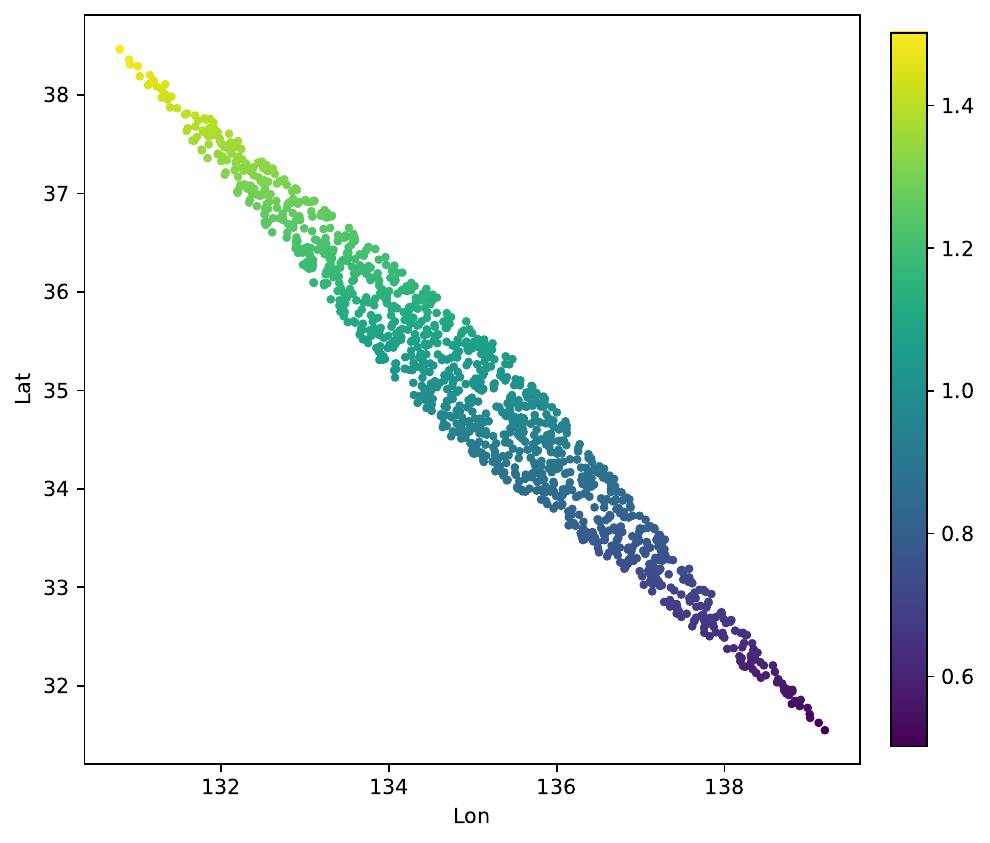}
        
        {(c) $\hat{\beta}_1(s)$}\vspace{4pt}
    \end{subfigure}\hfill
    \begin{subfigure}[t]{0.45\textwidth}
        \centering
        \includegraphics[width=\linewidth]{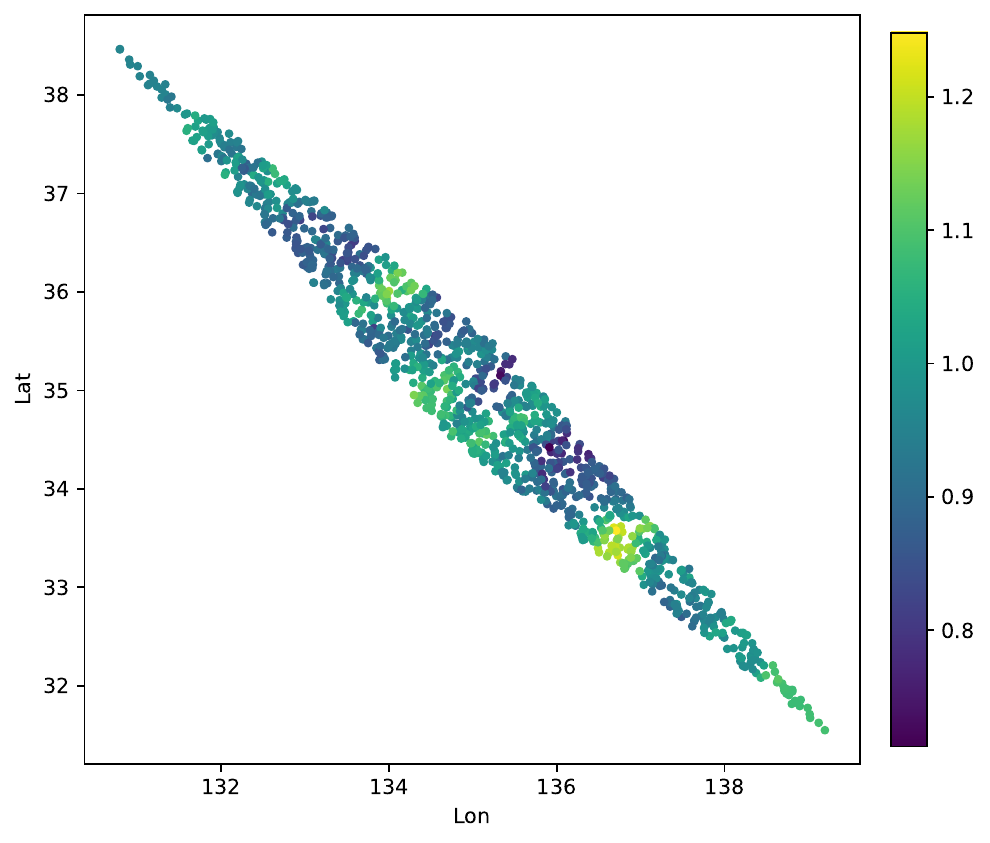}
        
        {(d) RMSE}\vspace{4pt}
    \end{subfigure}
    \begin{subfigure}[t]{0.45\textwidth}
        \centering
        \includegraphics[width=\linewidth]{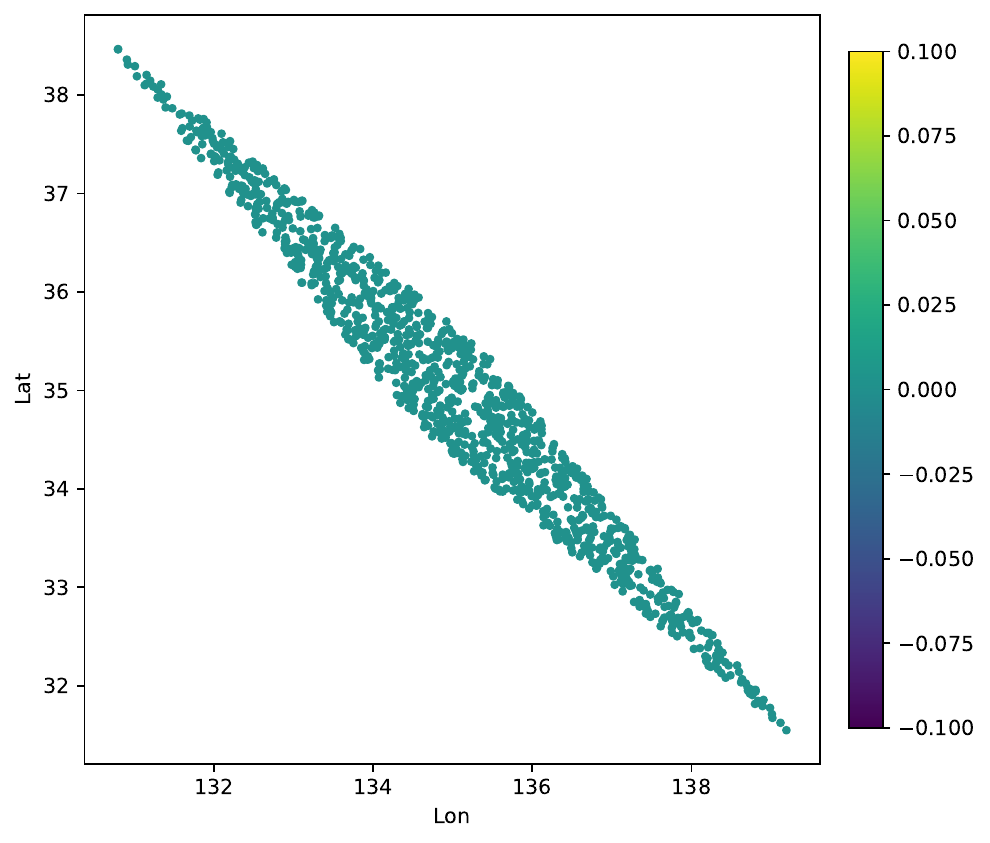}
        
        {(e) $\phi$}\vspace{4pt}
    \end{subfigure}\hfill
    \begin{subfigure}[t]{0.45\textwidth}
        \centering
        \includegraphics[width=\linewidth]{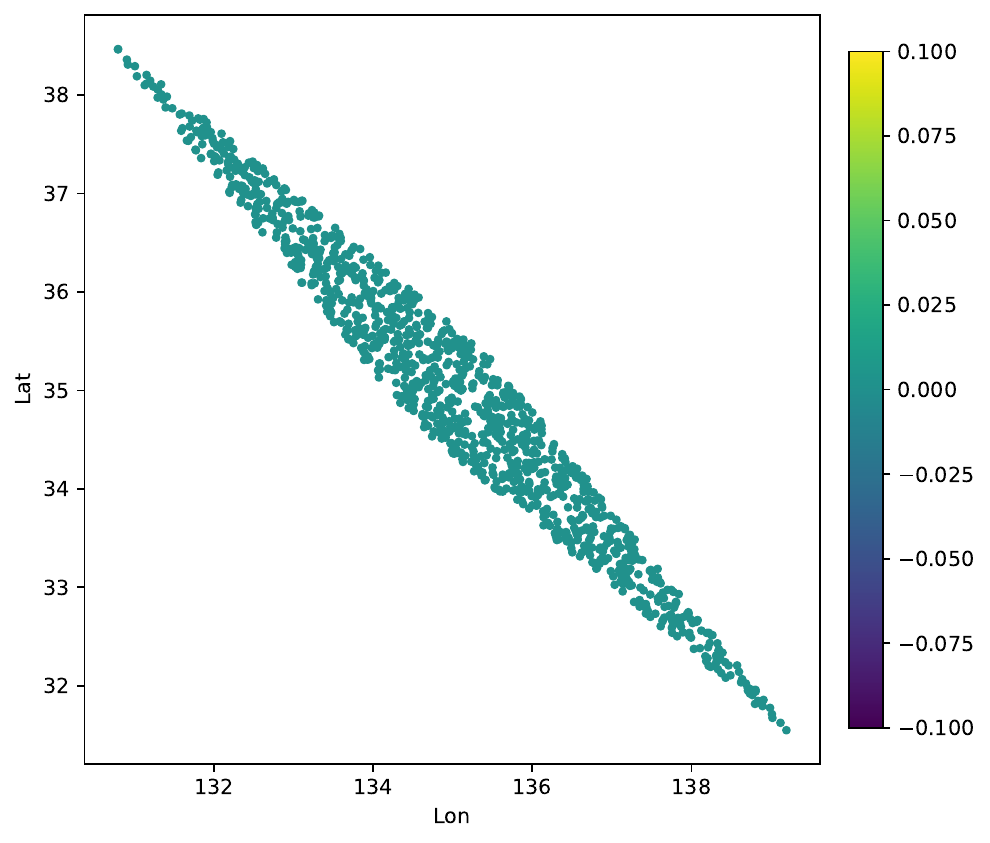}
        
        {(f) $\theta_{z}^\ast$}\vspace{4pt}
    \end{subfigure}
\end{figure}

\subsubsection{GR ($\rho$=10.0)}
\begin{figure}[H]
    \centering
    \begin{subfigure}[t]{0.45\textwidth}
        \centering
        \includegraphics[width=\linewidth]{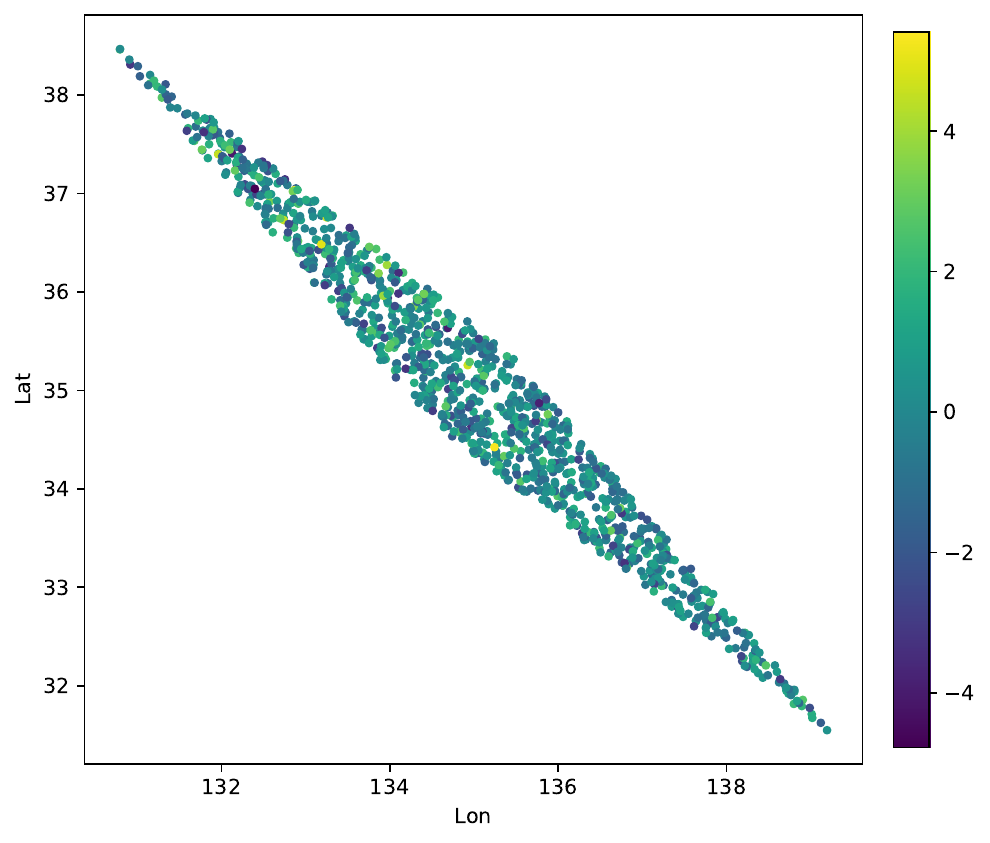}
        
        {(a) $y$}\vspace{4pt}
    \end{subfigure}\hfill
    \begin{subfigure}[t]{0.45\textwidth}
        \centering
        \includegraphics[width=\linewidth]{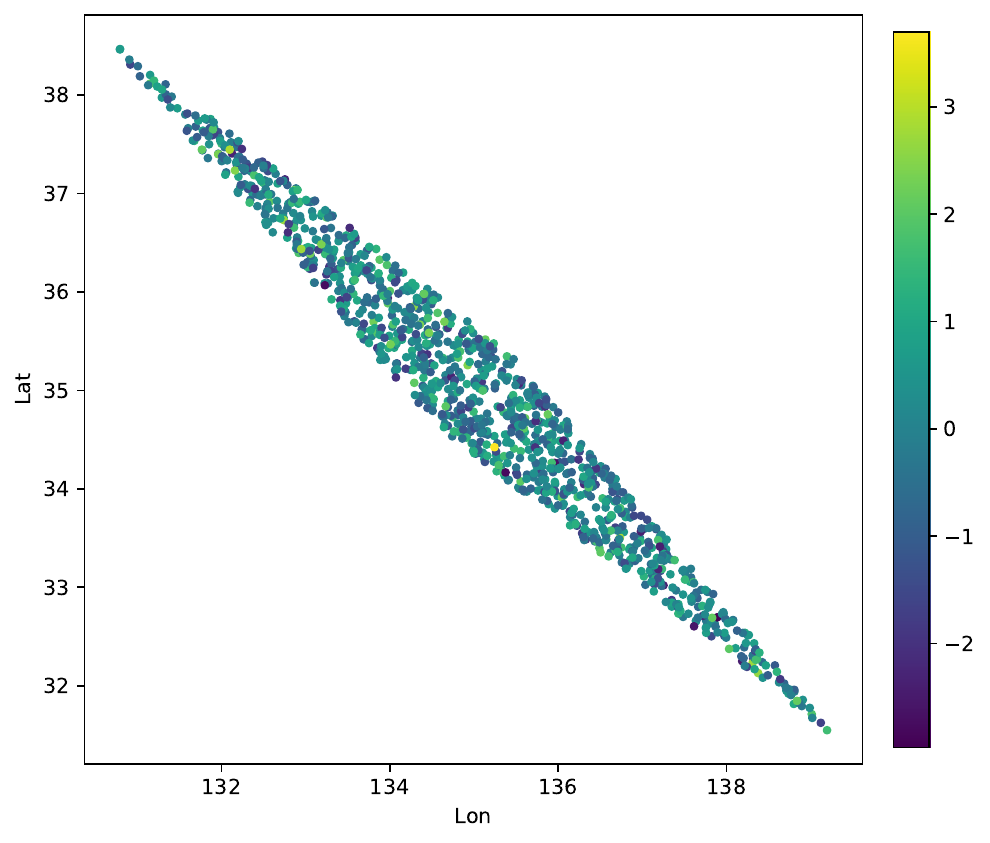}
        
        {(b) $X$}\vspace{4pt}
    \end{subfigure}
    \begin{subfigure}[t]{0.45\textwidth}
        \centering
        \includegraphics[width=\linewidth]{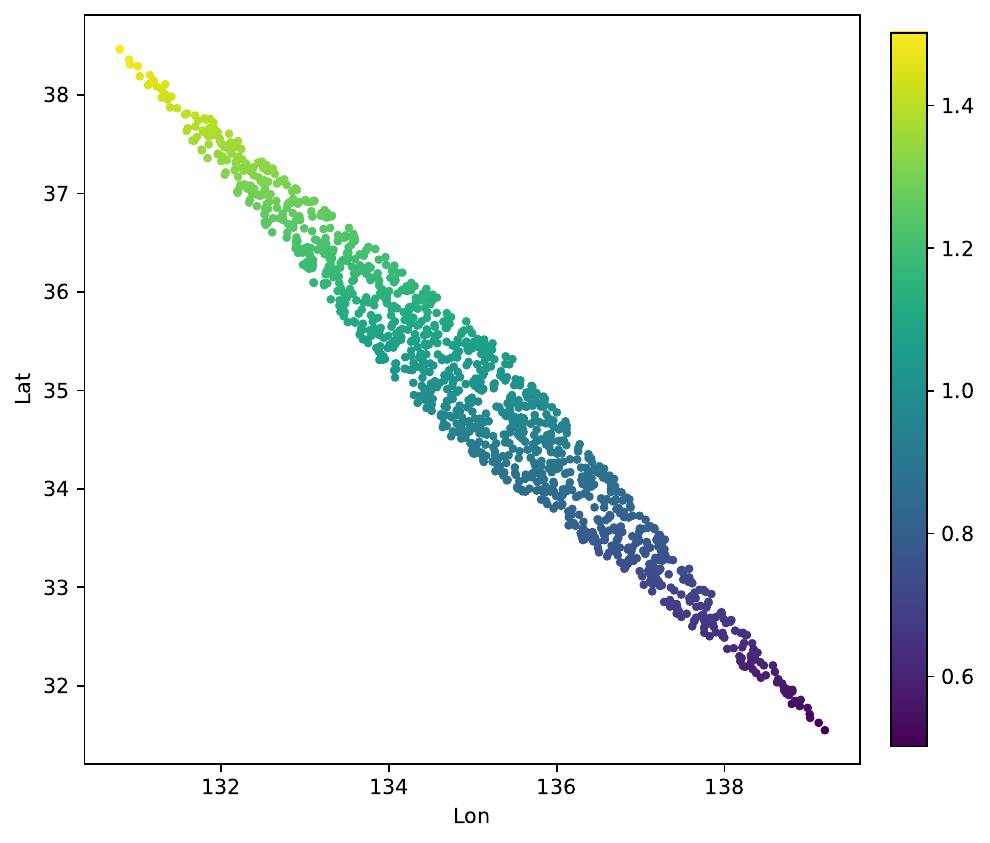}
        
        {(c) $\hat{\beta}_1(s)$}\vspace{4pt}
    \end{subfigure}\hfill
    \begin{subfigure}[t]{0.45\textwidth}
        \centering
        \includegraphics[width=\linewidth]{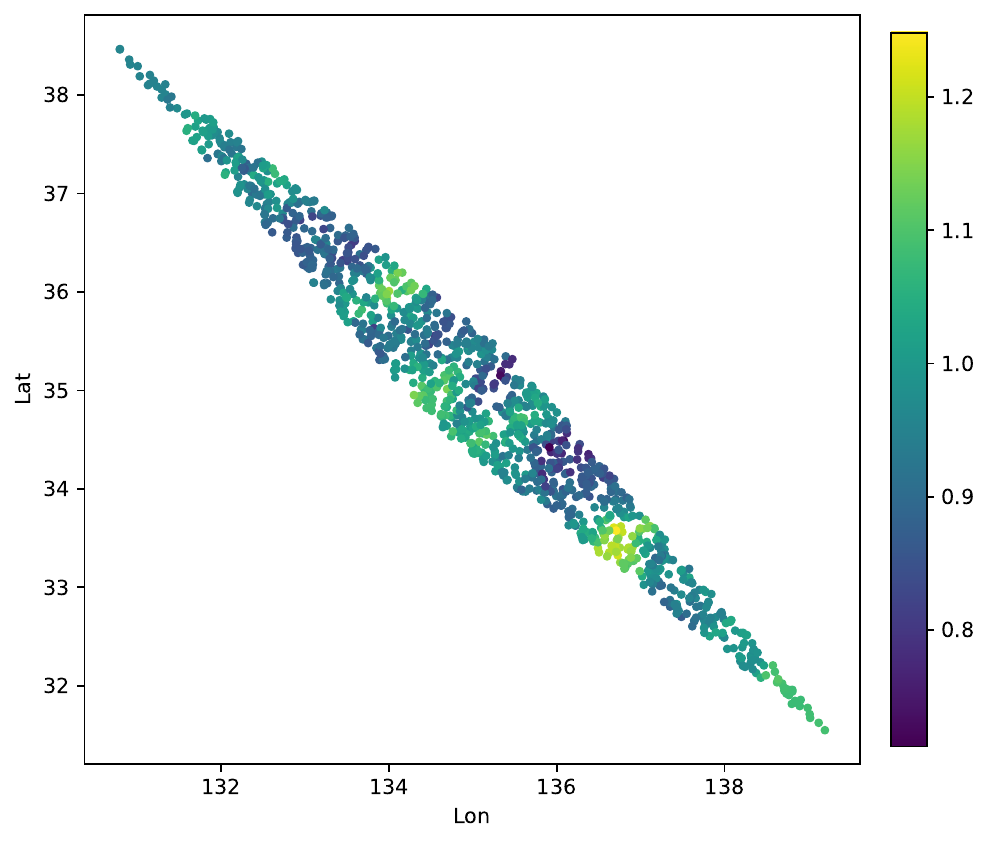}
        
        {(d) RMSE}\vspace{4pt}
    \end{subfigure}
    \begin{subfigure}[t]{0.45\textwidth}
        \centering
        \includegraphics[width=\linewidth]{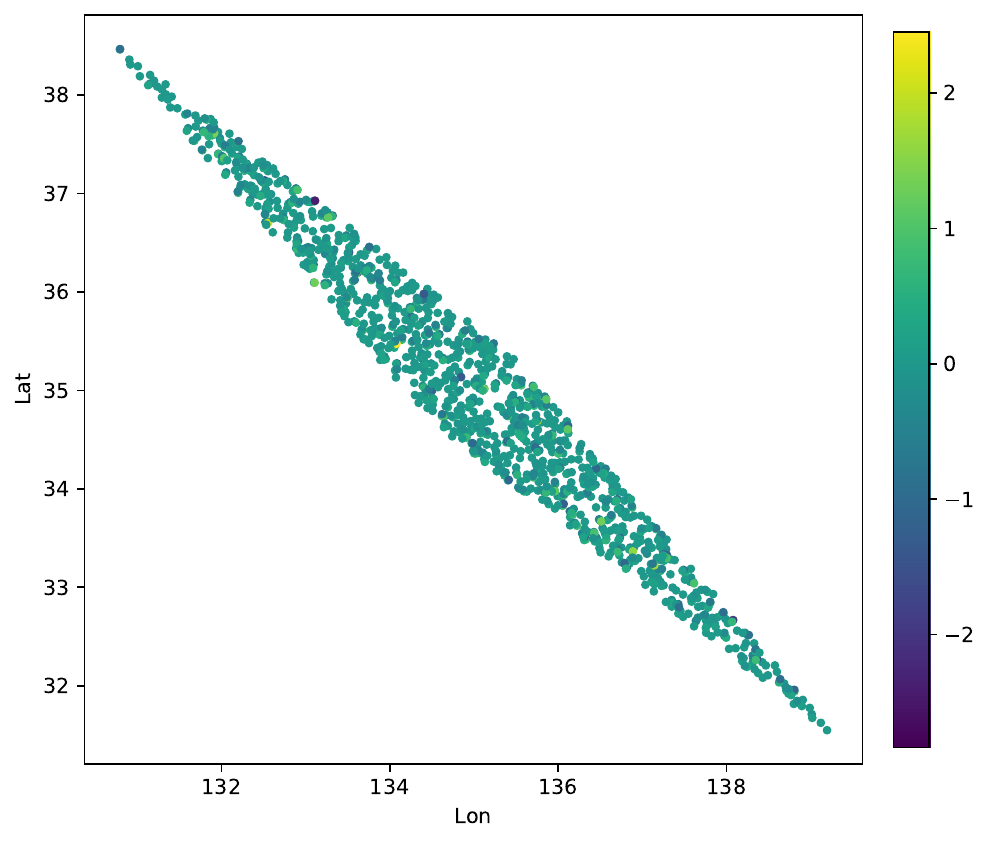}
        
        {(e) $\phi$}\vspace{4pt}
    \end{subfigure}\hfill
    \begin{subfigure}[t]{0.45\textwidth}
        \centering
        \includegraphics[width=\linewidth]{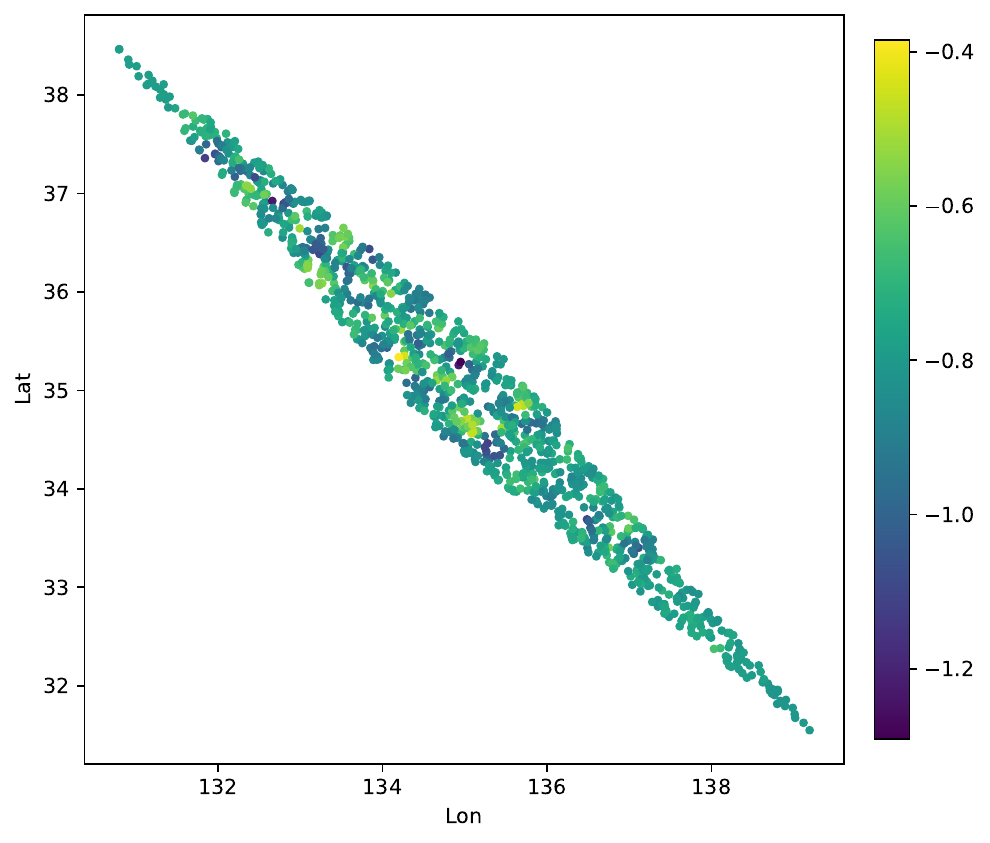}
        
        {(f) $\theta_{z}^\ast$}
    \end{subfigure}
\end{figure}

\section{Additional figures for Experiment 7.3 (one-shot ESS stress)}\label{app:D}
\begin{figure}[H]
    \centering
    \begin{subfigure}[t]{0.49\textwidth}
        \centering
        \includegraphics[width=\linewidth]{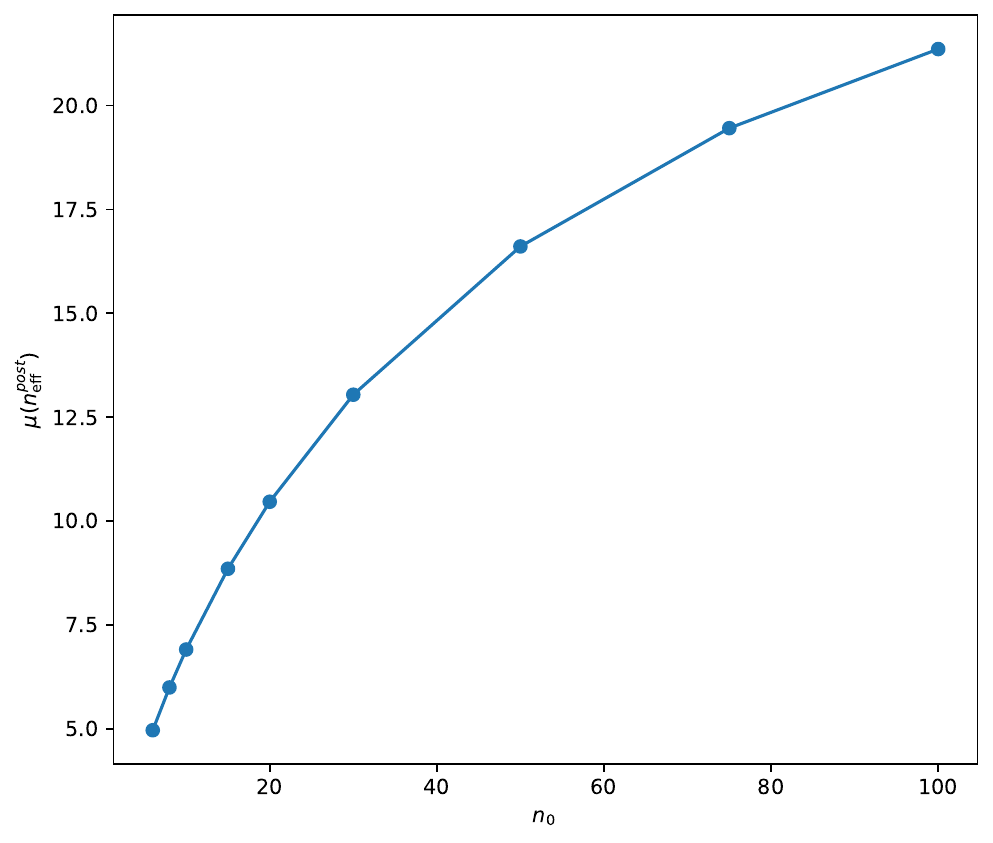}
        
        {(a) $\mu(n_{\mathrm{eff}}^{\mathrm{post}})$}\vspace{4pt}
        \label{fig:ex73-neff_post_mean}
    \end{subfigure}\hfill
    \begin{subfigure}[t]{0.49\textwidth}
        \centering
        \includegraphics[width=\linewidth]{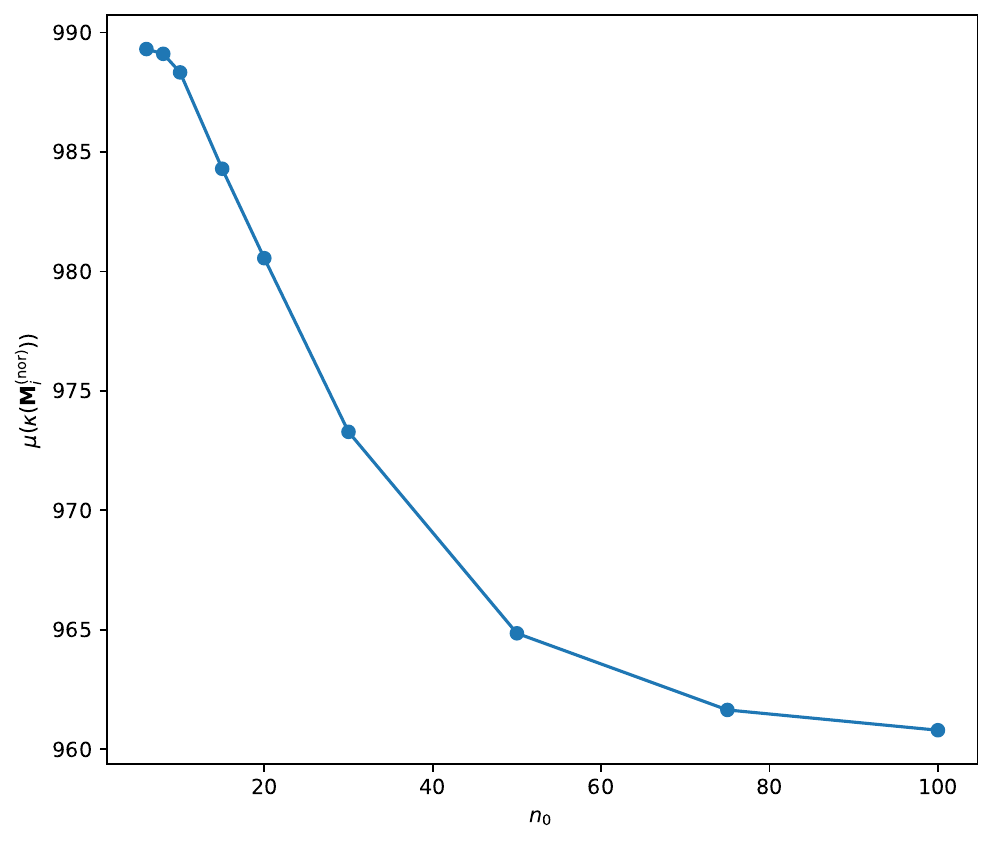}
        
        {(b) $\mu(\kappa(\mathbf M_i^{(\mathrm{nor})}))$}\vspace{4pt}
        \label{fig:ex73-condM_nor_mean}
    \end{subfigure}
    \begin{subfigure}[t]{0.49\textwidth}
        \centering
        \includegraphics[width=\linewidth]{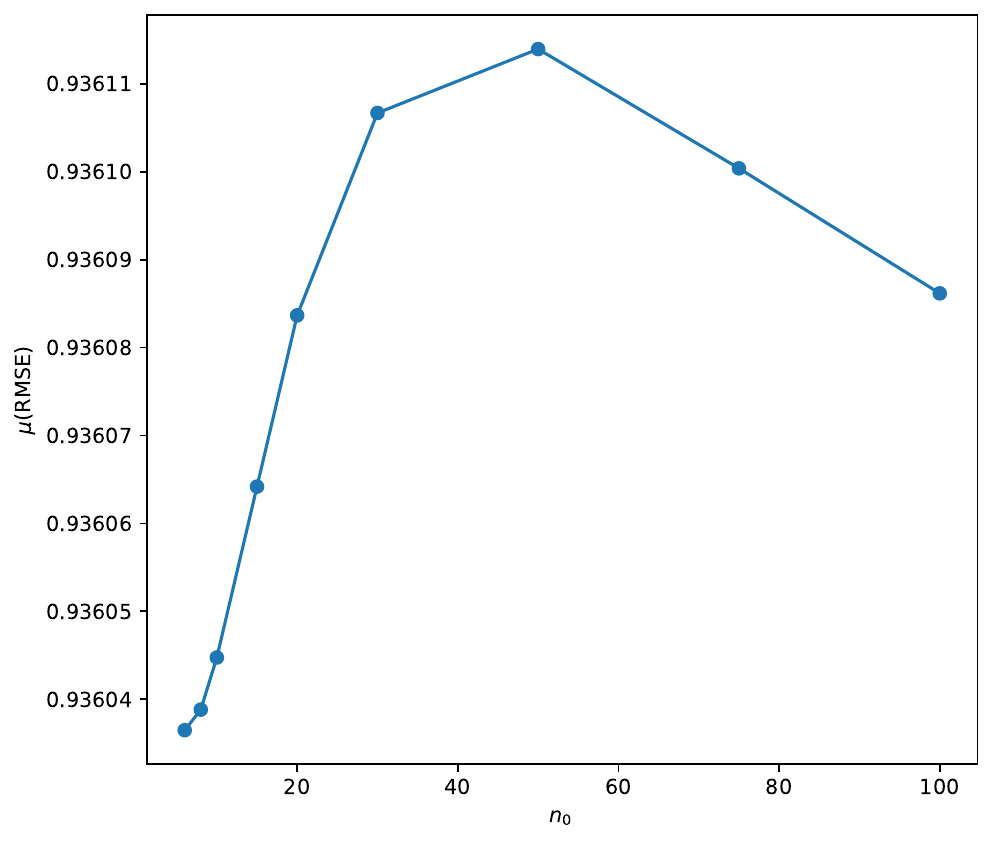}
        
        {(c) $\mu(\mathrm{RMSE})$}\vspace{4pt}
        \label{fig:ex73-RMSE_mean}
    \end{subfigure}\hfill
    \begin{subfigure}[t]{0.49\textwidth}
        \centering
        \includegraphics[width=\linewidth]{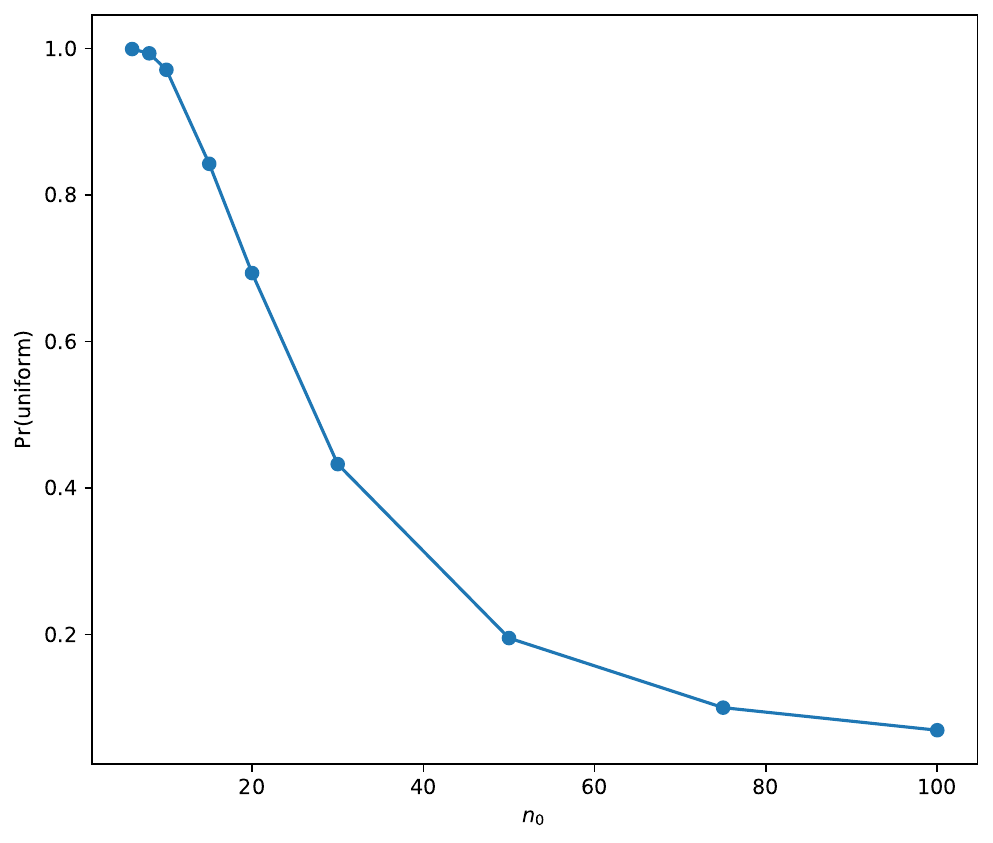}
        
        {(d) $\Pr(\mathrm{uniform})$}\vspace{4pt}
        \label{fig:ex73-uniform_rate}
    \end{subfigure}\hfill
    \caption{
        Mean diagnostic and fallback summaries across $n_0$.
    } 
    \label{fig:ex73}
\end{figure}

This appendix provides the full per-metric log (Table~\ref{tab:exp73-full}) and additional figures (Figure~\ref{fig:ex73}) for Exp 7.3. The figures visualize how the one-shot ESS safeguard (Appendix~\ref{app:one-shot-ESS-correction-and-uniform-fallback}) responds to the target ESS parameter $n_0$, including monotone changes in $n_{\mathrm{eff}}^{\mathrm{post}}$ and attenuation of uniform fallback usage. 
\subsection{Full per-metric log}
\begin{table}[H]
    \centering
    \begin{threeparttable}    
    \caption{Full map-level log across $n_0$.}
    \label{tab:exp73-full}
    \footnotesize
    \setlength{\tabcolsep}{3pt}
    \renewcommand{\arraystretch}{1.15}
    \begin{tabularx}{0.95\textwidth}{@{\hspace{0.5em}} X r r r r r r r r r @{\hspace{0.5em}}}
    \toprule
    Quantity
    & $n_0{=}6$ & $8$ & $10$ & $15$ & $20$ & $30$ & $50$ & $75$ & $100$ \\
    \midrule
    \multicolumn{10}{@{}l}{\textit{ESS-related quantities}}\\
    $\mu(n_{\mathrm{eff}}^{\mathrm{post}})$ & 4.963 & 5.993 & 6.906 & 8.850 & 10.464 & 13.041 & 16.610 & 19.458 & 21.362 \\
    $\sigma(n_{\mathrm{eff}}^{\mathrm{post}})$ & 1.488 & 1.975 & 2.412 & 3.302 & 3.970 & 4.857 & 5.663 & 5.901 & 5.821 \\
    $\mu(n_{\mathrm{eff}}^{\mathrm{raw}})$ & 3.229 & 3.229 & 3.229 & 3.229 & 3.229 & 3.229 & 3.229 & 3.229 & 3.229 \\
    $\sigma(n_{\mathrm{eff}}^{\mathrm{raw}})$ & 1.635 & 1.635 & 1.635 & 1.635 & 1.635 & 1.635 & 1.635 & 1.635 & 1.635 \\
    
    \addlinespace
    \multicolumn{10}{@{}l}{\textit{Conditioning diagnostics}}\\
    $\mu(\kappa(\mathbf M_i^{(\mathrm{nor})}))$ & 989.309 & 989.109 & 988.332 & 984.298 & 980.555 & 973.285 & 964.851 & 961.643 & 960.793 \\
    $p_{50}(\kappa(\mathbf M_i^{(\mathrm{nor})}))$ & 382.341 & 382.341 & 382.117 & 376.199 & 371.497 & 361.899 & 358.921 & 366.047 & 368.812 \\
    $p_{95}(\kappa(\mathbf M_i^{(\mathrm{nor})}))$ & 2799.711 & 2799.711 & 2799.711 & 2799.711 & 2799.711 & 2799.711 & 2765.878 & 2732.560 & 2711.607 \\
    $p_{99}(\kappa(\mathbf M_i^{(\mathrm{nor})}))$ & 9687.286 & 9687.286 & 9687.286 & 9687.286 & 9687.286 & 9687.286 & 9687.286 & 9687.286 & 9687.286 \\
    $\sigma(\kappa(\mathbf M_i^{(\mathrm{nor})}))$ & 3778.593 & 3778.626 & 3778.750 & 3779.342 & 3779.833 & 3780.710 & 3779.795 & 3777.964 & 3776.220 \\
    
    \addlinespace
    \multicolumn{10}{@{}l}{\textit{Fit summaries (context only)}}\\
    $\mu(\mathrm{RMSE})$ & 0.936 & 0.936 & 0.936 & 0.936 & 0.936 & 0.936 & 0.936 & 0.936 & 0.936 \\
    $\sigma(\mathrm{RMSE})$ & 0.133 & 0.133 & 0.133 & 0.133 & 0.133 & 0.133 & 0.133 & 0.133 & 0.133 \\
    $\mu(R^2)$ & 0.511 & 0.511 & 0.511 & 0.511 & 0.511 & 0.511 & 0.511 & 0.511 & 0.511 \\
    $\sigma(R^2)$ & 0.160 & 0.160 & 0.160 & 0.160 & 0.160 & 0.160 & 0.160 & 0.160 & 0.160 \\
    
    \addlinespace
    \multicolumn{10}{@{}l}{\textit{Fallback proxies}}\\
    $\Pr(\mathrm{uniform})$ & 0.999 & 0.993 & 0.971 & 0.843 & 0.693 & 0.433 & 0.195 & 0.100 & 0.069 \\
    $N_{\mathrm{uniform}}$ & 1199 & 1192 & 1165 & 1011 & 832 & 519 & 234 & 120 & 83 \\
    \bottomrule
    \end{tabularx}
    
    \begin{tablenotes}[flushleft]
        \footnotesize
        \item Notes. Here, $\mu(\cdot)$ and $\sigma(\cdot)$ denote mean and standard deviation across target locations $i$. $p_{50}(\cdot)$, $p_{95}(\cdot)$ and $p_{99}(\cdot)$ denote median, the 95th and 99th percentiles across $i$. All values are rounded to three decimals.
    \end{tablenotes}
    \end{threeparttable}
\end{table}

\subsection{Means of reported quantities}

\section{Additional figures for Experiment 7.4 (value-orientation dependence)}\label{app:E}

\begin{figure}[H]
    \centering
    \begin{subfigure}[t]{0.49\textwidth}
        \centering
        \includegraphics[width=\linewidth]{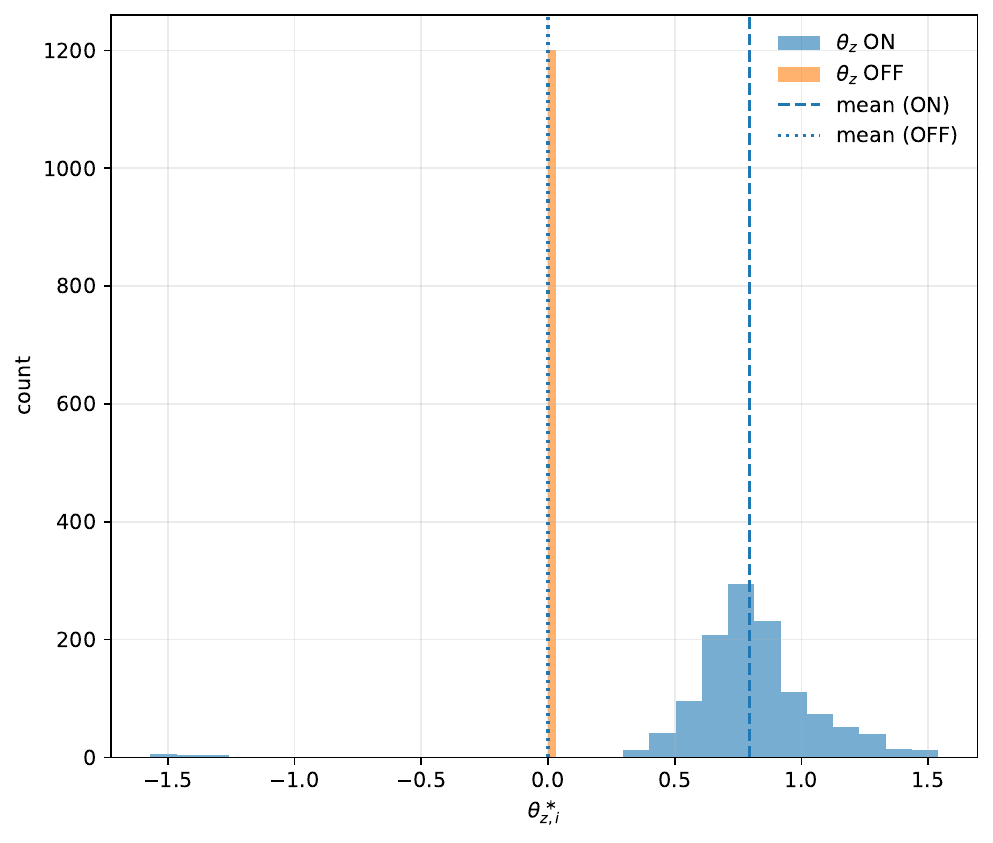}
        
        {(a) Distribution of $\theta_{z,i}^\ast$}\vspace{4pt}
        \label{fig:ex74-theta_z}
    \end{subfigure}\hfill
    \begin{subfigure}[t]{0.49\textwidth}
        \centering
        \includegraphics[width=\linewidth]{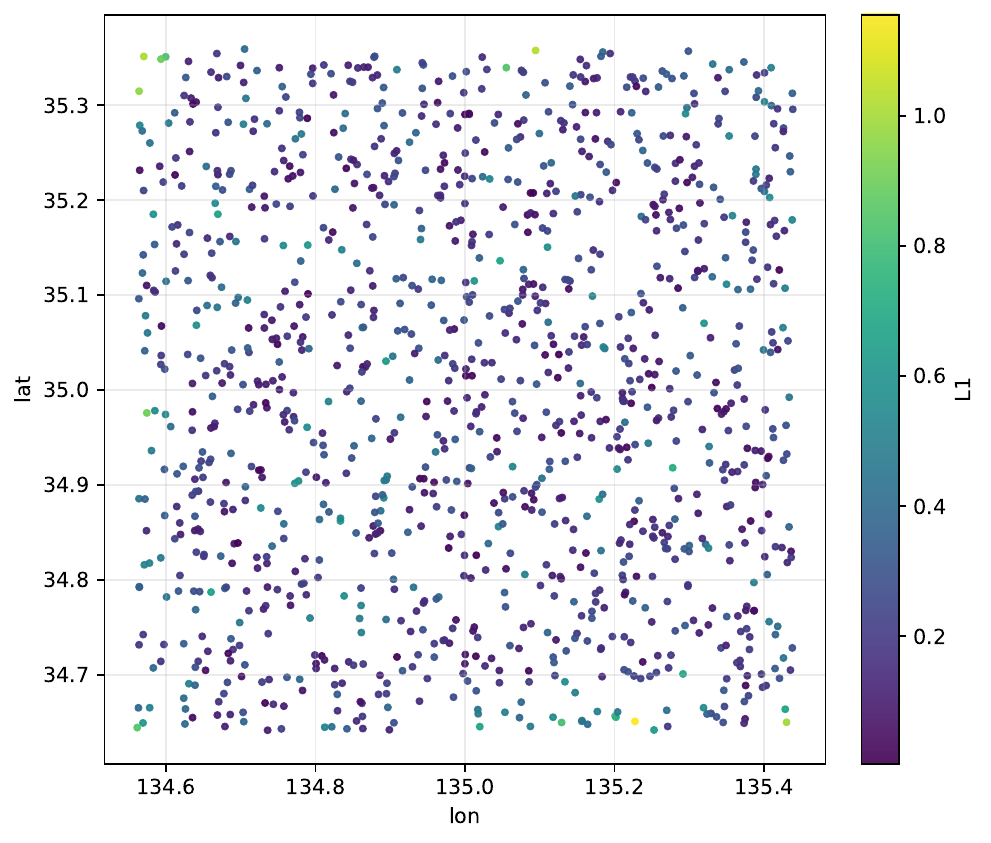}
        
        {(b) $l_1$}\vspace{4pt}
        \label{fig:ex74-L1}
    \end{subfigure}
    \caption{Distribution of $\theta_{z,i}^\ast$ and weight difference map ($l_1$).}  
    \label{fig:ex74}
\end{figure}

This appendix provides the full per-metric log (Table~\ref{tab:exp74-full-log}) and additional figures (Figure~\ref{fig:ex74}) for Exp 7.4. The figures visualize (i) activation of $\theta_{z,i}^\ast$ under the ON variant and (ii) the spatial distribution of weight-field changes between ON and OFF. No additional methodological claims are introduced.

\subsection{Full per-metric log}
\begin{table}[H]
    \centering
    \begin{threeparttable}
        \caption{Full per-metric log.}
        \label{tab:exp74-full-log}
        \footnotesize
        \setlength{\tabcolsep}{6pt}
        \renewcommand{\arraystretch}{1.15}
            \begin{tabularx}{0.75\textwidth}{@{\hspace{0.5em}} X r r @{\hspace{0.5em}}}
            \toprule
            Quantity
            & GR ($\theta_z$ OFF)
            & GR ($\theta_z$ ON; uses realized $y$) \\
                \midrule
                \multicolumn{3}{@{}l}{\textit{Fit summaries (context only)}}\\
                    $N$ & 1200 & 1200 \\
                    $\mu(\mathrm{RMSE})$ & 1.268 & 1.268 \\
                    $\sigma(\mathrm{RMSE})$ & 0.177 & 0.177 \\
                    $\mu(R^2)$ & 0.367 & 0.367 \\
                    $\sigma(R^2)$ & 0.124 & 0.124 \\
                \addlinespace
                \multicolumn{3}{@{}l}{\textit{Conditioning diagnostics}}\\
                    $\mu(\kappa(\mathbf M_i^{(\mathrm{nor})}))$ & 61.565 & 61.585 \\
                    $\sigma(\kappa(\mathbf M_i^{(\mathrm{nor})}))$ & 21.985 & 22.106 \\
                    $\mathrm{median}(\kappa(\mathbf M_i^{(\mathrm{nor})}))$ & 58.260 & 58.359 \\
                    $p_{95}(\kappa(\mathbf M_i^{(\mathrm{nor})}))$ & 102.730 & 103.593 \\
                    $p_{99}(\kappa(\mathbf M_i^{(\mathrm{nor})}))$ & 128.405 & 127.872 \\
                \addlinespace
                \multicolumn{3}{@{}l}{\textit{ESS-related quantities}}\\
                    $\mu(n_{\mathrm{eff}}^{\mathrm{raw}})$ & 6.769 & 6.770 \\
                    $\sigma(n_{\mathrm{eff}}^{\mathrm{raw}})$ & 2.097 & 2.118 \\
                    $\mu(n_{\mathrm{eff}}^{\mathrm{post}})$ & 17.235 & 17.238 \\
                    $\sigma(n_{\mathrm{eff}}^{\mathrm{post}})$ & 2.558 & 2.546 \\
                \addlinespace
                \multicolumn{3}{@{}l}{\textit{Geometry / orientation diagnostics}}\\
                    $\mu(\eta_i)$ & 1.445 & 1.445 \\
                    $\sigma(\eta_i)$ & 0.333 & 0.333 \\
                    $\mu(r_{\phi,i})$ & 0.430 & 0.430 \\
                    $\sigma(r_{\phi,i})$ & 0.205 & 0.205 \\
                    $\mu(g_{\mathrm{ident},i})$ & 2.156 & 2.156 \\
                    $\sigma(g_{\mathrm{ident},i})$ & 0.784 & 0.784 \\
                    $\mu(\theta_{z,i}^{\ast})$ & 0.000 & 0.795 \\
                    $\sigma(\theta_{z,i}^{\ast})$ & 0.000 & 0.319 \\
                \addlinespace
                \multicolumn{3}{@{}l}{\textit{Branch rates / fallback proxies}}\\
                    $\Pr(\phi_i=0)$ & 0.000 & 0.000 \\
                    $\Pr(\theta_{z,i}^{\ast}=0)$ & 1.000 & 0.000 \\
                    $\Pr(\mathrm{uniform})$ & 0.000 & 0.000 \\
                \bottomrule
            \end{tabularx}
           \begin{tablenotes}[flushleft]
                \footnotesize
                \item Notes. Here, $\mu(\cdot)$ and $\sigma(\cdot)$ are computed across target locations $i$. $p_{95}(\cdot)$ and $p_{99}(\cdot)$ denote median, the 95th and 99th percentiles across $i$. All values are rounded to three decimals.
            \end{tablenotes}
    \end{threeparttable}
\end{table}

\subsection{Supporting figures}

\section{Runtime environment and timing methodology}
\label{app:runtime-env}

This appendix documents the execution environment and timing conventions used for the runtime values reported in Chapter~8.

\subsection{Hardware and OS}
All runs were executed on:
\begin{itemize}
  \item CPU: \texttt{Intel(R) Xeon(R) CPU E3-1241 v3 @ 3.50GHz / 8 threads}
  \item RAM: \texttt{32 GB}
  \item OS: \texttt{Ubuntu 22.04.3 LTS}
\end{itemize}

\subsection{Software stack}
\begin{itemize}
  \item Python: \texttt{3.10.12}
  \item Core numerical libraries: \texttt{NumPy 1.24.2}, \texttt{SciPy 1.11.4}, \texttt{pandas 2.2.3}
  \item Spatial / ML libraries: \texttt{scikit-learn 1.4.2}, \texttt{statsmodels 0.13.5}, \texttt{PySAL 25.1}, \texttt{libpysal 4.13.0}, \texttt{esda 2.7.0}, \texttt{mgwr 2.2.1}, \texttt{geopandas 1.0.1}, \texttt{shapely 2.0.7}, \texttt{pyproj 3.6.0}, \texttt{rasterio 1.4.3}
\end{itemize}

\subsection{Timing methodology}
\begin{itemize}
  \item \textbf{What is timed.} Reported times correspond to end-to-end wall-clock time for the benchmark pipeline stage being reported (e.g., model fit + prediction over a fold), including neighborhood search when it is part of the method.
  \item \textbf{Warm-up and caching.} To reduce one-time compilation and caching artifacts (e.g., JIT warm-up, disk cache), timings are taken after an initial warm-up run where applicable, or are averaged over repeated runs with the same inputs.
  \item \textbf{Threading control.} If BLAS/LAPACK multi-threading is enabled, the thread count is fixed for all methods (e.g., via \texttt{OMP\_NUM\_THREADS} / \texttt{MKL\_NUM\_THREADS}) to ensure comparability.
  \item \textbf{Randomness.} Any randomized benchmark components use fixed seeds. Timing variability due to randomness is therefore minimized; remaining variability is attributed to system noise.
  \item \textbf{Reporting.} Chapter~8 tables report a representative summary statistic (e.g., mean time over folds). When a single hold-out run is used, the reported value corresponds to that run under the stated environment.
\end{itemize}

Because absolute runtimes depend on hardware, software versions, and implementation details (including neighbor-search backends), the reported times should be interpreted as \emph{empirical feasibility evidence} under the documented environment rather than as universal complexity claims.

\section{Evaluation protocol and leakage control}
\label{app:cv-protocol}

This appendix specifies the data-splitting, preprocessing, and out-of-sample evaluation protocol used for the benchmark and hold-out results reported in Chapter~8. The intent is to ensure comparability across methods while preventing information leakage through spatial duplicates or training-moment reuse.

\subsection{Data splits and coordinate-duplicate handling}
Let $\{(s_i,x_i,y_i)\}_{i=1}^n$ be the dataset with coordinates $s_i=(\mathrm{lat}_i,\mathrm{lon}_i)$.

\begin{itemize}
  \item \textbf{Split type.} Benchmark evaluation uses dataset-appropriate fixed-seed protocols. For Meuse, benchmark tables are reported under conventional $K$-fold cross-validation because of the small sample size and the dataset’s role as a classical local-regression benchmark. For the rice paddies benchmark, spatial block cross-validation is used to reduce optimistic performance arising from near-neighbor leakage in dense spatial data. Separate single-split hold-out checks are also reported for Meuse and rice paddies, as described in Chapter~8.
  \item \textbf{Unit of splitting.} Splits are generated at the level of unique coordinate pairs. All samples sharing an identical coordinate pair are assigned to the same fold (or the same train/test side) to prevent duplicate-location leakage.
  \item \textbf{Leakage accounting.} For each split, we record the number of test points whose coordinates exactly coincide with at least one training coordinate. If any such duplicates exist, they are not moved across folds; they remain grouped by the rule above and are reported for transparency.
\end{itemize}

\subsection{Training-only preprocessing and standardization}
All preprocessing steps that depend on data moments are computed on the training portion only.

\begin{itemize}
  \item \textbf{Standardization for diagnostics.} For $\kappa(\cdot)$ comparisons (Appendix~\ref{app:condwls2}), the local design is standardized to $X_i=[\mathbf 1, x]$ where $x$ is standardized using training-sample moments under the corresponding split. This standardization is used \emph{only} for the diagnostic $\kappa(\cdot)$ to make condition numbers dimensionless and comparable.
  \item \textbf{Model-fitting scale.} Unless stated otherwise in the chapter text, GR local diagnostic tables in Sections~8.2--8.3 are reported on the original data scale (no standardization) to support interpretability of coefficients and residual diagnostics.
\end{itemize}

\subsection{Comparable solver diagnostics for local-regression benchmarks}
The $\kappa(\cdot)$ values reported in Chapter~8 are benchmark-side comparable diagnostics rather than estimator-identifying quantities for every method. For GR and GWR, the reported diagnostic is based on the standardized local weighted normal matrix under the two-regressor design $[\mathbf 1, x]$. For LRR, the ridge-augmented counterpart is reported. For MGWR, because bandwidths are covariate-specific and the method does not admit a single estimator-defined local weight matrix in the same sense as GR or GWR, the reported $\kappa(\cdot)$ is computed using the pointwise aggregation rule implemented in the benchmark code for solver-stability comparison only. It is therefore a comparison-oriented diagnostic and not a redefinition of the MGWR estimator.

\subsection{Out-of-sample prediction for local methods}
For a test location $s$ and training pool $\mathcal T$:

\begin{itemize}
  \item \textbf{Neighborhood selection.} For all local methods that require neighborhoods (GR/LRR/GWR/MGWR and any KNN-based components), neighbors for a test location are selected from the training pool only (KNN in geographic space).
  \item \textbf{GR prediction at a test location.} GR weights and local coefficients are computed using only the training neighbors of the test location. Prediction uses the test covariate value $x(s)$. When the local design includes a distance-trend regressor $z_{ij}$, it is evaluated at the target as zero (i.e., target distance is $0$), so the trend term does not introduce test-to-train leakage.
  \item \textbf{Residual-KNN correction.} When GR+Residual-KNN is reported, the residuals used for correction are training residuals only. The residual correction at a test location is computed by KNN aggregation over training residuals (with the stated KNN settings), and added to the GR point prediction.
\end{itemize}

\subsection{Metrics and reporting}
For each split (or each fold), we compute RMSE, MAE (when applicable), and $R^2$ on the test set. Reported benchmark tables in Chapter~8 are aggregated across folds (mean over folds) or reported as single hold-out summaries when a hold-out protocol is used. Accordingly, single-split hold-out summaries and cross-validation benchmark summaries are reported for different purposes and should not be interpreted as directly interchangeable.

\subsection{Residual spatial diagnostic (Moran's I)}
Residual Moran's $I$ is reported as a \emph{diagnostic output} rather than a modeled dependence quantity.

\begin{itemize}
  \item \textbf{Adjacency structure.} The adjacency/weight matrix used for Moran's $I$ is fixed per dataset and does not depend on the regression weights $W_i$ used by GR/GWR/MGWR/LRR.
  \item \textbf{Computation.} Moran's $I$ is computed on residuals under the evaluation protocol (e.g., fold-wise residuals for CV benchmarks), using the same adjacency normalization across methods for comparability.
\end{itemize}

All benchmark methods use identical splits and evaluation pipelines so that performance differences reflect the modeling approach rather than differences in preprocessing or data partitioning.
\section{Dataset details}
\label{app:ch8-data-details}
This appendix records dataset details used in Chapter~8.
\subsection{Meuse dataset (heavy metals; small-$n$ benchmark)}
\label{app:ch8-data-meuse}
\begin{table}[H]
    \centering
    \begin{threeparttable}
        \caption{Descriptive statistics for Meuse variables ($n=155$).}
        \label{tab:meuse-descriptive}
        \footnotesize
        \setlength{\tabcolsep}{6pt}
        \renewcommand{\arraystretch}{1.15}
        \begin{tabularx}{0.75\textwidth}{@{\hspace{0.5em}} X r r @{\hspace{0.5em}}}
        \toprule
        Statistic & Cadmium & Lead \\
        \midrule
        Count & 155.000 & 155.000 \\
        Mean  & 3.246 & 153.361 \\
        Std. Dev. & 3.524 & 111.320 \\
        Min   & 0.200 & 37.000 \\
        25\%  & 0.800 & 72.500 \\
        Median  & 2.100 & 123.000 \\
        75\%  & 3.850 & 207.000 \\
        Max   & 18.100 & 654.000 \\
        \bottomrule
        \end{tabularx}
            \begin{tablenotes}[flushleft]
            \footnotesize
            \item Notes. Here, cadmium denotes the observed cadmium concentration at each sampling location,  and lead denotes the corresponding lead concentration used as the covariate. Values are reported for the full Meuse dataset ($n=155$) at the recorded observation coordinates. All values are rounded to three decimals.
        \end{tablenotes}
    \end{threeparttable}
\end{table}

\begin{figure}[H]
    \centering
    \begin{subfigure}[t]{0.45\textwidth}
        \centering
        \includegraphics[width=\linewidth]{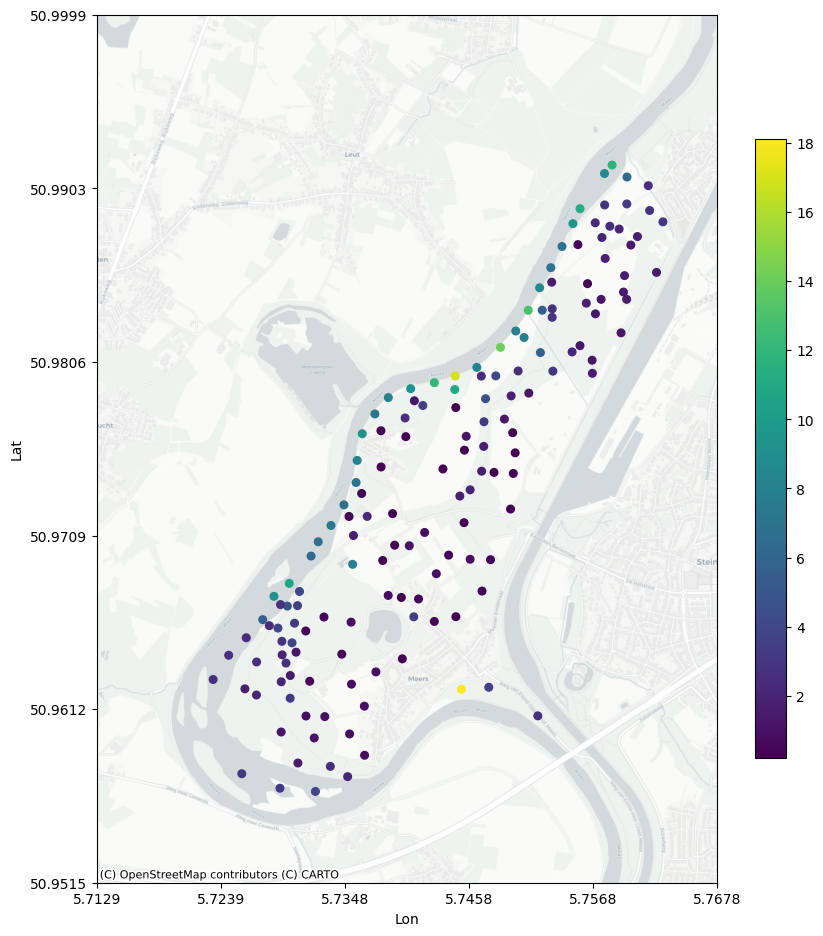}
        
        {(a) Cadmium concentration (response)}\vspace{4pt}
        \label{fig:meuse-cadmium}
    \end{subfigure}\hfill
    \begin{subfigure}[t]{0.45\textwidth}
        \centering
        \includegraphics[width=\linewidth]{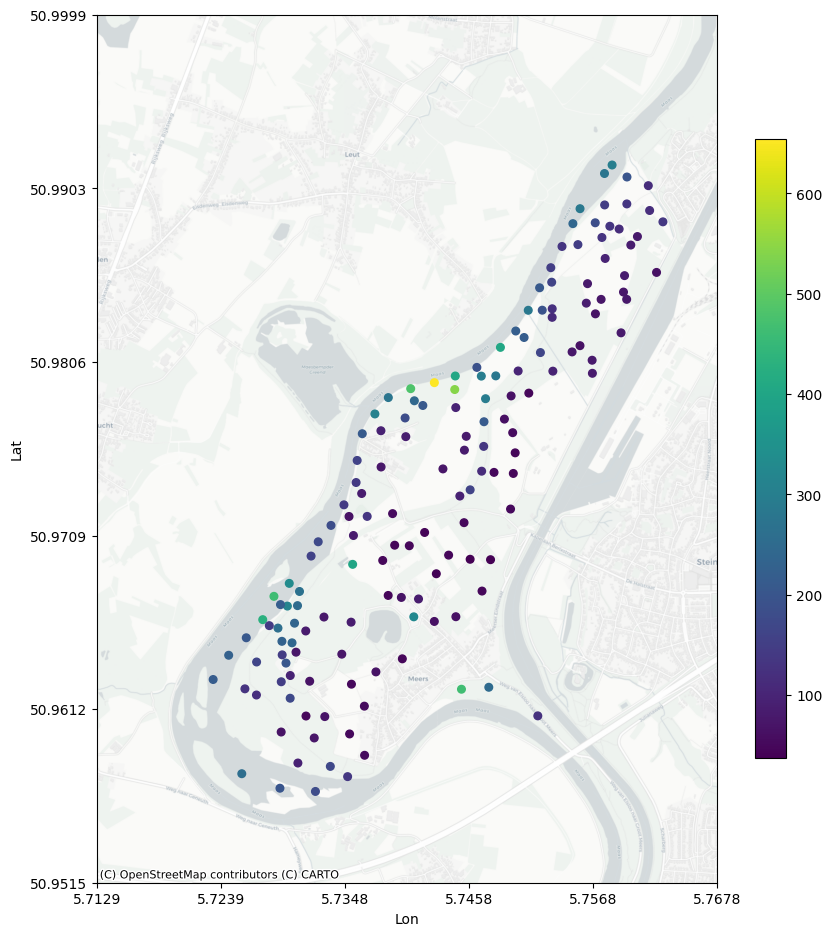}
        
        {(b) Lead (predictor)}\vspace{4pt}
        \label{fig:meuse-lead}
    \end{subfigure}
    \caption{Meuse dataset: spatial distribution of Cadmium (response) and Lead (predictor) at observation locations (WGS84; EPSG:4326).}    
    \label{fig:meuse-cadmium-lead}
\end{figure}

\subsection{Rice paddies dataset (large-$n$ application)}
\label{app:ch8-data-rice}
\begin{table}[H]
    \centering
    \begin{threeparttable}
        \caption{Descriptive statistics for rice paddies variables (n = 10{,}000).}
        \label{tab:ricepaddies-descriptive}
        \footnotesize
        \setlength{\tabcolsep}{6pt}
        \renewcommand{\arraystretch}{1.15}
        \begin{tabularx}{0.75\textwidth}{@{\hspace{0.5em}} X r r @{\hspace{0.5em}}}
        \toprule
        Statistic & NDVI & RVI \\
        \midrule
        Count & 10000 & 10000 \\
        Mean & 0.223 & 0.586 \\
        Std. Dev. & 0.155 & 0.235 \\
        Min & -0.150 & 0.009 \\
        25\% & 0.089 & 0.420 \\
        Median & 0.185 & 0.574 \\
        75\% & 0.346 & 0.735 \\
        Max & 0.677 & 1.888 \\
        \bottomrule
        \end{tabularx}
        \begin{tablenotes}[flushleft]
            \footnotesize
            \item Notes. Here, NDVI is computed from Sentinel-2 surface reflectance (red and NIR bands). RVI is computed from Sentinel-1 dual-polarization (VV, VH) backscatter. Values are reported at parcel centroids (WGS84; EPSG:4326). All values are rounded to three decimals.
        \end{tablenotes}
    \end{threeparttable}
\end{table}

\begin{figure}[H]
    \begin{subfigure}[t]{0.45\textwidth}
        \centering
        \includegraphics[width=\linewidth]{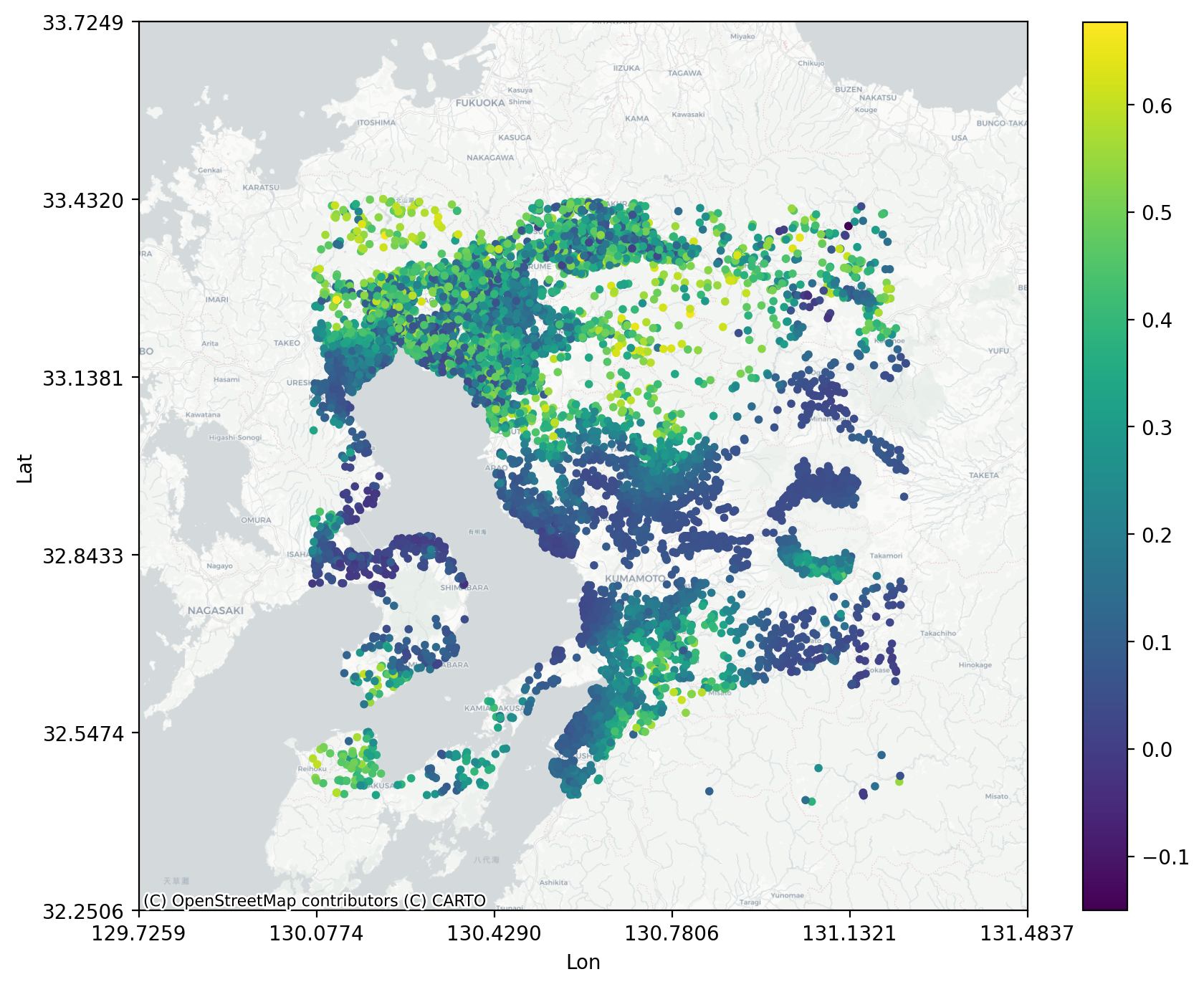}
        
        {(a) NDVI (Sentinel-2; response)}\vspace{4pt}
        \label{fig:ricepaddies-ndvi}
    \end{subfigure}\hfill
    \begin{subfigure}[t]{0.45\textwidth}
        \centering
        \includegraphics[width=\linewidth]{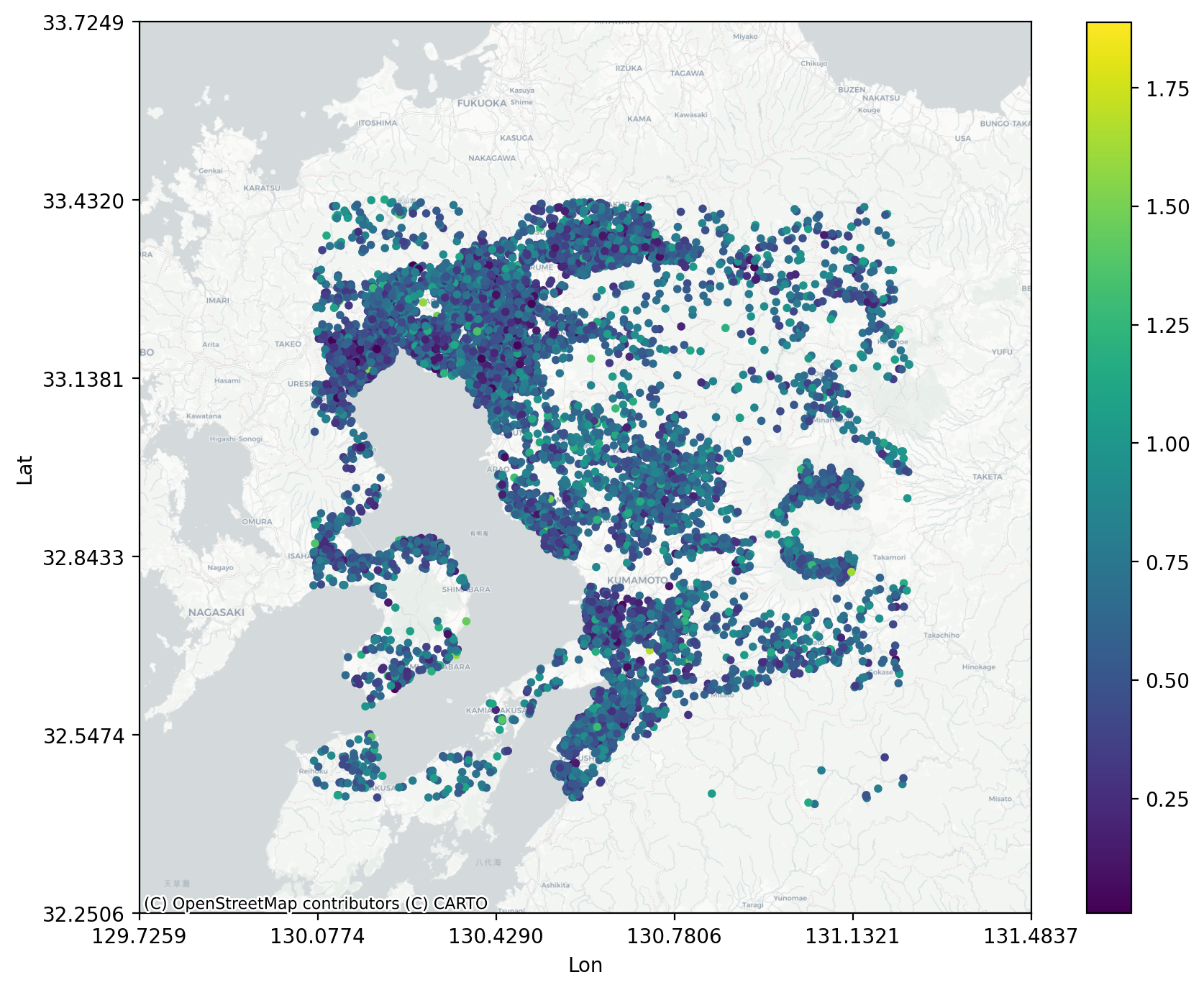}
        
        {(b) RVI (Sentinel-1; predictor)}\vspace{4pt}
        \label{fig:ricepaddies-rvi}
    \end{subfigure}
    \caption{Rice paddies dataset: spatial distribution of NDVI (response) and RVI (predictor) over parcel centroids (WGS84; EPSG:4326).}
    \label{fig:ricepaddies-ndvi-rvi}
\end{figure}

\section{Condition-number diagnostics for local WLS ($\kappa(\cdot)$)}
\label{app:condwls2}

This appendix reports $\kappa(\cdot)$ figures for the benchmark stability diagnostics. The figures visualize the $\kappa(\cdot)$ distributions using: (i) Histogram (full range), (ii) ECDF (full range), (iii) CCDF.

\subsection{Meuse}
\label{app:condwls2-Meuse}
\begin{figure}[H]
    \centering
    \begin{subfigure}[t]{0.33\textwidth}
        \centering
        \includegraphics[width=\linewidth]{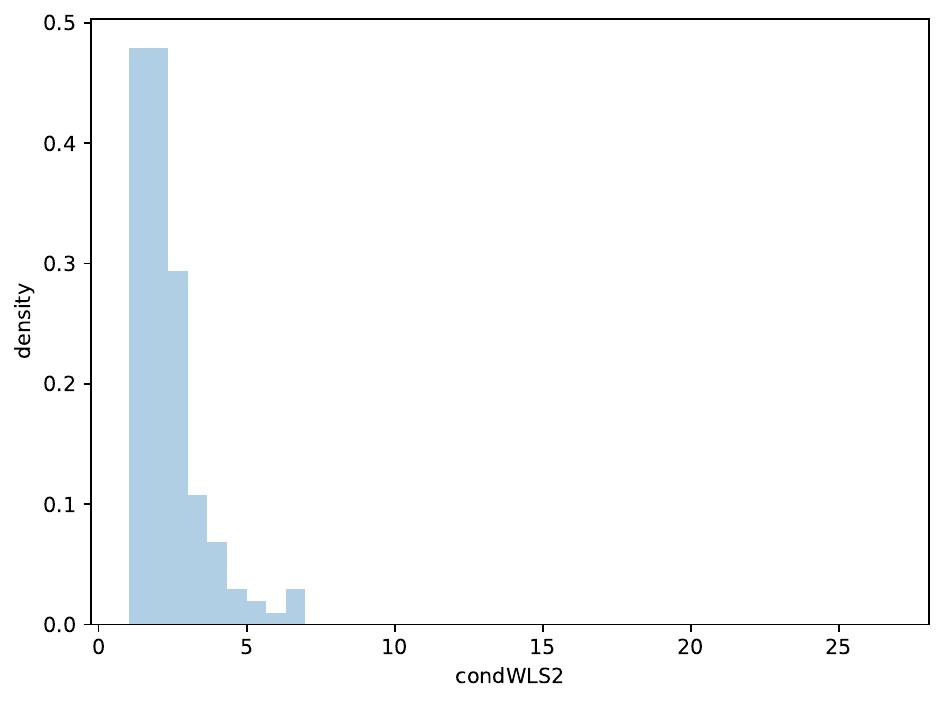}
        
        {(a) GR: Histogram}\vspace{4pt}
        \label{fig:ch8-Meuse-histogram-gr}
    \end{subfigure}\hfill
    \begin{subfigure}[t]{0.33\textwidth}
        \centering
        \includegraphics[width=\linewidth]{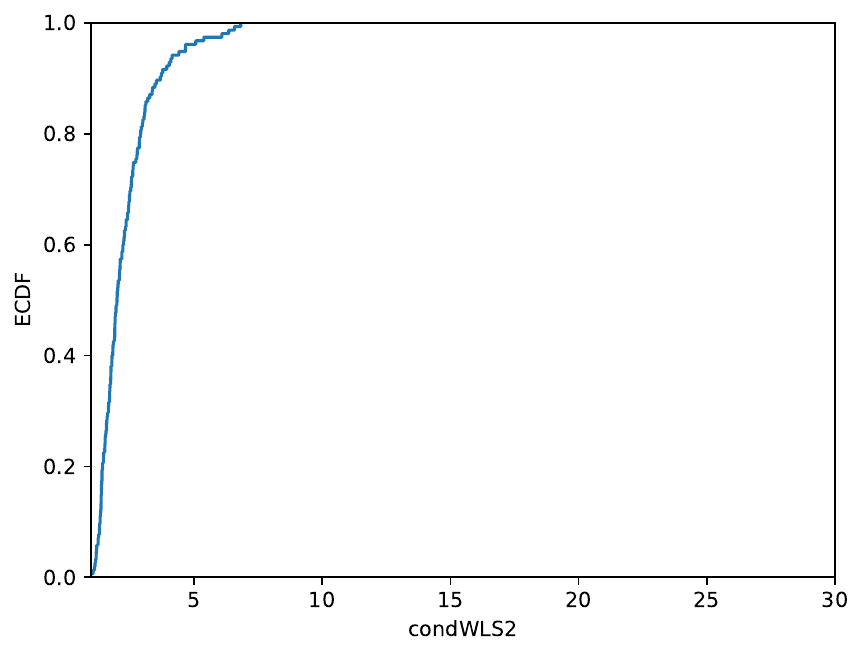}    
        
        {(b) GR: ECDF}\vspace{4pt}
        \label{fig:ch8-Meuse-ecdf-gr}
    \end{subfigure}\hfill
    \begin{subfigure}[t]{0.33\textwidth}
        \centering
        \includegraphics[width=\linewidth]{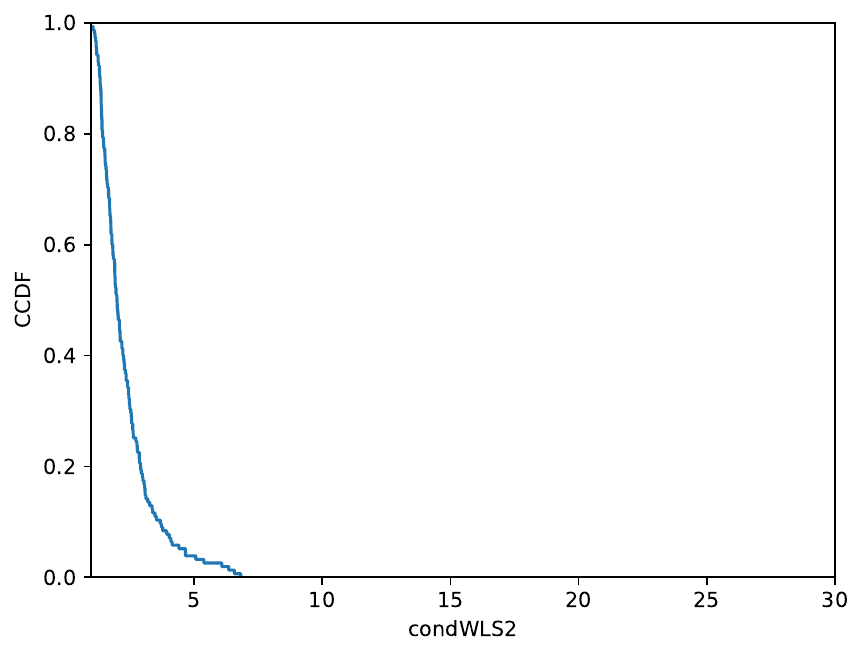}    
        
        {(c) GR: CCDF}\vspace{4pt}
        \label{fig:ch8-Meuse-ccdf-gr}
    \end{subfigure}
    \begin{subfigure}[t]{0.33\textwidth}
        \centering
        \includegraphics[width=\linewidth]{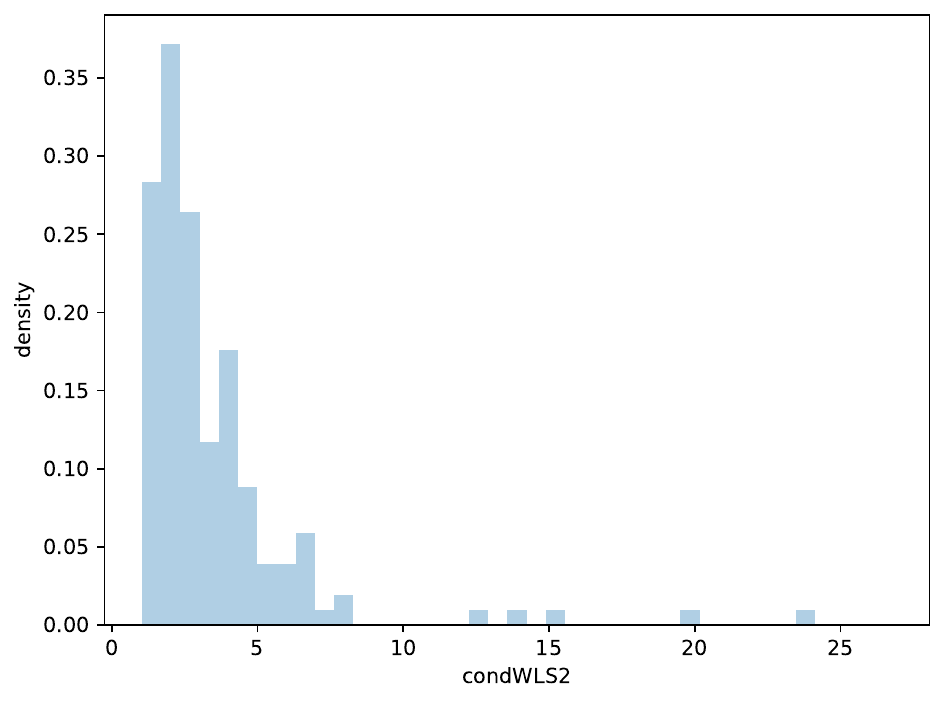}
        
        {(d) LLR: Histogram}\vspace{4pt}
        \label{fig:ch8-Meuse-histogram-llr}
    \end{subfigure}\hfill
    \begin{subfigure}[t]{0.33\textwidth}
        \centering
        \includegraphics[width=\linewidth]{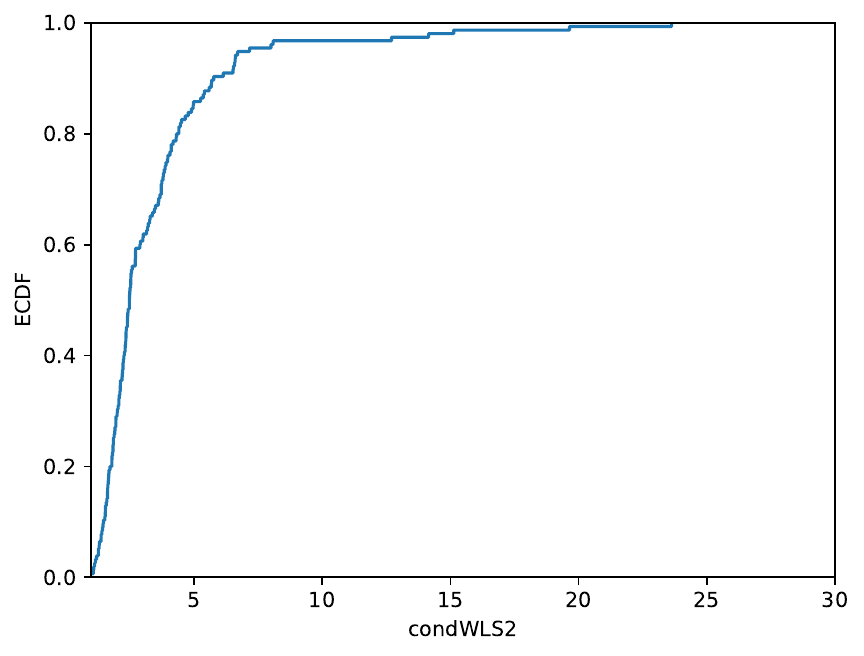}    
        
        {(e) LLR: ECDF}\vspace{4pt}
        \label{fig:ch8-Meuse-ecdf-llr}
    \end{subfigure}\hfill
    \begin{subfigure}[t]{0.33\textwidth}
        \centering
        \includegraphics[width=\linewidth]{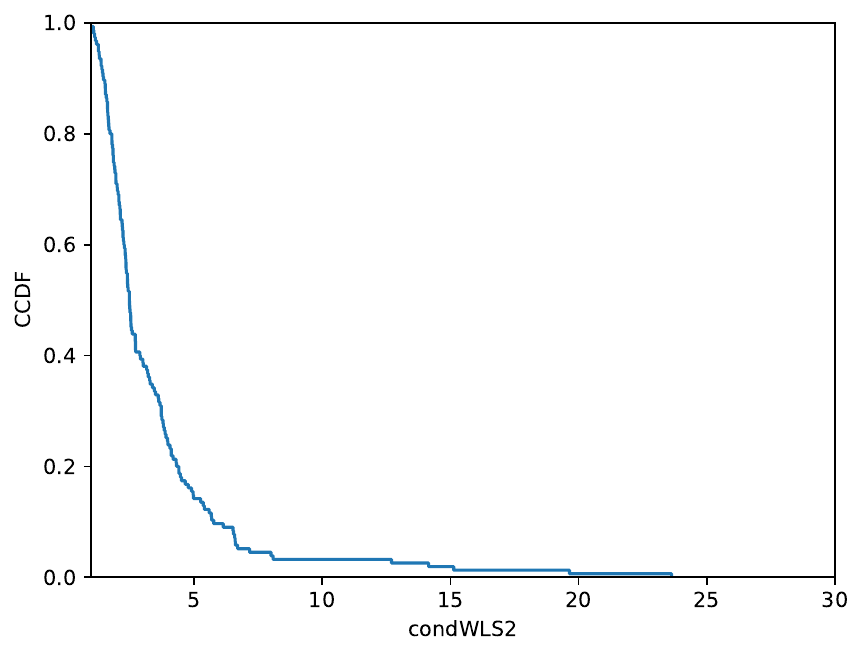}    
        
        {(f) LLR: CCDF}\vspace{4pt}
        \label{fig:ch8-Meuse-ccdf-llr}
    \end{subfigure}
    \begin{subfigure}[t]{0.33\textwidth}
        \centering
        \includegraphics[width=\linewidth]{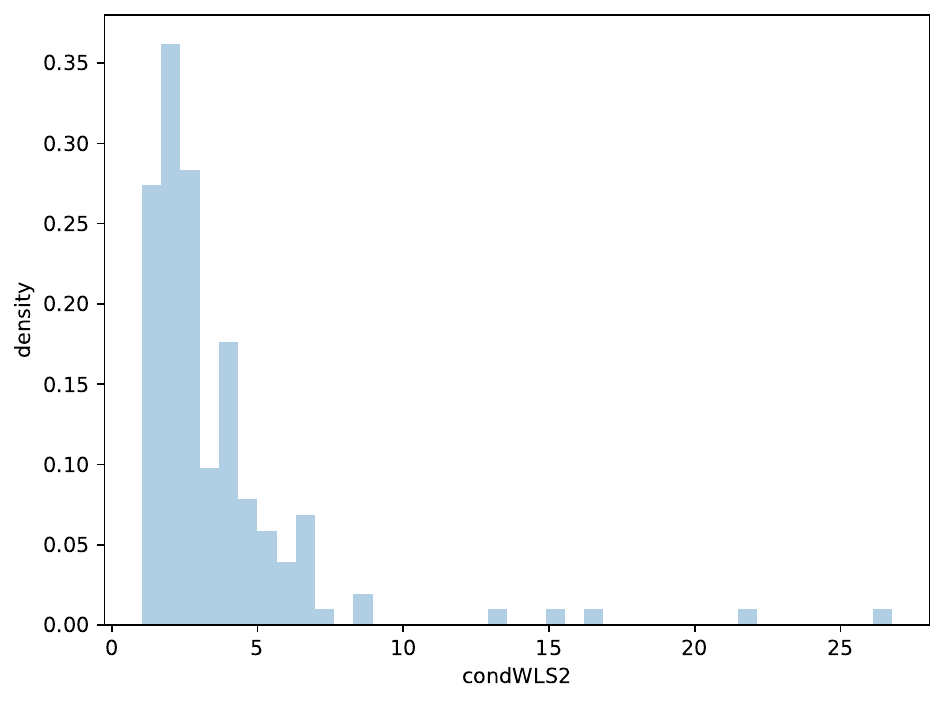}
        
        {(g) GWR: Histogram}\vspace{4pt}
        \label{fig:ch8-Meuse-histogram-gwr}
    \end{subfigure}\hfill
    \begin{subfigure}[t]{0.33\textwidth}
        \centering
        \includegraphics[width=\linewidth]{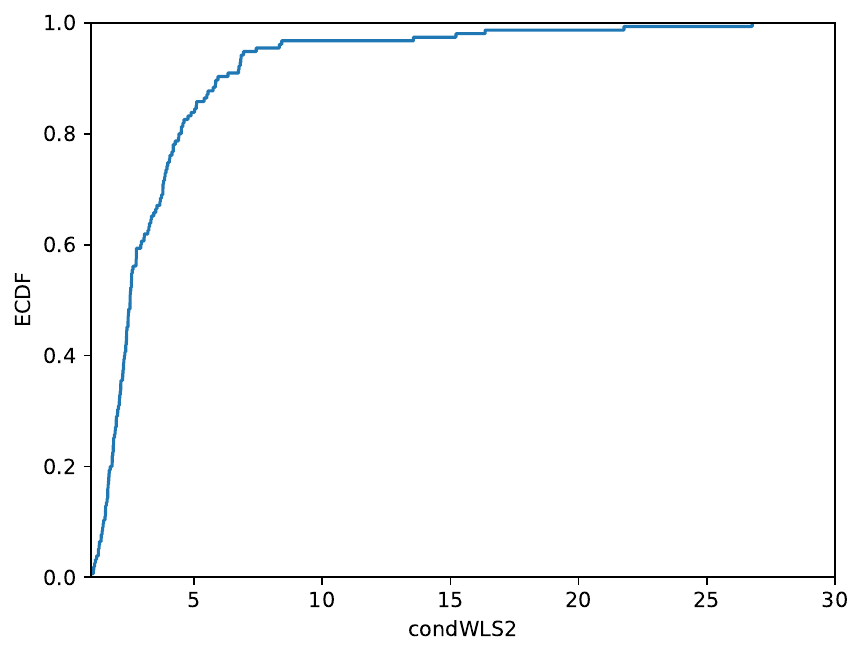}    
        
        {(h) GWR: ECDF}\vspace{4pt}
        \label{fig:ch8-Meuse-ecdf-gwr}
    \end{subfigure}\hfill
    \begin{subfigure}[t]{0.33\textwidth}
        \centering
        \includegraphics[width=\linewidth]{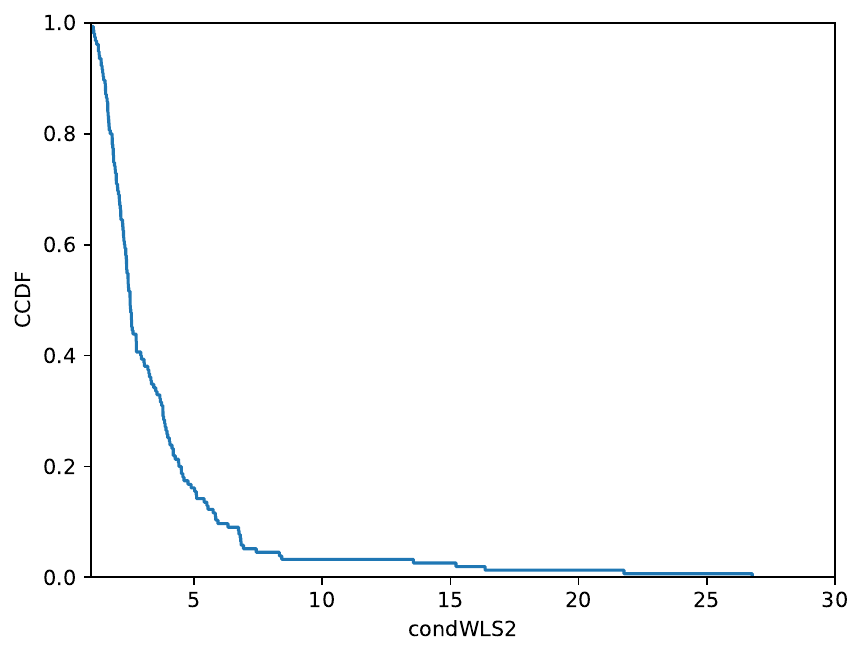}    
        
        {(i) GWR: CCDF}\vspace{4pt}
        \label{fig:ch8-Meuse-ccdf-gwr}
    \end{subfigure}
    \begin{subfigure}[t]{0.33\textwidth}
        \centering
        \includegraphics[width=\linewidth]{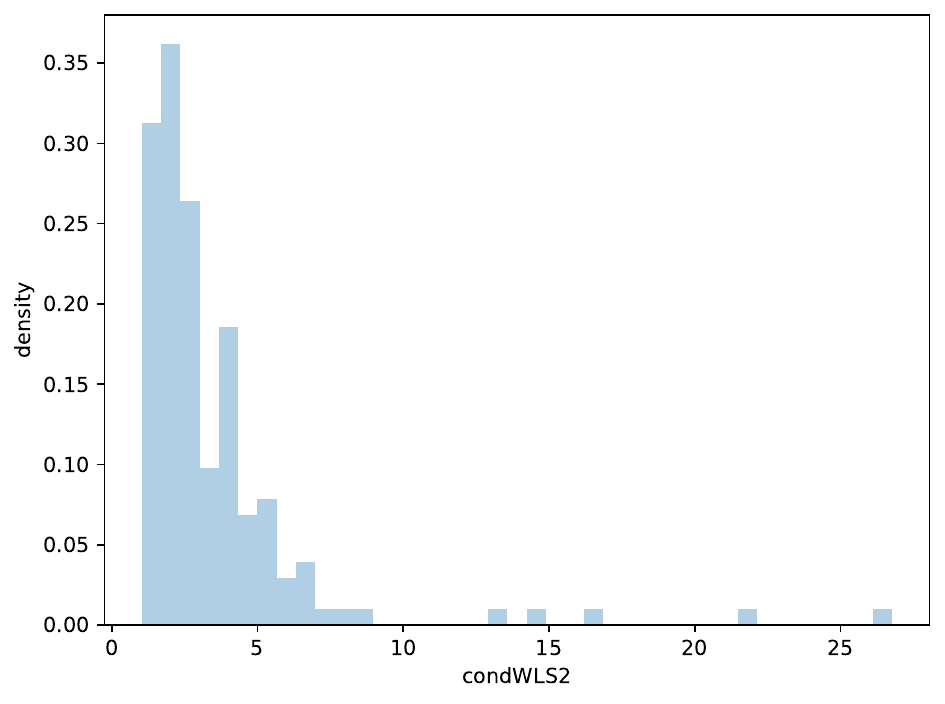}
        
        {(j) MGWR: Histogram}\vspace{4pt}
        \label{fig:ch8-Meuse-histogram-mgwr}
    \end{subfigure}\hfill
    \begin{subfigure}[t]{0.33\textwidth}
        \centering
        \includegraphics[width=\linewidth]{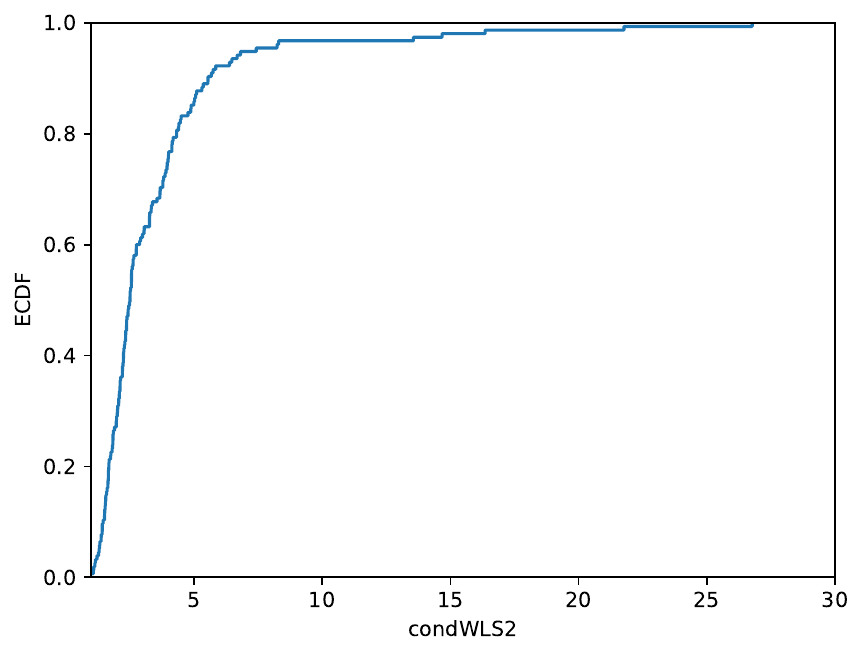}    
        
        {(k) MGWR: ECDF}\vspace{4pt}
        \label{fig:ch8-Meuse-ecdf-mgwr}
    \end{subfigure}\hfill
    \begin{subfigure}[t]{0.33\textwidth}
        \centering
        \includegraphics[width=\linewidth]{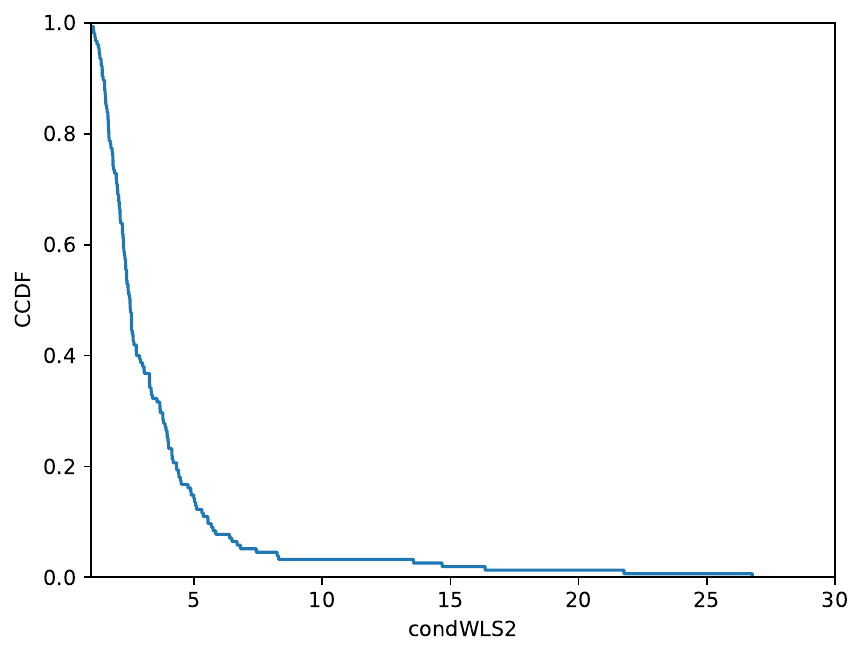}    
        
        {(l) MGWR: CCDF}\vspace{4pt}
        \label{fig:ch8-Meuse-ccdf-mgwr}
    \end{subfigure}
    \label{fig:ch8-Meuse}
    \caption{Histogram, ECDF, CCDF of $\texttt{condWLS2}$ for Meuse}
\end{figure}

\subsection{Rice paddies}
\label{app:condwls2-RicePaddies}
\begin{figure}[H]

    \centering
    \begin{subfigure}[t]{0.33\textwidth}
        \centering
        \includegraphics[width=\linewidth]{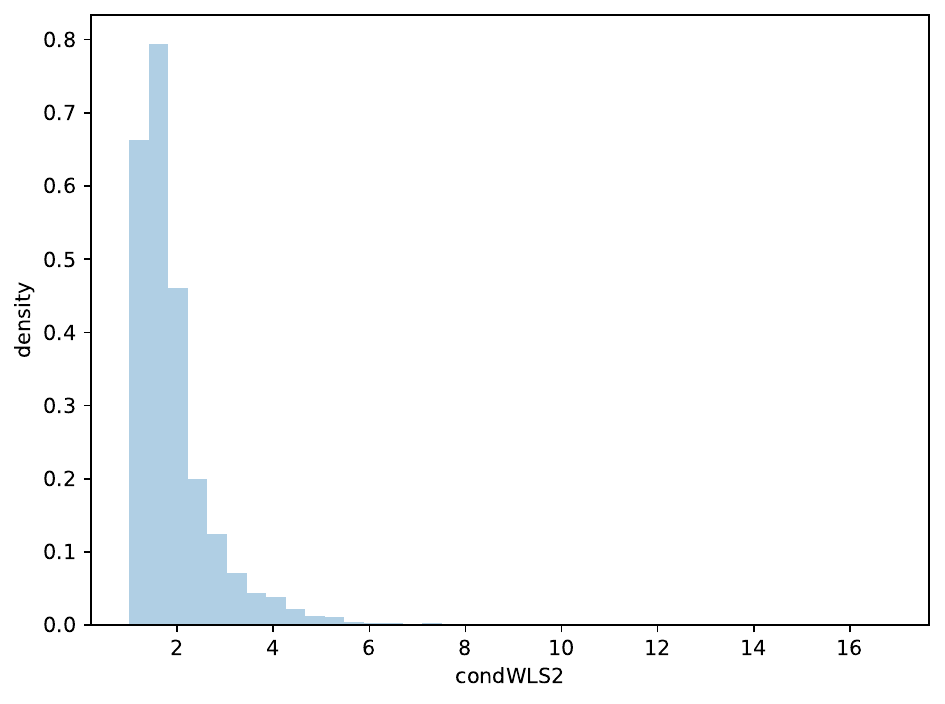}
        
        {(a) GR: Histogram}\vspace{4pt}
        \label{fig:ch8-RicePaddies-histogram-gr}
    \end{subfigure}\hfill
    \begin{subfigure}[t]{0.33\textwidth}
        \centering
        \includegraphics[width=\linewidth]{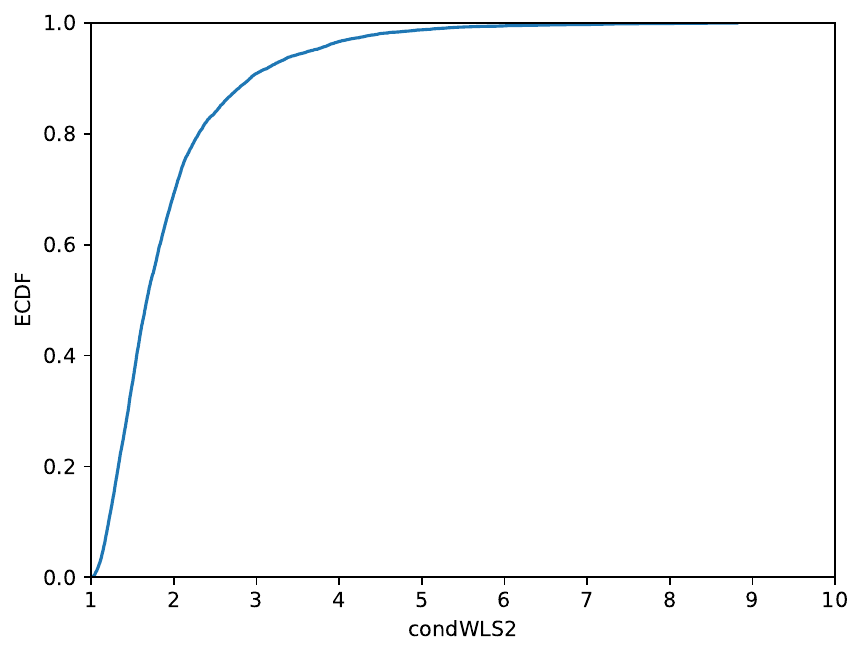}    
        
        {(b) GR: ECDF}\vspace{4pt}
        \label{fig:ch8-RicePaddies-ecdf-gr}
    \end{subfigure}\hfill
    \begin{subfigure}[t]{0.33\textwidth}
        \centering
        \includegraphics[width=\linewidth]{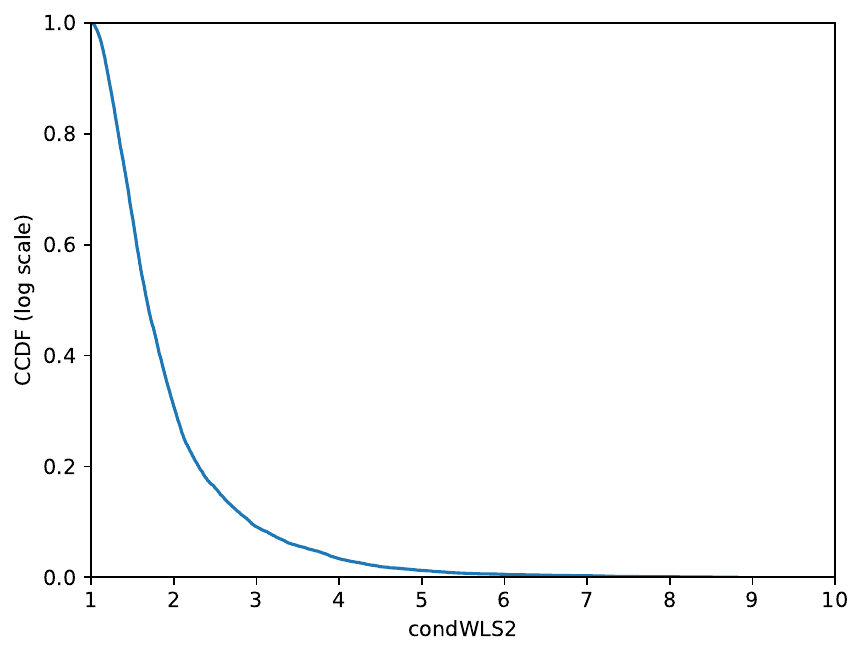}    
        
        {(c) GR: CCDF}\vspace{4pt}
        \label{fig:ch8-RicePaddies-ccdf-gr}
    \end{subfigure}
    \begin{subfigure}[t]{0.33\textwidth}
        \centering
        \includegraphics[width=\linewidth]{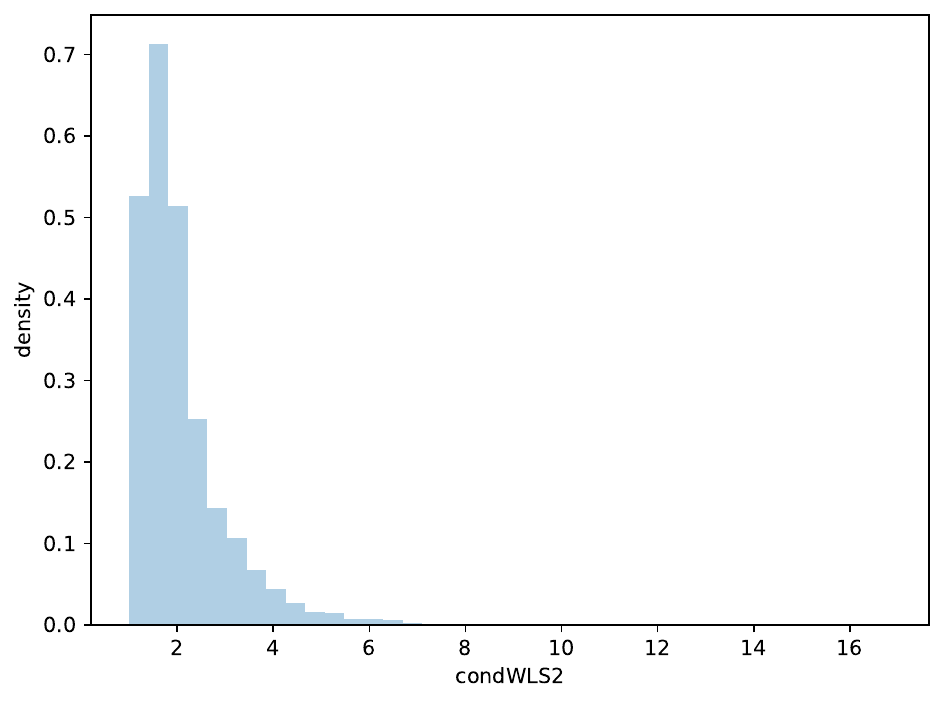}
        
        {(d) LLR: Histogram}\vspace{4pt}
        \label{fig:ch8-RicePaddies-histogram-lrr}
    \end{subfigure}\hfill
    \begin{subfigure}[t]{0.33\textwidth}
        \centering
        \includegraphics[width=\linewidth]{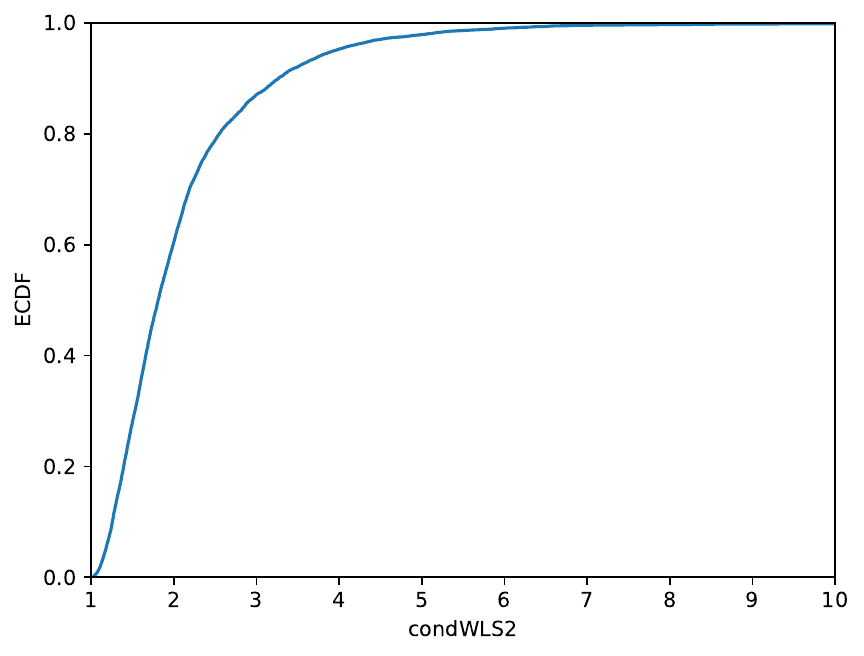}    
        
        {(e) LLR: ECDF}\vspace{4pt}
        \label{fig:ch8-RicePaddies-ecdf-lrr}
    \end{subfigure}\hfill
    \begin{subfigure}[t]{0.33\textwidth}
        \centering
        \includegraphics[width=\linewidth]{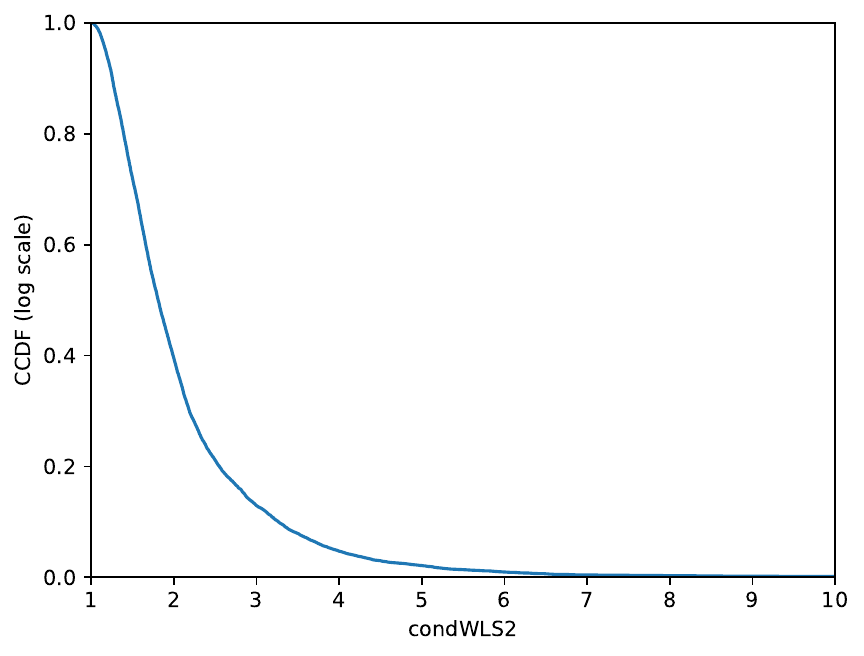}    
        
        {(f) LLR: CCDF}\vspace{4pt}
        \label{fig:ch8-RicePaddies-ccdf-llr}
    \end{subfigure}
    \begin{subfigure}[t]{0.33\textwidth}
        \centering
        \includegraphics[width=\linewidth]{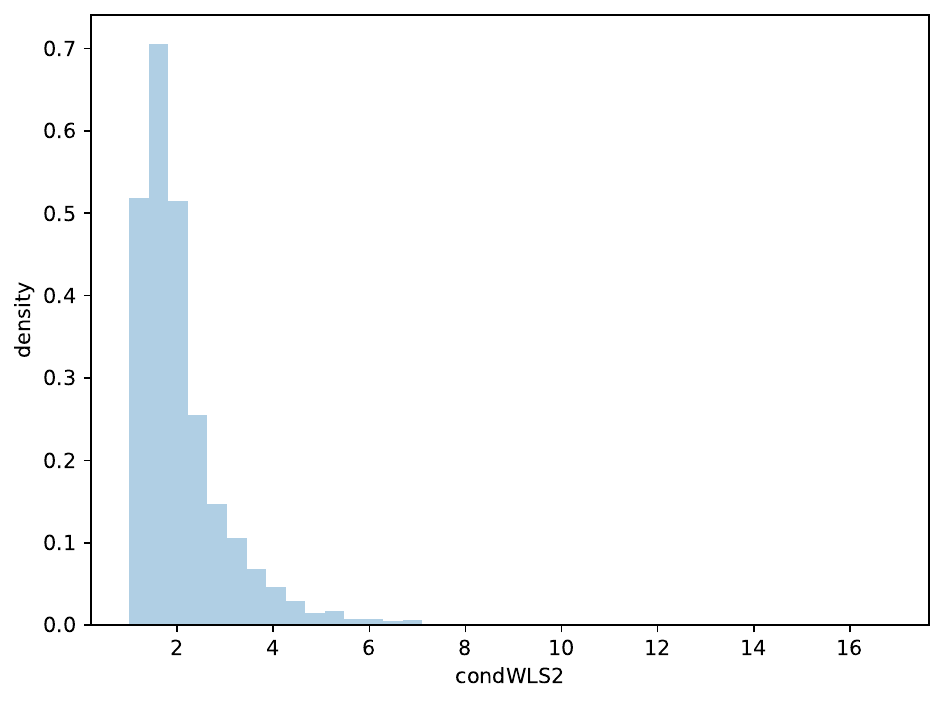}
        
        {(g) GWR: Histogram}\vspace{4pt}
        \label{fig:ch8-RicePaddies-histogram-gwr}
    \end{subfigure}\hfill
    \begin{subfigure}[t]{0.33\textwidth}
        \centering
        \includegraphics[width=\linewidth]{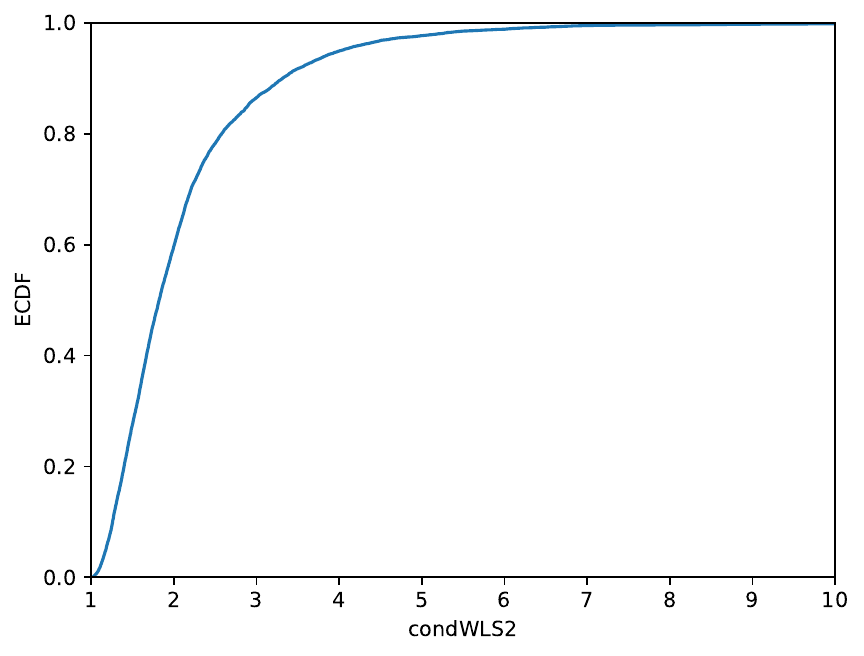}    
        
        {(h) GWR: ECDF}\vspace{4pt}
        \label{fig:ch8-RicePaddies-ecdf-gwr}
    \end{subfigure}\hfill
    \begin{subfigure}[t]{0.33\textwidth}
        \centering
        \includegraphics[width=\linewidth]{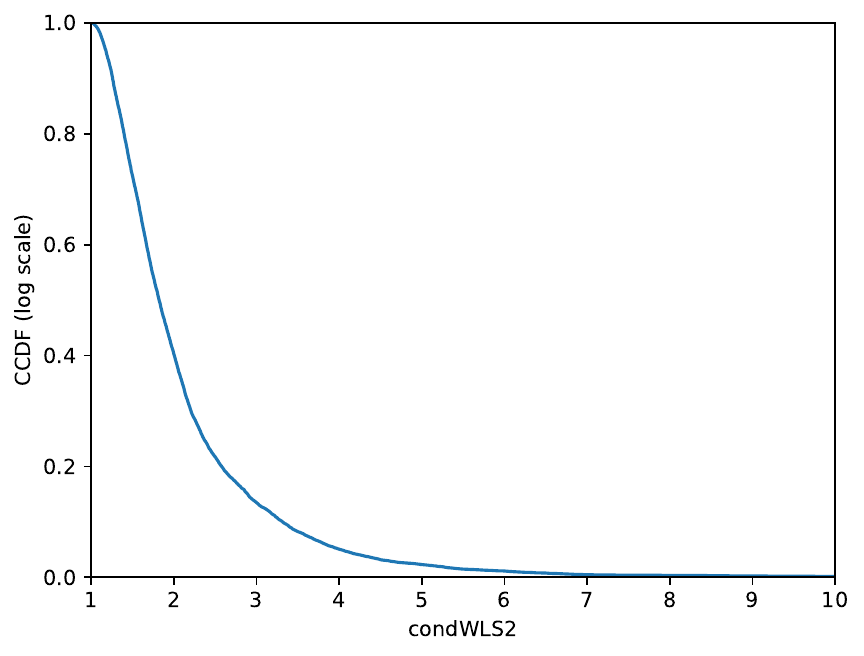}    
        
        {(i) GWR: CCDF}\vspace{4pt}
        \label{fig:ch8-RicePaddies-ccdf-gwr}
    \end{subfigure}
    \begin{subfigure}[t]{0.33\textwidth}
        \centering
        \includegraphics[width=\linewidth]{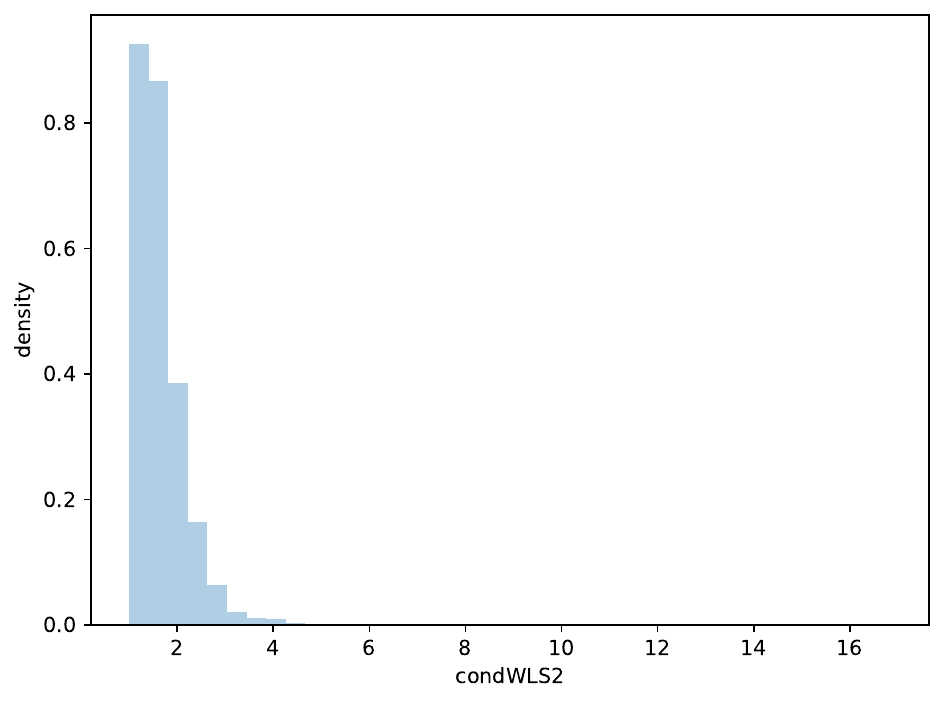}
        
        {(j) MGWR: Histogram}\vspace{4pt}
        \label{fig:ch8-RicePaddies-histogram-mgwr}
    \end{subfigure}\hfill
    \begin{subfigure}[t]{0.33\textwidth}
        \centering
        \includegraphics[width=\linewidth]{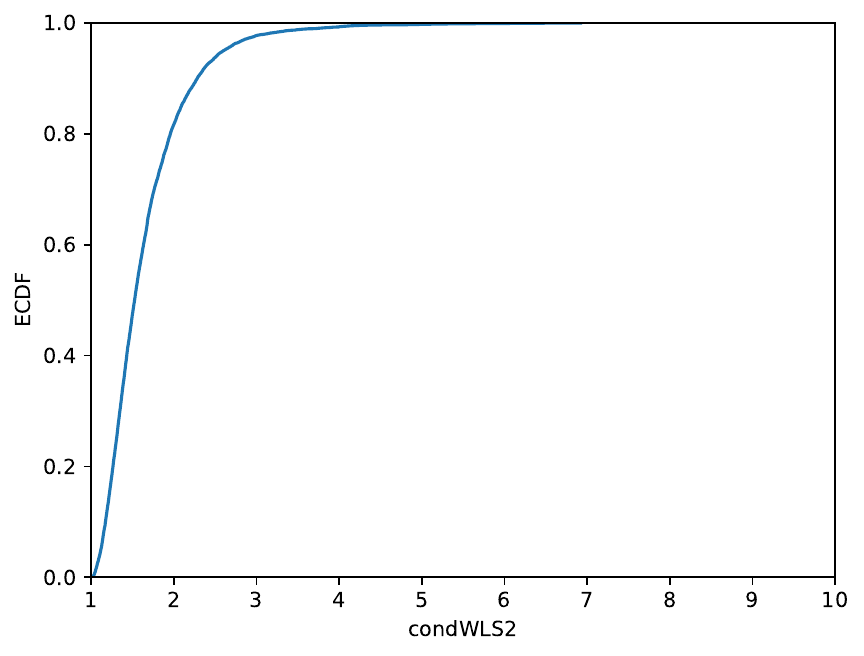}    
        
        {(k) MGWR: ECDF}\vspace{4pt}
        \label{fig:ch8-RicePaddies-ecdf-mgwr}
    \end{subfigure}\hfill
    \begin{subfigure}[t]{0.33\textwidth}
        \centering
        \includegraphics[width=\linewidth]{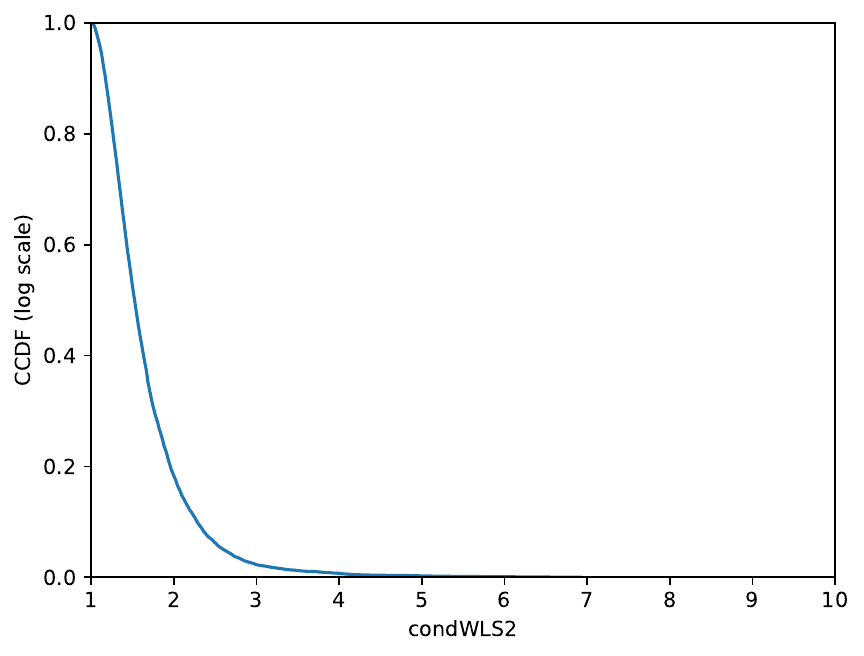}    
        {(l) MGWR: CCDF}\vspace{4pt}
        \label{fig:ch8-RicePaddies-ccdf-mgwr}
    \end{subfigure}
    \label{fig:ch8-RicePaddies}
    \caption{Histogram, ECDF, CCDF of $\kappa(\cdot)$ for Rice Paddies}
\end{figure}

\section{Diagnostic maps and reliability-layer interpretation}
\label{app:diagnostic-maps}

This appendix describes the diagnostic map panels referenced in Chapter~8. The purpose is to provide a practical ``reliability layer'' for interpreting local coefficient surfaces. 

\subsection{What is mapped}
For each target location $s_i$ where a local estimate is reported, we visualize selected realized outputs from Chapters~4--7.

For the method-comparison benchmark tables in Chapter~8, the reported conditioning diagnostic follows the standardized design convention described in Appendix~\ref{app:cv-protocol}. By contrast, the GR diagnostic maps shown here visualize the realized GR normal-matrix conditioning $\kappa(\mathbf M_i^{(\mathrm{nor})})$ computed from the realized local weight map.
\begin{itemize}
  \item \textbf{Solver stability:} for the GR diagnostic maps in this appendix, we visualize the realized GR normal-matrix conditioning $\kappa(\mathbf M_i^{(\mathrm{nor})})$. This is distinct from the benchmark-side comparable conditioning diagnostic reported in Chapter~8 under the standardized design convention of Appendix~\ref{app:cv-protocol}.
  \item \textbf{Effective support:} $n_{\mathrm{eff}}^{\mathrm{post}}(i)$ and, where informative, $n_{\mathrm{eff}}^{\mathrm{raw}}(i)$.
  \item \textbf{Directional concentration:} $r_{\phi,i}$ (bearing resultant magnitude), indicating whether neighbors are directionally one-sided.
  \item \textbf{Geometry anisotropy:} $\eta_i$ and (optionally) the realized dominant bearing $\phi_i$.
  \item \textbf{Value-orientation (when enabled):} $\theta_{z,i}^\ast$ and its identifiability/fallback indicators (e.g., $\mathbb{1}\{\theta_{z,i}^\ast=0\}$ when this corresponds to the non-identifiable branch).
  \item \textbf{Safeguard flags:} uniform fallback indicators and any deterministic exclusion flags used when a local solve is undefined under the well-posedness convention.
\end{itemize}

\subsection{How to read the maps}
The maps are intended as conditional interpretation aids:

\begin{itemize}
    \item \textbf{Masking / down-weighting.} Locations with extreme $\kappa(\mathbf M_i^{(\mathrm{nor})})$ (heavy tail) or very small $n_{\mathrm{eff}}^{\mathrm{post}}$ can be marked as numerically fragile and excluded from substantive coefficient interpretation under the current neighborhood geometry.
    \item \textbf{Diagnosing geometry-induced fragility.} Large $r_{\phi,i}$ indicates strongly one-sided support; combined with large $\kappa(\mathbf M_i^{(\mathrm{nor})})$, this suggests the realized local normal equations may be sensitive to small neighborhood perturbations.
    \item \textbf{Hand-off logic.} Regions flagged as fragile under the above diagnostics are candidates for hand-off to alternative workflows (e.g., dependence-modeling approaches such as kriging when explaining spatial dependence is central, or nonparametric learners when prediction is the primary objective).
\end{itemize}

\subsection{Plotting conventions}
\begin{itemize}
  \item \textbf{Color scaling.} When distributions are heavy-tailed (especially condition numbers), plots may use clipped color scales (e.g., at a high quantile) to preserve spatial structure visibility while reporting the unclipped quantiles in tables.
  \item \textbf{Consistency across methods.} For method comparisons of solver diagnostics, the standardized design convention is used so that $\texttt{condWLS2}$ is comparable across local-regression solvers that admit $X_i^\top W_i X_i$.
  \item \textbf{Adjacency independence.} Residual Moran's $I$ maps (when shown) use the fixed adjacency structure and are not altered by regression weights.
\end{itemize}

No additional methodological claims are introduced by these maps; they visualize realized outputs already defined in Chapters~4--7 and are provided to support transparent, conditional interpretation.

\subsection{How to read the maps}
The maps are intended as conditional interpretation aids:

\begin{itemize}
    \item \textbf{Masking / down-weighting.} Locations with extreme $\kappa(\mathbf M_i^{(\mathrm{nor})})$ (heavy tail) or very small $n_{\mathrm{eff}}^{\mathrm{post}}$ can be marked as numerically fragile and excluded from substantive coefficient interpretation under the current neighborhood geometry.
    \item \textbf{Diagnosing geometry-induced fragility.} Large $r_{\phi,i}$ indicates strongly one-sided support; combined with large $\kappa(\mathbf M_i^{(\mathrm{nor})})$, this suggests the realized local normal equations may be sensitive to small neighborhood perturbations.
    \item \textbf{Hand-off logic.} Regions flagged as fragile under the above diagnostics are candidates for hand-off to alternative workflows (e.g., dependence-modeling approaches such as kriging when explaining spatial dependence is central, or nonparametric learners when prediction is the primary objective).
\end{itemize}

\subsection{Plotting conventions}
\begin{itemize}
  \item \textbf{Color scaling.} When distributions are heavy-tailed (especially condition numbers), plots may use clipped color scales (e.g., at a high quantile) to preserve spatial structure visibility while reporting the unclipped quantiles in tables.
  \item \textbf{Consistency across methods.} For method comparisons of solver diagnostics, the standardized design convention is used so that $\texttt{condWLS2}$ is comparable across local-regression solvers that admit $X_i^\top W_i X_i$.
  \item \textbf{Adjacency independence.} Residual Moran's $I$ maps (when shown) use the fixed adjacency structure and are not altered by regression weights.
\end{itemize}

No additional methodological claims are introduced by these maps; they visualize realized outputs already defined in Chapters~4--7 and are provided to support transparent, conditional interpretation.

\subsection{Diagnostic Maps: Meuse}
\label{app:diagnostic-maps-meuse}
\begin{figure}[H]
    \centering
    \begin{subfigure}[t]{0.49\textwidth}
        \centering
        \includegraphics[width=\linewidth]{Figures/Appendix.Meuse.condM_nor.png}
        
        {(a) $\kappa(\mathbf M_i^{(\mathrm{nor})})$}\vspace{4pt}
    \end{subfigure}\hfill
    \begin{subfigure}[t]{0.49\textwidth}
        \centering
        \includegraphics[width=\linewidth]{Figures/Appendix.Meuse.r_phi.png}
        
        {(b) $r_{\phi}$}\vspace{4pt}
    \end{subfigure}
    \begin{subfigure}[t]{0.49\textwidth}
        \centering
        \includegraphics[width=\linewidth]{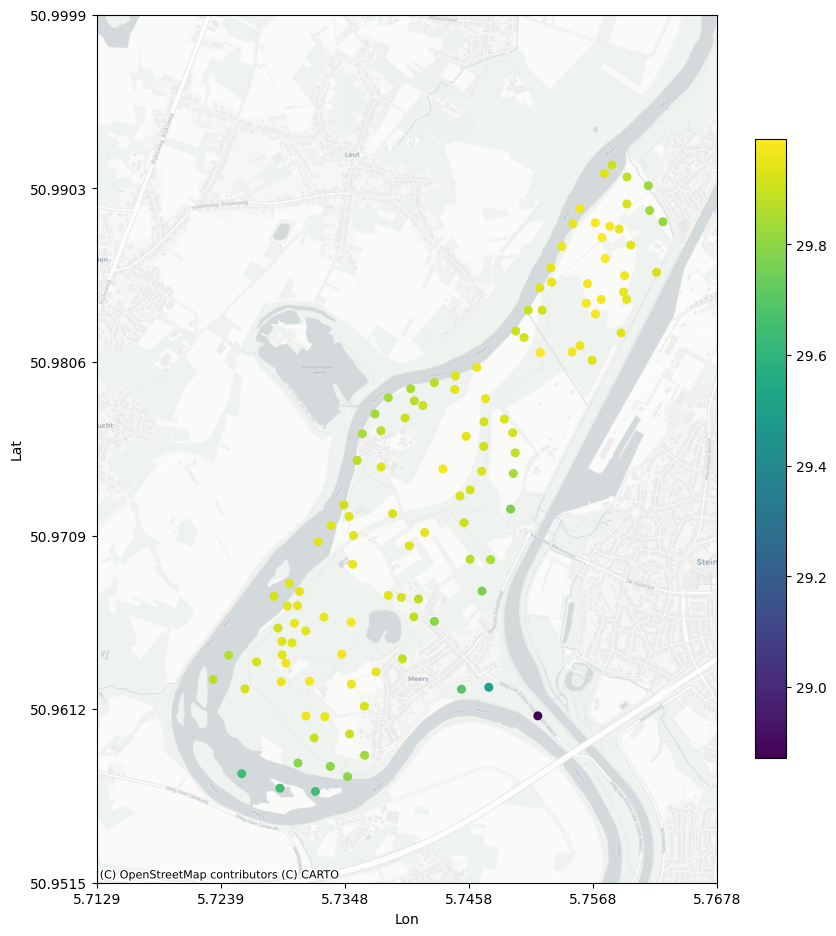}
        
        {(c) $n_{\mathrm{eff}}^{\mathrm{post}}$}\vspace{4pt}
    \end{subfigure}\hfill
    \begin{subfigure}[t]{0.49\textwidth}
        \centering
        \includegraphics[width=\linewidth]{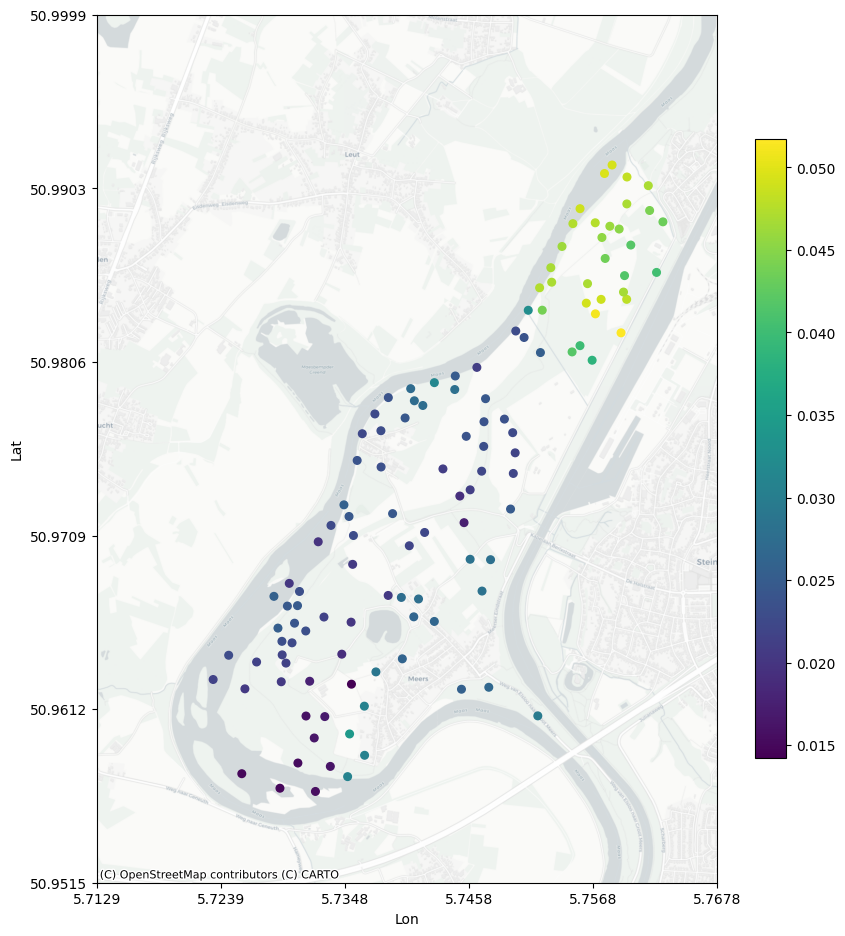}
        
        {(d) $\hat\beta_1(s)$}\vspace{4pt}
    \end{subfigure}
    \caption{Diagnostic Maps: Meuse}
    \label{fig:diagnostic-maps-meuse}
\end{figure}

\subsection{Diagnostic Maps: Rice paddies}
\label{app:diagnostic-maps-RicePaddies}
\begin{figure}[H]
    \centering
    \begin{subfigure}[t]{0.49\textwidth}
        \centering
        \includegraphics[width=\linewidth]{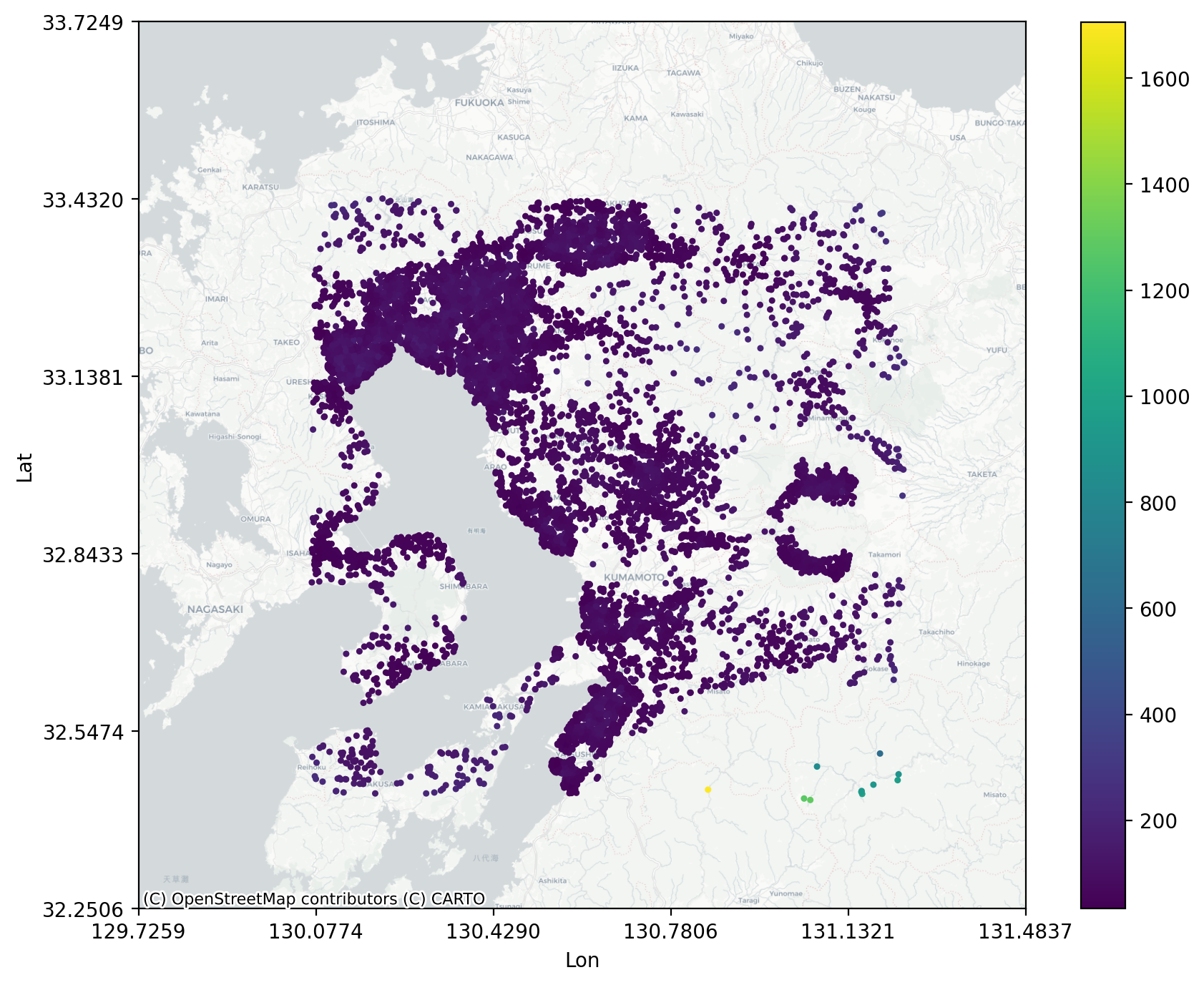}
        
        {(a) $\kappa(\mathbf M_i^{(\mathrm{nor})})$}\vspace{4pt}
    \end{subfigure}\hfill
    \begin{subfigure}[t]{0.49\textwidth}
        \centering
        \includegraphics[width=\linewidth]{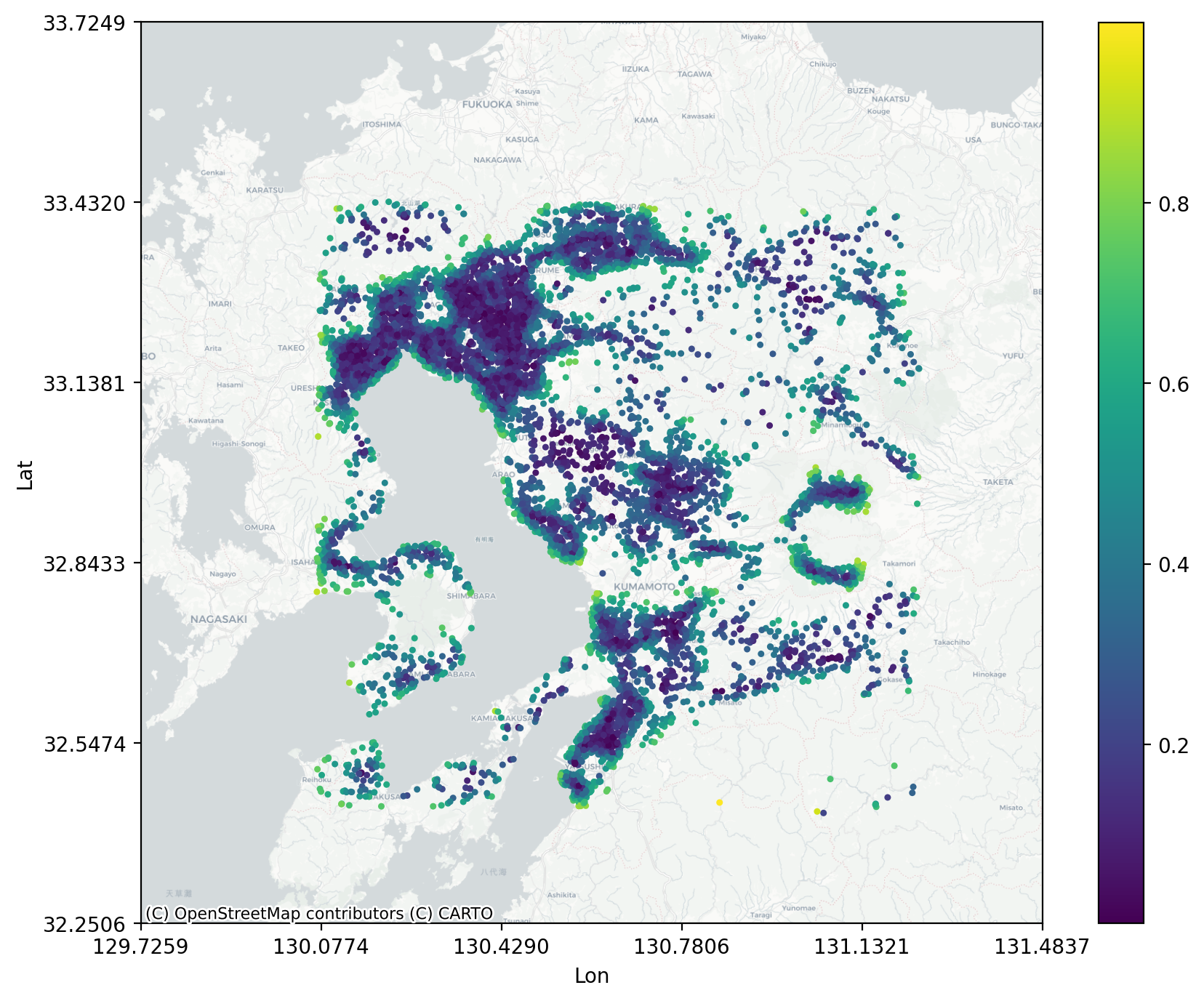}
        
        {(b) $r_{\phi}$}\vspace{4pt}
    \end{subfigure}
    \begin{subfigure}[t]{0.49\textwidth}
        \centering
        \includegraphics[width=\linewidth]{Figures/Appendix.RicePaddies.neff_post.png}
        
        {(c) $n_{\mathrm{eff}}^{\mathrm{post}}$}\vspace{4pt}
    \end{subfigure}\hfill
    \begin{subfigure}[t]{0.49\textwidth}
        \centering
        \includegraphics[width=\linewidth]{Figures/Appendix.RicePaddies.B1.png}
        
        {(d) $\hat\beta_1(s)$}
    \end{subfigure}
    \caption{Diagnostic Maps: Rice paddies}\vspace{4pt}
    \label{fig:diagnostic-maps-RicePaddies}
\end{figure}

\nocite{*}
\bibliography{References}

\end{document}